\documentclass[10pt,a4paper]{article}
\usepackage[numbers,sort&compress]{natbib}
\usepackage{authblk}
\usepackage{graphicx}
\usepackage[top=30truemm,bottom=30truemm,left=25truemm,right=25truemm]{geometry}
\usepackage[colorlinks=true,linktoc=page,citecolor=red,linkcolor=blue]{hyperref}
\graphicspath{{./figs/}}
\usepackage{color,comment}
\usepackage{amsmath,amssymb,amsfonts}
\usepackage{bm,braket,array}
\usepackage[all]{xy}
\usepackage{amsthm,amscd}
\usepackage{mdframed}
\usepackage{slashed}
\usepackage[format=hang,justification=raggedright]{caption}

\theoremstyle{definition}

\theoremstyle{remark}

\newcommand{\p}{\partial}

\newcommand{\wh}{\widehat}

\newcommand{\Hom}{{\rm Hom}}

\newcommand{\Mat}{{\rm Mat}}

\newcommand{\s}{{\sigma}}

\newcommand{\C}{\mathbb{C}}

\newcommand{\R}{\mathbb{R}}
\newcommand{\Z}{\mathbb{Z}}

\def\widebar{\accentset{{\cc@style\underline{\mskip10mu}}}}
\def\wideubar{\underaccent{{\cc@style\underline{\mskip10mu}}}}
\newcommand{\pet}[1]{\left.\middle| #1 \right)}

\begin{document}
\title{Equivariant Parameter Families of Spin Chains: \\ A Discrete MPS Formulation~\thanks{YITP-25-114}}
\author{Ken Shiozaki}
\affil{Center for Gravitational Physics and Quantum Information, Yukawa Institute for Theoretical Physics, Kyoto University, Kyoto 606-8502, Japan}
\date{\today}
\maketitle
\begin{abstract}
We analyze topological phase transitions and higher Berry curvature in one-dimensional quantum spin systems, using a framework that explicitly incorporates the symmetry group action on the parameter space. Based on a $G$-compatible discretization of the parameter space, we incorporate both group cochains and parameter-space differentials, enabling the systematic construction of equivariant topological invariants. We derive a fixed-point formula for the higher Berry invariant in the case where the symmetry action has isolated fixed points. This reveals that the phase transition point between Haldane and trivial phases acts as a monopole-like defect where higher Berry curvature emanates. We further discuss hierarchical structures of topological defects in the parameter space, governed by symmetry reductions and compatibility with subgroup structures.
\end{abstract}
\tableofcontents

\section{Introduction}

In an isolated quantum many-body system, symmetry is expressed by the requirement that the Hamiltonian $\hat H$, which governs the time evolution of the system, satisfies
\begin{align}
\hat g\, \hat H\, \hat g^{-1} = \hat H,\quad g \in G,
\label{eq:ham_sym}
\end{align}
for some group $G$, where $\hat g$ denotes a unitary or antiunitary representation of $G$ on the Hilbert space. In this work, we relax this symmetry condition by parameter-dependent Hamiltonians $\hat H(\tau)$, where $\tau \in \mathcal{M}$ and the parameter space $\mathcal{M}$ carries a $G$-action. A family of Hamiltonians $\hat H(\tau)$ is said to be \emph{$G$-equivariant} if it satisfies
\begin{align}
\hat g\, \hat H(\tau)\, \hat g^{-1} = \hat H(g \tau),\quad g \in G,
\label{eq:gr_action}
\end{align}
where $\tau \mapsto g \tau$ defines a left action of $G$ on $\mathcal{M}$, i.e., $(gh)\tau = g(h\tau)$.

For example, consider the one-dimensional Heisenberg model with a uniform magnetic field $\bm{h}$:
\begin{align}
\hat H(J,\bm{h}) = J \sum_{j\in \Z} \hat{\mathbf{S}}_j \cdot \hat{\mathbf{S}}_{j+1} + \bm{h} \cdot \sum_{j\in \Z} \hat{\mathbf{S}}_j\,.
\end{align}
This model has time-reversal symmetry $\hat T = \big(\otimes_j e^{i \pi S^y_j}\big)\mathcal{K}$ (where $\mathcal{K}$ denotes complex conjugation) only at $\bm{h} = \bm{0}$. However, even for $\bm{h} \neq \bm{0}$, it satisfies 
\begin{align}
\hat T\, \hat H(J,\bm{h})\, \hat T^{-1} = \hat H(J, -\bm{h}),
\end{align}
meaning that the family of Hamiltonians is $\Z_2$-equivariant under $(J,\bm{h}) \mapsto (J,-\bm{h})$ in the space of coupling constants.

When a Hamiltonian $\hat H$ has a finite energy gap between the ground state and the first excited state (i.e., when the system is gapped), it belongs to a certain topological phase, characterized by a topological invariant. Similarly, for a family of gapped Hamiltonians $\hat H(\tau)$, one can define a topological invariant associated with the family itself. In particular, for so-called invertible states, gapped systems with a unique ground state and no spontaneous symmetry breaking (SSB), the topological classification of families is well understood through the classification of adiabatic cycles and their generalization, known as the Ω-spectrum structure~\cite{TeoTopological2010,KitaevTopological2011,KitaevClassification2013,KitaevHomotopytheoretic2015,KubotaStable2025,GaiottoSymmetry2019,XiongMinimalist2018,RoyFloquet2017,KikuchiGlobal2017,CordovaAnomalies2020,CordovaAnomalies2020a,HsinBerry2020,KapustinHigherdimensional2020a,ShiozakiAdiabatic2022,YaoGappability2022,ChoiHigher2022,BachmannClassification2024,OhyamaDiscrete2024,DebrayLong2024,BeaudryHomotopical2024,BeaudryClassifying2025,ManjunathAnomalous2025,JonesCharge2025}.

For example, in one-dimensional spin chains with the parameter space $S^1$, if the Hamiltonian satisfies periodicity $\hat H(\theta+2\pi) = \hat H(\theta)$, a nontrivial topological invariant of the family implies the adiabatic pumping of a $G$-symmetry charge during time evolution. This phenomenon, known as the Thouless pump~\cite{ThoulessQuantization1983}, describes the quantized transport of a physical quantity (such as electric charge or spin) per cycle when a periodic parameter is varied adiabatically and topologically nontrivially.

Even in the absence of symmetry, it is known that for one-dimensional invertible systems with a three-dimensional, oriented, closed parameter space such as the 3-sphere $S^3$, one can define an integer-valued topological invariant from the family $\hat H(\tau)$, given by the integral of a so-called higher Berry curvature~\cite{KapustinHigherdimensional2020,KapustinLocal2022,WenFlow2023,QiCharting2025,OhyamaHigher2024,ShiozakiHigher2025,OhyamaHigher2025,OhyamaHigher2025a,SommerHigher2025,SommerHigher2025a,WenSpace2025,ChoiHigher2025}. Mathematically, this invariant corresponds to the Dixmier--Douady class associated with a gerbe connection. The concept of higher Berry curvature was introduced by Kapustin and Spodyneiko~\cite{KapustinHigherdimensional2020}. See also \cite{KitaevDifferential2019} for an earlier proposal related to this idea. In this paper, we refer to this topological invariant as the Dixmier--Douady--Kapustin--Spodyneiko (DDKS) number\footnote{See also \cite{ViennotNew2011} for a prior proposal of gerbe structures in open quantum systems.}.

In this paper, we systematically investigate the relationship between topological invariants of families of Hamiltonians and the role of the $G$-equivariant structure in one-dimensional quantum spin systems, focusing on invertible states. In particular, when the group action on the parameter space has fixed points, the Hamiltonians at these high-symmetry points satisfy the standard symmetry condition \eqref{eq:ham_sym} and can be assigned well-defined symmetry-protected topological (SPT) invariants~\cite{PollmannSymmetry2012,ChenClassification2011,SchuchClassifying2011,PollmannDetection2012}. We demonstrate that nontrivial relations can exist between the topological invariant associated with the entire family of Hamiltonians and the SPT invariants defined at such high-symmetry points. In this paper, we formalize these relations. As an application, we prove that the phase transition point between the Haldane phase and the trivial phase, protected by either time-reversal symmetry or $\mathbb{Z}_2 \times \mathbb{Z}_2$ symmetry, acts as a source of higher Berry curvature.

As a methodological framework, we follow Ref.~\cite{ShiozakiHigher2025} and adopt a discrete formulation based on matrix product state (MPS) representations~\cite{Perez-GarciaMatrix2007}, assuming translational invariance. The parameter space is discretized into a sufficiently fine mesh compared to the variation of the Hamiltonian, and topological invariants are constructed from MPS data assigned to the vertices. A key advantage of this formulation is that gauge invariance is manifest, and it can be straightforwardly implemented numerically, for instance using DMRG.

We note that the assumption of translational symmetry is a technical simplification to facilitate the formulation; a generalization to non-translationally invariant systems remains an important direction for future work.

The structure of this paper is as follows.  
Sec.~2, as a preliminary to the formulation for one-dimensional spin systems, reviews the relationship between topological invariants of families of pure states in quantum mechanical systems and group actions, particularly describing a discrete formulation based on a triangulation of the parameter space.  
In Sec.~3, using the injective MPS, we carry out a discrete formulation with group action for parameter families of MPSs.  
In Sec.~4, as an application of Sec. 3, we derive a fixed-point formula for the DDKS number and some new invariants introduced by group action.  
Sec. 5~summarizes general aspects of defect structures in theory space.  
Finally, Sec.~6 provides a summary and outlook of this work.  
Appendix~\ref{app:double_complex} summarizes necessary technical background on simplicial $G$-complexes.

\section{Warm up: Equivariant families of pure states}
\label{sec:0d}

Before moving on to the main discussion, we review a well-known example in the context of a 0-dimensional system to illustrate the relationship between the topological number of a family and high-symmetry points. In addition, we describe its discrete formulation with manifest gauge invariance, and further mention the relationship between level crossings and the sources of Berry curvature in the parameter space~\footnote{Such $G$-equivariant families of quantum mechanical systems are also widely known as a setting of symmetry constraint for Bloch Hamiltonians in band theory.}.

Consider a Hamiltonian $\hat H(\tau)$ of an $N$-level system that depends on a point $\tau$ in the parameter space ${\cal M}$,  that is, a family of $N \times N$ Hermitian matrices parametrized by $\tau \in {\cal M}$. Let $G$ be a discrete group, and denote its left action on ${\cal M}$ by $(g, \tau) \mapsto g \tau$, which is associative: $(gh)\tau = g(h\tau)$. We further fix a unitary linear representation $\hat g$ of $G$ on the $N$-level system:
\begin{align}
\hat g \hat h = \widehat{gh},\quad g,h \in G. 
\label{eq:gr_str}
\end{align}
We allow $\hat g$ to be antiunitary as well, by specifying the homomorphism
\begin{align}
\phi: G \to \mathbb{Z}_2 = \{\pm 1\}
\label{eq:0d_phi_g}
\end{align}
so that $\hat g$ is unitary when $\phi_g = 1$, and antiunitary when $\phi_g = -1$. Namely, for the imaginary unit $i$, we have $\hat g i = \phi_g i \hat g$. Here, we assume that $\hat g$ does not depend on the point in the parameter space ${\cal M}$~\footnote{More generally, $\hat g$ may depend on the point in ${\cal M}$, and one may also allow projective representations with two cocycle $z_{g,h} \in U(1)$ such that $\hat g \hat h = z_{g,h} \widehat{gh}$. Furthermore, even without assuming the non-degeneracy of the ground state, a topological invariant can still be defined, and its classification is given by a certain type of $K$-theory~\cite{FreedTwisted2013,ThiangKTheoretic2016,ShiozakiTopological2017}.}.

We assume that the Hamiltonian $\hat H(\tau)$ satisfies the following $G$-equivariant constraint:
\begin{align}
\hat g \, \hat H(\tau) \, \hat g^{-1} = \hat H(g \tau), \quad \tau \in {\cal M}, \quad g \in G.
\label{eq:0d_gstr}
\end{align}
We focus on the ground state $\ket{\psi(\tau)}$ of $\hat H(\tau)$, which satisfies $\hat H(\tau)\ket{\psi(\tau)} = E_{\min}(\tau)\ket{\psi(\tau)}$ and assume that no level crossing (i.e., degeneracy with excited states) occurs throughout the parameter space ${\cal M}$. (The same formulation does work for any excited states unless a level crossing occurs.) Alternatively, one may consider that ${\cal M}$ is a punctured space obtained by removing regions that include level crossing points, i.e., the defect regions. 

From the viewpoint of symmetries in quantum mechanics, the $U(1)$ phase factor of the unitary linear representation $\hat g$ cannot be fixed. In general, there is an ambiguity
\begin{align}
\hat g \mapsto e^{i\theta_g} \hat g,\quad g \in G.
\label{eq:0d_hat_g_amb}
\end{align}
in $\hat g$ up to 1-dimensional representations $e^{i\theta} \in \Hom(G,U(1))$ of $G$, satisfying $e^{i(\theta_g+\phi_g\theta_h)} = e^{i\theta_{gh}}$ for $g,h \in G$.

In what follows, we denote by $d$ the differential along the simplicial complex ${\cal M}$, and by $\delta$ the differential twisted by $\phi$ in the direction of group cochains. See App.~\ref{app:double_complex} for details.

\subsection{Discrete formulation}

Below, we describe the discrete formulation for the topological number of the family of Hamiltonians $\hat H(\tau)$. (In the next subsection~\ref{sec:0d_group_action}, we will introduce the group action.) We divide the parameter space ${\cal M}$ into small simplices (triangulate it), denoting the vertices by $\tau_j$, etc. In the following, for a $q$-simplex $\Delta^q = (\tau_0,\dots,\tau_q)$, we abbreviate the action of $g$ as $g\Delta^q := (g\tau_0,\dots,g\tau_q)$. Similarly, we write $gc$ for the action of $g$ on any $q$-chain $c$.

Assign a state $|\psi(\tau_0)\rangle$ to each vertex $\tau_0$. The $U(1)$ phase of $|\psi(\tau_0)\rangle$ can be chosen arbitrarily and independently at each vertex. For any edge (1-simplex) $\Delta^1 = (\tau_0,\tau_1)$, define the (discrete) Berry connection along that edge as the $U(1)$ phase of the inner product between the states at its endpoints:
\begin{align}
{\cal A}(\Delta^1) := \arg \Big( \langle \psi(\tau_0) \mid \psi(\tau_1) \rangle \Big) \in \R/2\pi\Z,\quad \Delta^1 = (\tau_0,\tau_1)\,.
\label{eq:0d_DBC}
\end{align}
Note that reversing the orientation of an edge $\Delta^1=(\tau_0,\tau_1)$ (i.e., considering $-\Delta^1 = (\tau_1,\tau_0)$) flips the sign of the Berry connection~\footnote{Hereafter, for any quantity valued in $\R/2\pi\Z$, the equality sign “$=$” is always understood as an equality modulo $2\pi$.}:${\cal A}(-\Delta^1) = -{\cal A}(\Delta^1)$. Under the gauge transformation 
\begin{align}
|\psi(\tau_0)\rangle \mapsto |\psi(\tau_0)\rangle\, e^{i\chi(\tau_0)}\,,
\label{eq:0d_gauge_tr}
\end{align}
the Berry connection changes as ${\cal A}(\Delta^1) \mapsto {\cal A}(\Delta^1) + d\chi(\Delta^1)$, where $d\chi(\Delta^1) = \chi(\partial \Delta^1) = \chi(\tau_1) - \chi(\tau_0)$. For any loop (1-cycle) $\ell = (\tau_1,\tau_2)+(\tau_2,\tau_3)+\cdots+(\tau_N,\tau_1)$, define the $\R/2\pi \Z$-valued (discrete) Berry phase $\gamma(\ell)$ as the sum of Berry connection along the loop:
\begin{align}
\gamma(\ell) := \sum_{\Delta^1 \in \ell} {\cal A}(\Delta^1)\,.
\label{eq:dis_BP}
\end{align}
Nice properties of this definition are that invariance under the gauge transformation (\ref{eq:0d_gauge_tr}) is manifest, and it is easily implemented numerically. Reversing the orientation of a closed loop $\ell$ flips the sign of the Berry phase: $\gamma(-\ell) = -\gamma(\ell)$. For any triangle (2-simplex) $\Delta^2 = (\tau_0,\tau_1,\tau_2)$, the Berry phase around its boundary loop $\partial \Delta^2 = (\tau_1,\tau_2) - (\tau_0,\tau_2) + (\tau_0,\tau_1)$ is, if the triangle $\Delta^2$ is sufficiently small, close to $0$ mod $2\pi$, so there is no branch ambiguity and one can take a logarithm. We thus define the Berry flux (the integral of Berry curvature) on the triangle $\Delta^2$ as a real number by lifting the $\R/2\pi\Z$-valued Berry phase to a branch $(-\pi,\pi]$: writing this lift as 
\begin{align}
\R/2\pi\Z \ni \theta \mapsto \tilde{\theta} \in \R, \quad -\pi \le \tilde{\theta} < \pi. 
\end{align}
With this, we define the Berry flux 
\begin{align}
{\cal F}(\Delta^2) := \widetilde{\,d{\cal A}(\Delta^2)\,}, 
\end{align}
where $d{\cal A}(\Delta^2) = {\cal A}(\partial \Delta^2) = {\cal A}(\tau_0,\tau_1)+{\cal A}(\tau_1,\tau_2)+{\cal A}(\tau_2,\tau_0)$. By definition, note that ${\cal F}(\Delta^2) \equiv d{\cal A}(\Delta^2)$ modulo $2\pi$ holds. For any oriented closed 2D surface (2-cycle) $\Sigma \subset {\cal M}$, the Chern number (a topological number of the family) is defined as the sum of Berry fluxes on $\Sigma$~\cite{FukuiChern2005}:
\begin{align}
\nu(\Sigma) := \frac{1}{2\pi} \sum_{\Delta^2 \in \Sigma} {\cal F}(\Delta^2) \in \Z\,.
\end{align}
The quantization of this value is evident from the equality 
\begin{align}
\sum_{\Delta^2 \in \Sigma} {\cal F}(\Delta^2)
\equiv \sum_{\Delta^2 \in \Sigma} d{\cal A}(\Delta^2) \equiv \sum_{\Delta^1 \in \partial \Sigma} {\cal A}(\Delta^1) \equiv 0 \quad \bmod\, 2\pi. 
\end{align}
Note that the Chern number flips sign under orientation reversing of $\Sigma$: $\nu(-\Sigma) = -\,\nu(\Sigma)$.

\subsection{Group action}
\label{sec:0d_group_action}

Next, let us implement the $G$-equivariant structure. We choose the triangulation so that it is compatible with the $G$-action on the parameter space ${\cal M}$. Namely, we require:
\begin{itemize}
  \item The group $G$ acts on the vertex set $\{\tau_j\}_j$.
  \item For each $q$-simplex $\Delta^q_a$ labeled by $a$, the action of a group element $g \in G$ is either:
  \begin{itemize}
    \item to fix the interior of $\Delta^q_a$ pointwise (i.e. $g\tau = \tau$ for all $\tau \in \mathring{\Delta}^q_a$), or
    \item to map $\Delta^q_a$ to another $q$-simplex (i.e. $g(\mathring{\Delta}^q_a) = \mathring{\Delta}^q_{b}$ with some $b$).
  \end{itemize}
\end{itemize}
A triangulation obtained in this way is compatible with a $G$-simplicial structure, and the set of $q$-simplices is invariant under $G$.

The $G$-equivariance (\ref{eq:gr_action}) means that for any parameter point $\tau_0$, the state $\hat g|\psi(\tau_0)\rangle$ is the same up to a $U(1)$ phase as the state $|\psi(g\tau_0)\rangle$ at the point $g\tau_0$. Therefore, we have the equality 
\begin{align}
\hat g\,|\psi(\tau_0)\rangle = |\psi(g\tau_0)\rangle\, e^{i\alpha_g(\tau_0)}, \quad g \in G\,,
\label{eq:0d_G_phase}
\end{align}
thereby introducing a $U(1)$ phase $e^{i\alpha_g(\tau_0)} \in U(1)$. Under the gauge transformation (\ref{eq:0d_gauge_tr}), $\alpha_g(\tau_0)$ changes as
\begin{align}
\alpha_g(\tau_0) \mapsto \alpha_g(\tau_0) + (\delta \chi)_g(\tau_0) = \alpha_g(\tau_0) + \phi_g\,\chi(\tau_0) - \chi(g\tau_0)\,,
\end{align}
and it satisfies the cocycle condition
\begin{align}
(\delta \alpha)_{g,h}(\tau_0) = \phi_g\,\alpha_h(\tau_0) - \alpha_{gh}(\tau_0) + \alpha_g(g\tau_0) = 0\,.
\label{eq:0d_10_cocycle}
\end{align}
In particular, for group elements of the stabilizer $G_{\tau_0} = \{g \in G \mid g\tau_0 = \tau_0\}$ of $\tau_0$ (so that $g$ leaves the point $\tau_0$ invariant) which are unitary ($\phi_g=1$), the $U(1)$ phase 
\begin{align}
e^{i\alpha_g(\tau_0)} = \langle \psi(\tau_0)\mid \hat g \mid \psi(\tau_0)\rangle,\quad 
g \in G_{\tau_0},\ \phi_g=1,
\end{align}
is a one-dimensional representation (“charge”) of $G_{\tau_0}$ and is gauge-invariant.

The Berry connection ${\cal A}(\Delta^1)$ and $\alpha_g(\tau_0)$ satisfy the following “descendant equation”:
\begin{align}
\delta {\cal A} \equiv d \alpha\,.
\end{align}
In components:
\begin{align}
(\delta {\cal A})_g(\Delta^1)
&= \phi_g\, {\cal A}(\Delta^1) - {\cal A}(g\Delta^1) \nonumber \\
&= \alpha_g(\tau_1) - \alpha_g(\tau_0) = d\alpha_g(\Delta^1)\,.
\label{eq:0d_A_theta_rel}
\end{align}
Furthermore, from (\ref{eq:0d_A_theta_rel}) we find for the Berry flux that $(\delta {\cal F})_g(\Delta^2) = \phi_g\, {\cal F}(\Delta^2) - {\cal F}(g\Delta^2) = 0$, i.e.
\begin{align}
{\cal F}(g\Delta^2) = \phi_g\, {\cal F}(\Delta^2)\,.
\end{align}
Therefore, the action of $g$ on the Chern number is
\begin{align}
\nu(g\Sigma) = \phi_g\, \nu(\Sigma)\,.
\label{eq:0d_Ch_group_action}
\end{align}
It follows that if there exists an antiunitary element $g \in G$ such that a closed surface $\Sigma$ is invariant as a set (not necessarily pointwise) under $g$, or if there exists a unitary $g$ that reverses the orientation of $\Sigma$, then the Chern number is zero:
\begin{align}
\exists\, g \in G \ \text{s.t.}\quad 
g\Sigma = -\,\phi_g \Sigma \quad \text{then} \quad 
\nu(\Sigma) = 0\,.
\end{align}

Similarly, integrating both sides of (\ref{eq:0d_A_theta_rel}) over a closed loop $\ell$, the contributions on the right-hand side cancel, yielding the action of $g$ on the Berry phase:
\begin{align}
\gamma(g\ell) = \phi_g\, \gamma(\ell),\quad g \in G.
\end{align}
Therefore, the same constraint arises as the Chern number:
\begin{align}
\exists\, g \in G \ \text{s.t.}\quad 
g\ell = -\,\phi_g \ell \quad \text{then} \quad 
\gamma(\ell) \in \{0,\pi\}.
\label{eq:0d_BP_quantized_OP}
\end{align}

\subsection{Fixed-point formulas}

In the following, we describe examples of fixed-point formulas for the quantized Berry phase and Chern number. 

\begin{figure}[!t]
\centering
\includegraphics[width=\linewidth]{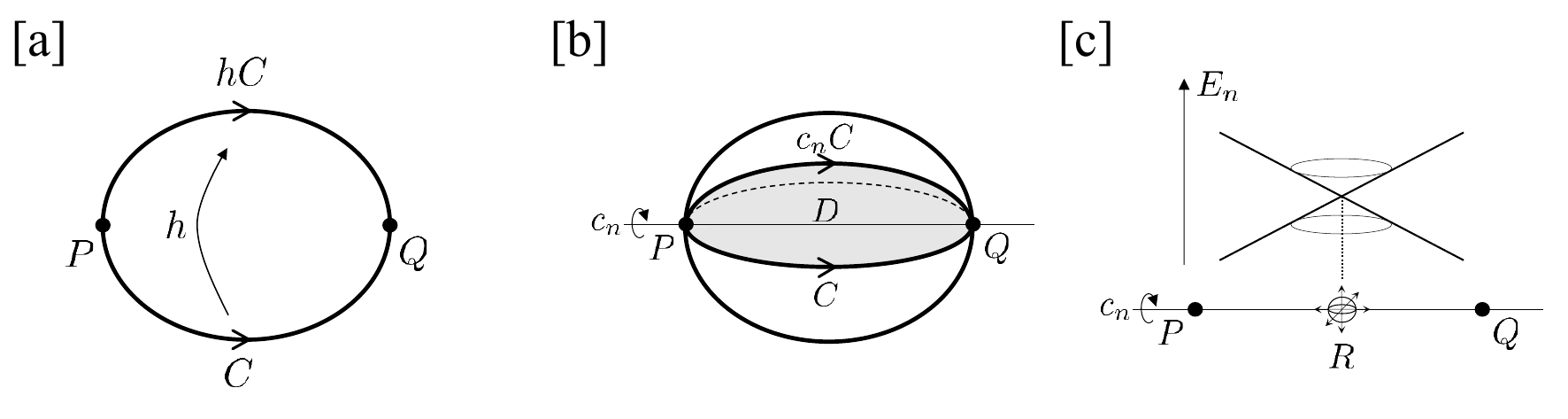}
\caption{[a] An arc $C$ together with its $h$-image forms a closed path. 
[b] A $\Z_n$ rotation $c_n$ on a 2-sphere $S^2$ has fixed points $P,Q$. $D$ represents a disk region independent under the $\Z_n$ action. 
[c] Horizontal axis represents the parameter space, while the vertical axis represents energy eigenvalues. A level crossing corresponds to a source (monopole) of Berry curvature in the parameter space.}
\label{fig:0d_fixed_pt}
\end{figure}

\subsubsection{Fixed-point formula for the Berry phase}
\label{sec:BP_fixed_pt}

Consider the case in Eq.~(\ref{eq:0d_BP_quantized_OP}) where $g$ is unitary, i.e. suppose there exists $h \in G$ with $\phi_h=1$ such that $h\ell = -\ell$. Then there exist two fixed points $P, Q$ of the group element $h$ (see Fig.~\ref{fig:0d_fixed_pt}[a]). In general, the integration path $\ell$ need not be smooth; here, more generally, consider the case where the integration path is given by $\ell = C - hC$, the linear sum of an independent arc region $C$ and its $h$-image. Then, from the descendant equation (\ref{eq:0d_A_theta_rel}),
\begin{align}
\gamma(\ell) 
&= \sum_{\Delta^1 \in C} \big( {\cal A}(\Delta^1) - {\cal A}(h\Delta^1) \big) \nonumber \\
&= \sum_{\Delta^1 \in C} \delta A(h\Delta^1) \nonumber \\
&= \sum_{\Delta^1 \in C} d\alpha_h(\Delta^1) \nonumber \\
&= \alpha_h(Q) - \alpha_h(P)\,. 
\end{align}
That is, 
\begin{align}
e^{i\gamma(\ell)} 
= \frac{\langle \psi(Q)|\, \hat h\, | \psi(Q)\rangle}{\langle \psi(P)|\, \hat h\, | \psi(P)\rangle}\,.
\label{eq:0d_BP_fixed_pt_formula}
\end{align}
Therefore, the quantized Berry phase is determined solely by the expectation value of $\hat h$ at the fixed points. Note that, being a ratio, there is no ambiguity in the 1-dimensional representation of $\hat h$ (see Eq.~(\ref{eq:0d_hat_g_amb})). If $h^n = e$, then the Berry phase is quantized to $\Z_n$-values.

\subsubsection{Fixed-point formula for the Chern number}
\label{sec:Chern_mod_n}

Suppose that $\Z_n$ acts on part of the parameter space $S^2 \in {\cal M}$ as an “$C_n$” rotation. That is, using spherical coordinates $(\theta,\phi)$, consider the action $c_n(\theta,\phi) = (\theta,\phi+\frac{2\pi}{n})$, where $c_n$ denote the generator of $\Z_n$. The points $\theta=0,\pi$ are fixed points of the $\Z_n$ action; call them $P,Q$. Assume the generator $c_n$ is unitary, so $\phi_{c_n}=1$. The Berry curvature does not flip the sign under the $\Z_n$ action: $F(c_n \Delta^2) = {\cal F}(\Delta^2)$, so the Chern number on the sphere equals $n$ times the integral over an independent domain $D$. The boundary of the independent domain $D$ consists of an arc $C$ and its image $c_n C$, and taking orientation into account, we can write $\partial D = C - c_n C$ (see Fig.~\ref{fig:0d_fixed_pt}[b]). Then, modulo $n$, the Chern number can be expressed as the Berry phase along the boundary of $D$, to which the fixed-point formula (\ref{eq:0d_BP_fixed_pt_formula}) can be applied. Therefore, we obtain:
\begin{align}
\frac{1}{n} \times 2\pi\, \nu(S^2) 
&\underset{\bmod 2\pi}{\equiv} \sum_{\Delta^2 \in D} d{\cal A}(\Delta^2) \nonumber \\
&= \sum_{\Delta^1 \in \partial D} {\cal A}(\Delta^1) \\
&= (\alpha_{c_n}(Q) - \alpha_{c_n}(P))\,,
\end{align}
and thus the mod $n$ relation characterizing the Chern number $\nu(S^2)$~\cite{FangBulk2012}:
\begin{align}
e^{\frac{2\pi i\, \nu(S^2)}{n}}
= \frac{\langle \psi(Q)|\, \hat c_n\, | \psi(Q)\rangle}{\langle \psi(P)|\, \hat c_n\, | \psi(P)\rangle}\,.
\label{eq:Ch_mod_n}
\end{align}

\subsection{Relationship with level crossings}

At a point in the parameter space where the $\hat g$-charge changes, a level crossing may appear, and such level crossing points should accompany the source of the Berry curvature (monopole) in the parameter space, which is consistent with Eq.~(\ref{eq:Ch_mod_n}) as discussed below. For the inner products of the states at the fixed points $P, Q$ of $c_n$, note the following relation:
\begin{align}
\big( e^{i\alpha_{c_n}(Q)} - e^{i\alpha_{c_n}(P)} \big)\, 
\langle \psi(P)|\psi(Q)\rangle = 0\,.
\end{align}
Therefore, if the $c_n$-eigenvalues at $P$ and $Q$ are different, the states are orthogonal, and there must exist a level crossing point $R$ on any $c_n$-invariant path connecting $P$ to $Q$. On the other hand, by breaking the $c_n$ symmetry, it is possible to construct a path that avoids the level crossing point $R$. In this case, Eq.~(\ref{eq:Ch_mod_n}) indicates that if there exists a 2-sphere $S^2$ surrounding the level crossing point $R$ along the path connecting $P$ and $Q$, then the integral of the Berry curvature over $S^2$ is
\begin{align}
\frac{n}{2\pi}\big( \alpha_{c_n}(Q) - \alpha_{c_n}(P) \big) \mod n\,.
\end{align}
In other words, a level crossing point $R$ protected by $c_n$ symmetry is a source of Berry curvature. (See Fig.~\ref{fig:0d_fixed_pt}[c].)

As an example, using the spin operator $\hat{\mathbf{S}} = (\hat S_x, \hat S_y, \hat S_z)$ with spin-$S$ representation ($S \in \{\frac{1}{2},1,\frac{3}{2},2,\dots\}$), consider the Hamiltonian depending on a three-dimensional magnetic field $\bm{h}$:
\begin{align}
\hat H(\bm{h}) = \bm{h} \cdot \hat{\mathbf{S}}.
\label{eq:0d_spin_S_model}
\end{align}
Introduce the $c_n$ matrix by $\hat c_n = e^{-i \frac{2\pi}{n} \hat S_z}$ corresponding to a $2\pi/n$ rotation about the $z$-axis. Then $\Z_n$-equivariant relation is given by 
\begin{align}
\hat c_n\, \hat H(\bm{h})\, \hat c_n^{-1} = \hat H(R_n \bm{h})\,,
\end{align}
where $R_n$ is the rotation matrix for a $2\pi/n$ rotation in the $xy$-plane:
\begin{align}
R_n =
\begin{pmatrix}
\cos \frac{2\pi}{n} & -\sin \frac{2\pi}{n} & 0,\\
\sin \frac{2\pi}{n} &  \cos \frac{2\pi}{n} & 0,\\
0 & 0 & 1
\end{pmatrix}.
\end{align}
The fixed points of this $c_n$ action are those with zero $xy$-components, i.e. $\bm{h} = (0,0,h_z)$. In particular, $\bm{h} = \bm{0}$ is a point where all levels are degenerate (a level crossing point). In the vicinity of this point, $\bm{h} = (0,0,\pm \epsilon)$ with $\epsilon>0$, the difference in the $c_n$-eigenvalues of the lowest energy states depends on the spin $S$:
\begin{align}
\frac{n}{2\pi} \Big( \alpha_{c_n}(0, 0, \epsilon) - \alpha_{c_n}(0, 0, -\epsilon) \Big) = 2S\,.
\end{align}
On the other hand, this is consistent with the fact that in the model~(\ref{eq:0d_spin_S_model}), the Chern number obtained by integrating the Berry curvature of the ground state on the sphere $|\bm{h}| = \epsilon$ is $2S$.

\subsection{Example of a topological number defined by a group action}

Neither the Berry phase $\gamma(\ell)$ nor the Chern number $\nu(\Sigma)$ requires a group action in their definitions. In this subsection, as an example of a topological invariant that can only be defined in the presence of a group action, we introduce a topological invariant defined under a free $\mathbb{Z}_2$ action.

\begin{figure}[!t]
\centering
\includegraphics[width=\linewidth]{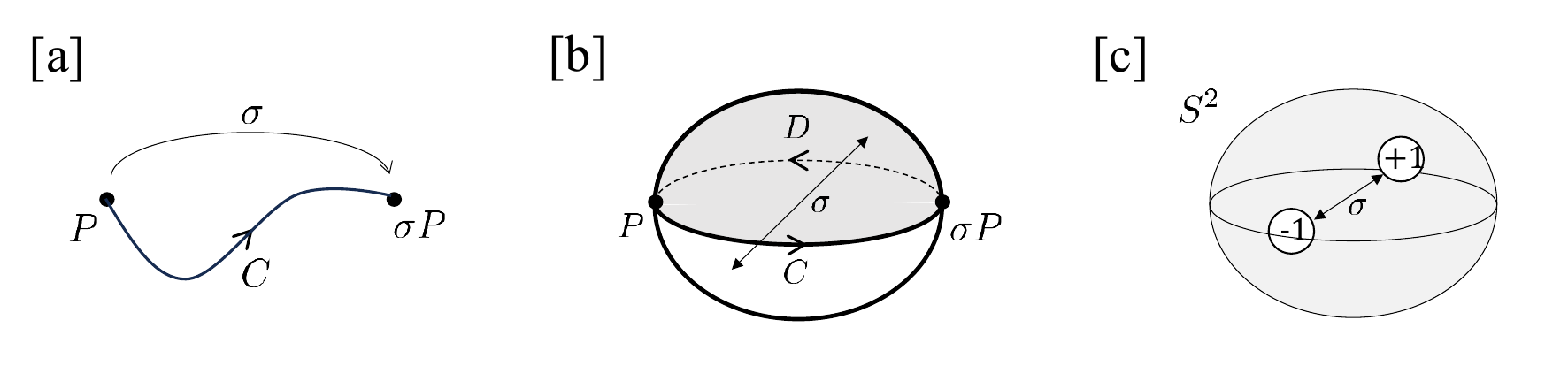}
\caption{[a] The endpoint of an arc $C$ is related by a group action $\sigma$. 
[b] A $\sigma$ action on a 2-sphere $S^2$ that sends a point $\bm{n} \in S^2$ to its antipodal point $-\bm{n}$. $D$ is the upper hemisphere, and its boundary $\partial D$ is the union of an arc $C$ and its image $\sigma C$ under the $\Z_2$ action. $P$ denotes the starting point of arc $C$. 
[c] Inside the sphere $S^2$, a pair of unit monopoles with opposite charge cannot annihilate due to the $\Z_2$-equivariant structure.}
\label{fig:0d_z2_number}
\end{figure}

First, consider an arbitrary path $C$ connecting two points $P, \sigma P$ that are related by the group action $\s \in G$. If $\sigma$ is a unitary element (i.e. $\phi_\sigma = 1$), then even if the integration path is not closed, one can define a gauge-invariant Berry phase as:
\begin{align}
e^{i\gamma(C, \sigma)} := \left( \prod_{\Delta^1 \in C} e^{i {\cal A}(\Delta^1)} \right)\, \langle \psi(\sigma P) |\, \hat \sigma\, | \psi(P)\rangle\,.
\end{align}
(See Fig.~\ref{fig:0d_z2_number}[a] for the integration path $C$.) On the other hand, the symmetry operation $\hat g$ always has an ambiguity corresponding to a one-dimensional representation of $G$ (see Eq.~(\ref{eq:0d_hat_g_amb})), and the Berry phase $e^{i\gamma(C,g)}$ shares the same ambiguity. Therefore, the Berry phase $e^{i\gamma(C, g)}$ should be considered as a relative geometric quantity: it is only well-defined for a pair of families of pure states $\big(|\psi(\tau)\rangle,\, |\psi'(\tau)\rangle\big)$ over a common Hilbert space and group action. 

Now we proceed to construct a topological invariant. Take the parameter space to be the 2-sphere $S^2$, with the unit vector $\bm{n} \in \mathbb{R}^3$ ($|\bm{n}| = 1$) as the coordinate on the sphere. Define the action of the generator $\sigma$ of $\mathbb{Z}_2$ as the inversion in $\R^3$: $\bm{n} \mapsto \sigma \bm{n} := -\bm{n}$. Under this action, $\sigma$ is closed on $S^2$ and reverses the orientation $\sigma S^2 = -S^2$. Take $\hat \sigma$ to be unitary ($\phi_\sigma = 1$). In this situation, by Eq.~(\ref{eq:0d_Ch_group_action}), the Chern number on $S^2$ is zero $\nu(S^2) = 0$. Also, for any closed path $\ell$ on $S^2$ that is invariant under $\sigma$ (for example, a closed curve lying along the equator $n_3 = 0$), since $\sigma$ preserves the orientation of the path, the corresponding Berry phase $\gamma(\ell)$ is not quantized. 

Nevertheless, the $\mathbb{Z}_2$-equivariant structure allows a new $\mathbb{Z}_2$-valued topological number to be defined. Consider the upper hemisphere of $S^2$ as a fundamental domain independent under $\sigma$:
\begin{align}
D := \{\, \bm{n} \in S^2 \mid n_3 \ge 0 \,\}\,,
\end{align}
and decompose its boundary under $\sigma$ as $\partial D = C + \sigma C$. Let the starting point of arc $C$ be $P$, i.e. $\partial C = \sigma P - P$. (See Fig.~\ref{fig:0d_z2_number}[b].) We then define a $\mathbb{Z}_2$ number $\xi(S^2,\sigma) \in \{0,\pi\}$ as~\cite{FreedDeterminants1986,ShiozakiTopological2017}:
\begin{align}
e^{i\, \xi(S^2,\sigma)}
:= \left( \prod_{\Delta^2 \in D} e^{\frac{i}{2} {\cal F}(\Delta^2)} \right)\, \times\, e^{-i\, \gamma(C,\sigma)}\,.
\end{align}
Equivalently,
\begin{align}
\xi(S^2,\sigma) := \frac{1}{2} \sum_{\Delta^2 \in D} {\cal F}(\Delta^2)
- \sum_{\Delta^1 \in C} {\cal A}(\Delta^1) + \alpha_\sigma(P)\,. 
\end{align}
It is easily checked that $\xi(S^2,\sigma)$ is gauge-invariant. Furthermore, the $\mathbb{Z}_2$ quantization (i.e. $\xi(S^2,\sigma) \in \{0,\pi\}$) is ensured by the following:
\begin{align}
2\, \xi(S^2,\sigma)
&= \sum_{\Delta^2 \in D} {\cal F}(\Delta^2) 
- 2 \sum_{\Delta^1 \in C} {\cal A}(\Delta^1)
+ 2 \alpha_\sigma(P) \nonumber \\
&= \sum_{\Delta^2 \in D} {\cal F}(\Delta^2)
- \sum_{\Delta^1 \in C} \big( {\cal A}(\Delta^1) + {\cal A}(\sigma \Delta^1) - d\alpha_\sigma(\Delta^1) \big) \nonumber \\
&\underset{\bmod 2\pi}{\equiv} \alpha_\sigma(\sigma P) + \alpha_\sigma(P) = 0\,.
\end{align}
In the final line, we used the cocycle condition~(\ref{eq:0d_A_theta_rel}) for $\alpha_\sigma$. This $\mathbb{Z}_2$ invariant $\xi(S^2,\sigma)$ is a topological invariant that essentially depends on the $\Z_2$-equivariant structure, defined by combining the $\Z_2$ inversion action on $S^2$ with the Berry curvature.\footnote{By quotienting the 2-sphere $S^2$ by the $\Z_2$ action $\sigma$, one realizes a $U(1)$ bundle over the real projective plane $\R P^2$. $\xi(S^2,\sigma)$ is the topological invariant that detects the cohomology group $H^2(\R P^2,\Z) = \Z_2$ classifying complex line bundles over $\R P^2$.}

Note that the $\Z_2$ number $\xi(S^2,\sigma)$ changes as $\xi(S^2,\sigma) \mapsto \xi(S^2,\sigma) + \pi$ if we flip the phase of $\hat \sigma$ ($\hat \sigma \mapsto -\hat \sigma$), swapping the values of the $\Z_2$ invariant. Thus, $\xi(S^2,\sigma)$ detects the relative $\Z_2$ topological class between two Hamiltonians $\hat H_0(\bm{n})$ and $\hat H_1(\bm{n})$ defined on the same Hilbert space with the same $\Z_2$ action $\hat \s$. 

As a model example, consider the following $2 \times 2$ model~\cite{ShiozakiTopological2017}:
\begin{align}
\hat H(\bm{n}) = 1_2 - 2\, |\bm{n}\rangle\langle \bm{n}|,\quad 
|\bm{n}\rangle = 
\begin{pmatrix}
n_x + i n_y \\
n_z 
\end{pmatrix}, \quad 
\hat \sigma = 1_2\,.
\end{align}
The ground state of this Hamiltonian $\hat H(\bm{n})$ is $|\bm{n}\rangle$, and $\xi(S^2,\sigma) = \pi$. It cannot be adiabatically connected (while preserving the $\hat \sigma$-equivariant condition) to a trivial state independent of $\bm{n}$ (for example, $|\bm{n}\rangle = (1,0)^\top$).

Here, we remark on what kind of defect structure is stabilized by $\xi(S^2,\s)$. If no symmetry leaves any parameter point invariant, the only stable defect is the codimension-3 defect (monopole) characterized by the Chern number. Therefore, if $\xi(S^2,\sigma)$ is nonzero relative to a trivial model, we conclude that inside the sphere $S^2$ there appear a pair of codimension-3 defects with Chern numbers $1$ and $-1$ (see Fig.~\ref{fig:0d_z2_number}[c]). Such a defect pair does not annihilate due to the $\Z_2$-equivariant structure, and remains stable as long as no level crossing occurs over the sphere $S^2$.

\section{Equivariant family of MPS: discrete formulation}
\label{sec:1d}

In this section, we review the gauge redundancy of MPS and how the $G$-equivariant structure is encoded in the parameter space. Following Ref.~\cite{ShiozakiHigher2025}, we employ a discrete formulation based on a triangulation of the parameter space. In Sec.~\ref{sec:MPS_inv_with_G}, we derive fixed-point formulas for family topological invariants. 

Consider a parameter family of gapped Hamiltonians $\hat H(\tau \in {\cal M})$ in a one-dimensional quantum spin system with ${\cal M}$ the parameter space. Assume that over the whole parameter space (or the subspace of interest), the ground state of $\hat H(\tau)$ is unique and there is no SSB (i.e., no degeneracy of the macroscopic quantum state). As a technical assumption of this paper, we assume the Hamiltonian $\hat H(\tau)$ is translationally invariant and also that the ground state $|\psi(\tau)\rangle$ is translationally invariant. In other words, letting $\hat T_{\rm r}$ be the translation operator, we assume $\hat T_{\rm r}\, \hat H(\tau)\, \hat T_{\rm r}^{-1} = \hat H(\tau)$ and $\hat T_{\rm r}\, |\psi(\tau)\rangle = |\psi(\tau)\rangle$. Under these assumptions, the ground state wavefunction $\psi(\tau)$ can be described by an injective MPS, which we introduce shortly. 

\subsection{Fundamental theorem of MPS}

First, we summarize the gauge degrees of freedom in an MPS, which is analogous to the $U(1)$ gauge ambiguity (\ref{eq:0d_gauge_tr}) in pure states of a quantum mechanical system.

In a one-dimensional quantum spin system, let $\{\ket{i_x}\}_{i_x=1}^n$ denote a local basis at site $x$. A translationally invariant matrix product state (MPS) is specified by a set of $n$ matrices $\{A^i\}_{i=1}^n$ with $A^i \in \mathrm{Mat}_{D \times D}(\mathbb{C})$, where $D$ is called the bond dimension, as follows:
\begin{align}
\ket{\{A^i\}_i}
= \sum_{i_1,\dots,i_N=1}^n \mathrm{Tr}[A^{i_1} \cdots A^{i_N}]\, \ket{i_1 \cdots i_N}\,.
\end{align}
Here, $N$ is the system size. It is known that if a quantum state $\ket{\psi}$ satisfies the area law of entanglement entropy, meaning that the entanglement entropy remains bounded independently of the system size, then $\ket{\psi}$ can be approximated by an MPS in the following sense: if we define the approximation error between $\ket{\psi}$ and the MPS $\ket{\{A^i\}_i}$ as $\epsilon = 1 - |\langle \psi|\{A^i\}_i\rangle|$, the required bond dimension $D$ grows polynomially with the system size $N$ and the inverse error $\epsilon^{-1}$, that is, $D = \mathrm{poly}(N, \epsilon^{-1})$~\cite{CiracMatrix2021}. Therefore, since the bond dimension $D$ depends on both $N$ and $\epsilon$, any physically meaningful quantity such as correlation functions must be formulated in a way that is insensitive to variations in $D$.

An appropriate weighting to extract the dominant components in the bond Hilbert space is given by the Schmidt eigenvalues: An MPS can always be brought, without changing the physical state, into the following right-canonical form that satisfies 
\begin{align}
&\sum_{i=1}^n A^i A^{i\dagger} = 1_D, \quad 
\sum_{i=1}^n A^{i\dagger} \Lambda^2 A^{i} = \Lambda^2, 
\label{eq:1d_Schmidt_eigenvalue_right}
\end{align}
or the left-canonical form 
\begin{align}
&\sum_{i=1}^n B^{i\dagger} B^{i} = 1_D, \quad 
\sum_{i=1}^n B^{i} \Lambda^2 B^{i\dagger} = \Lambda^2, 
\label{eq:1d_Schmidt_eigenvalue_left}
\end{align}
such that $\ket{\{A^i\}_i} = \ket{\{B^i\}_i}$. Here $\Lambda$ is a diagonal matrix with positive real entries
\begin{align}
\Lambda = \mathrm{diag}(\lambda_1,\dots,\lambda_D), \quad 
\lambda_1 \ge \cdots \ge \lambda_D > 0,
\end{align}
and it satisfies the normalization condition 
\begin{align}
\mathrm{Tr}\,\Lambda^2 = \sum_{a=1}^D \lambda_a^2 = 1\,.
\end{align}
The set of eigenvalues $\{\lambda_a\}_a$ in an MPS in canonical form coincides with the Schmidt eigenvalues obtained by the Schmidt decomposition at a bond $(x,x+1)$ for an infinitely 1D quantum state: $\ket{\{A^i\}_i} = \sum_{a=1}^D \lambda_a\, \ket{L,a}\otimes \ket{R,a}$.
The right- and left-canonical forms are related by the following transformation~\footnote{$\Lambda^{-1}$ involves reciprocals of arbitrarily small Schmidt eigenvalues, so one should avoid using it in numerical computations.}:
\begin{align}
B^i = \Lambda\, A^i\, \Lambda^{-1},\quad i=1,\dots,n.
\label{eq:rl_rel_canonical}
\end{align}
In this paper, we proceed using the right-canonical form, i.e., $\{A^i\}_i$. 

Moreover, to exclude cat states (i.e., linear combinations of macroscopically distinct states), we assume the property called injectivity. Injectivity is defined in terms of the transfer matrix. Define the transfer matrix $T \in \mathrm{End}(\mathrm{Mat}_{D \times D}(\C))$ as 
\begin{align}
T(X) := \sum_{i=1}^n A^i\, X\, A^{i\dagger}\,.
\end{align}
We say $T$ is injective if the eigenvalue $\mu_{\rm max}$ of $T$ with the largest absolute value is unique and its eigenspace is one-dimensional~\footnote{Let ${\rm spr}(T) = \max\{|\mu| \mid \mu \in {\rm Spec}(T)\}$ denote the spectral radius of $T$. We assume that the eigenvalue $\mu \in {\rm Spec}(T)$ satisfying $|\mu| = {\rm spr}(T)$ is unique and that the eigenvector $X$ with eigenvalue $\mu$, $T(X)=\mu X$, is unique up to a scalar multiplication.}. In this case, by positivity of $T$, $\mu_{\rm max}$ is positive real ($\mu_{\rm max}>0$), and one can choose the eigenvector $X$ satisfying $T(X) = \mu_{\rm max} X$ to be positive definite ($X>0$). If the MPS $\{A^i\}_i$ is in canonical form, then $\mu_{\rm max} = 1$ and we can take $X = 1_D$ as the positive definite eigenvector.

For an injective MPS, the following gauge structure theorem holds~\cite{Perez-GarciaMatrix2007}. Let $\{A^i\}_i$ and $\{\tilde A^i\}_i$ be two injective MPSs in right-canonical form with bond dimension $D$, with Schmidt matrices $\Lambda$ and $\tilde{\Lambda}$, respectively. For a sufficiently large system size $N > O(D^4)$, if there exists some phase $e^{i\alpha}$ such that $|\{A^i\}_i\rangle = e^{i\alpha} |\{\tilde A^i\}_i\rangle$ (i.e., the two MPS represent the same physical state for a system size $N$), then there exist some $e^{i\theta} \in U(1)$ and some $V \in U(D)$ such that 
\begin{align}
&A^i = e^{i\theta}\, V^\dagger\, \tilde A^i\, V,\quad i=1,\dots,n, \label{eq:gauge_tr}\\
&V\, \Lambda = \tilde \Lambda\, V. \label{eq:La_gauge}
\end{align}
Furthermore, $e^{i\theta}$ is unique, and $V$ is unique up to an overall $U(1)$ phase~\cite{Perez-GarciaString2008}. Since $\Lambda, \tilde{\Lambda}$ are arranged in descending order, Eq.~(\ref{eq:La_gauge}) implies $\Lambda = \tilde \Lambda$, i.e., $\Lambda$ is gauge-invariant. Eq.~(\ref{eq:La_gauge}) means that the unitary matrix $V$ is block-diagonal in the eigensectors of the Schmidt eigenvalues.

Since $V$ is unique up to an overall phase, the equivalence class $[V] = \{z V \mid z \in U(1)\}$ is unique; in other words, one can regard it as an element of the projective unitary group $PU(D) = U(D)/(z \sim zU, z \in U(1))$. Thus, the gauge ambiguity of an MPS is given by the pair $(e^{i\theta}, [V]) \in U(1) \times PU(D)$. 

Moreover, for a pair of gauge-equivalent MPSs $\{A^i\}_i, \{\tilde A^i\}_i$, the gauge transformation $(e^{i\theta}, [V])$ can be computed numerically using the following mixed transfer matrix~\cite{Perez-GarciaString2008}:
\begin{align}
T_{A\tilde A}(X) := \sum_{i=1}^n A^i\, X\, \tilde A^{i\dagger}\,. 
\label{eq:mixed_tr}
\end{align}
From Eq.~(\ref{eq:gauge_tr}), the eigenvalue of $T_{A\tilde A}$ with the largest absolute value is $\mu = e^{i\theta}$, and the corresponding eigenvector is given by $X = V^\dagger$.

In the rest of this section, $\{A^i\}_i$ will always denote an injective MPS in the right-canonical form.

\subsection{Overlap matrix}

We introduce the \emph{overlap matrix}, which plays a central role in constructing family topological invariants.

Let $\{A_0^i\}_i$ and $\{A_1^i\}_i$ be two injective MPSs in right-canonical form, with bond dimensions $D_0$ and $D_1$, respectively. Define the mixed transfer matrix $T_{A_0 A_1} \in \mathrm{End}(\mathrm{Mat}_{D_0 \times D_1}(\mathbb{C}))$ similarly to Eq.~(\ref{eq:mixed_tr}) as:
\begin{align}
T_{A_0 A_1} := \sum_{i=1}^n A_0^i \otimes A_1^{i*}\,.
\end{align}
Explicitly,
\begin{align}
T_{A_0 A_1}(X) = \sum_{i=1}^n A_0^i\, X\, A_1^{i\dagger},\quad X \in \Mat_{D_0 \times D_1}(\C). 
\end{align}
We say that the two MPSs $\{A^i_0\}_i$, $\{A^i_1\}_i$ are “close” if, similarly to the injective condition, the eigenvalue of the mixed transfer matrix $T_{A_0 A_1}$ with the largest absolute value is unique and its eigenspace is one-dimensional. In that case, there exists a rectangular matrix $X_{01} \in \mathrm{Mat}_{D_0 \times D_1}(\mathbb{C})$ as the eigenvector corresponding to the eigenvalue $\mu_{01}$ with the largest absolute value:
\begin{align}
T_{A_0 A_1}(X_{01}) = \sum_{i=1}^n A_0^i\, X_{01}\, A_1^{i\dagger} = \mu_{\rm max,01}\, X_{01}\,. \label{eq:T_X_def}
\end{align}
We call this matrix $X_{01} \in \Mat_{D_0 \times D_1}(\C)$ the \emph{overlap matrix}, by analogy with the inner product of pure states. (Note that, being the solution of a linear eigenvalue equation, the complex coefficient of $X_{01}$ is undetermined.)

To align with the formulation for quantum mechanical systems, we introduce the bra-ket notation. Endow the matrix space $\Mat_{D_0 \times D_1}(\C)$ with the Hilbert=Schmidt inner product
\begin{align}
(Y|X) := \mathrm{Tr}[Y^\dagger X]\,, \quad 
X,Y \in \Mat_{D_0 \times D_1}(\C)\,. 
\end{align}
With this inner product, the matrix elements of the mixed transfer matrix in an operator basis $e_{ab}=|a)(b|$ are:
\begin{align}
[T_{A_0 A_1}]_{ab,cd} 
&= \mathrm{Tr}[e_{ab}^\dagger\, T_{A_0 A_1}(e_{cd})] \nonumber \\
&=\sum_{i=1}^n (a|A^i_0|c)\, (d|A_1^{i\dagger}|b)
= \sum_{i=1}^n [A^i_0]_{ac}\, [A_1^i]^*_{bd}\,.
\end{align}
The eigenvalue equation (\ref{eq:T_X_def}) is written in bra-ket notation as a right-eigenstate of the mixed transfer matrix:
\begin{align}
T_{A_0 A_1}\, |X_{01}) = \mu_{\rm max,01}\, |X_{01})\,.
\end{align}

We summarize the gauge redundancy of the overlap matrix. Under gauge transformations on the MPSs $\{A^i_0\}_i$ and $\{A^i_1\}_i$:
\begin{align}
A_m^i \mapsto e^{i\theta_m} W_m^\dagger\, A_m^i\, W_m,\quad 
e^{i\theta_m} \in U(1),\ W_m \in U(D_m), \quad m=0,1,
\end{align}
the mixed transfer matrix transforms as 
\begin{align}
T_{A_0 A_1} \mapsto e^{i(\theta_0 - \theta_1)}\, (W_0 \otimes W_1^*)^\dagger\, T_{A_0 A_1}\, (W_0 \otimes W_1^*)\,.
\end{align}
Therefore, the eigenvalue and eigenvector transform as
\begin{align}
&\mu_{\rm max,01} \mapsto e^{i(\theta_0 - \theta_1)}\, \mu_{\rm max,01}\,, \\
&|X_{01}) \mapsto W_0^\dagger \otimes W_1^\top\, |X_{01}) \nonumber\\
&\quad \Leftrightarrow \quad 
X_{01} \mapsto W_0^\dagger\, X_{01}\, W_1\,,
\label{eq:X01_gauge_tr}
\end{align}
respectively. Thus, we find that any product of the form 
\begin{align}
A^{i_0}_0\, X_{01}\, A^{i_1}_1 
\mapsto W_0^\dagger\, A^{i_0}_0\, X_{01}\, A^{i_1}_1\, W_1
\end{align}
transforms covariantly under gauge transformations (up to an overall $U(1)$ phase). Thanks to this property, the overlap matrix $X_{01}$ plays a role in constructing an MPS that transforms covariantly under independent gauge transformations defined at each site. In other words, $X_{01}$ plays the role of a “gauge field” defined on the bond between 0 and 1. This point was discussed at the end of subsection~\ref{sec:thouless}, in the context of constructing a topological invariant for an adiabatic pump.

\subsection{Discrete formulation with group action}
We approximate the parameter space $\mathcal{M}$ by a triangulation, and at each vertex $\tau_0$ assign an injective right-canonical MPS $\{A^i(\tau_0)\}_i$. Let $\Lambda(\tau_0)$ be the positive diagonal matrix of Schmidt eigenvalues (arranged in descending order) associated with the canonical condition (\ref{eq:1d_Schmidt_eigenvalue_right}) at vertex $\tau_0$. In general, note that the bond dimension $D(\tau_0)$ may depend on the vertex $\tau_0$. Each vertex MPS has the following vertex gauge transformation:
\begin{align}
&A^i(\tau_0) \mapsto 
e^{i\chi^{(0,0)}(\tau_0)}\, W(\tau_0)^\dagger\, A^i(\tau_0)\, W(\tau_0)\,, \quad 
e^{i\chi^{(0,0)}(\tau_0)} \in U(1),\quad W(\tau_0) \in U(D(\tau_0)), 
\label{eq:v_gauge}\\
&\text{with}\qquad W(\tau_0)\, \Lambda(\tau_0) = \Lambda(\tau_0)\, W(\tau_0)\,.
\end{align}
For any adjacent vertices $\Delta^1 = (\tau_0,\tau_1)$, define the mixed transfer matrix
\begin{align}
T(\Delta^1) := \sum_{i=1}^n A^i(\tau_0) \otimes A^i(\tau_1)^*\,.
\label{eq:mixed_transfer_mat_Delta^1}
\end{align}
We assume that the MPSs at neighboring vertices are “close,” i.e., the largest eigenvalue $\mu_{\rm max}(\Delta^1)$ of the mixed transfer matrix $T(\Delta^1)$ is unique and its eigenspace is nondegenerate. Under this assumption, for each edge $\Delta^1=(\tau_0,\tau_1)$ we can define a nonzero complex number $\mu_{\rm max}(\Delta^1)$ and a rectangular matrix $X(\Delta^1) \in \Mat_{D(\tau_0) \times D(\tau_1)}(\C)$ as the eigenvalue and the eigenvector (overlap matrix) corresponding to the largest eigenvalue of the mixed transfer matrix:
\begin{align}
&T(\Delta^1)\, \pet{X(\Delta^1)} = \mu_{\rm max}(\Delta^1)\, \pet{X(\Delta^1)}\nonumber\\
&\Leftrightarrow 
\sum_{i=1}^n A^i(\tau_0)\, X(\Delta^1)\, A^i(\tau_1)^\dagger =  \mu_{\rm max}(\Delta^1)\, X(\Delta^1)\,. 
\label{eq:X_def}
\end{align}
Taking the Hermitian conjugate of (\ref{eq:X_def}), for the reversed edge $-\Delta^1=(\tau_1,\tau_0)$ we may choose phases so that 
\begin{align}
\mu_{\rm max}(-\Delta^1) = \mu_{\rm max}(\Delta^1)^*, \quad 
X(-\Delta^1) = X(\Delta^1)^\dagger\,,
\label{eq:X01_dag}
\end{align}
and we always impose (\ref{eq:X01_dag}) throughout this paper. Define an $\R/2\pi\Z$-valued (0,1)-cochain for each edge $\Delta^1$ as 
\begin{align}
{\cal A}^{(0,1)}(\Delta^1) := \arg \Big( \mu_{\rm max}(\Delta^1) \Big)\,.
\end{align}
For any triangle (2-simplex) $\Delta^2 = (\tau_0,\tau_1,\tau_2)$, we define a $U(1)$-valued (discrete) higher Berry connection as the phase of the following Wilson loop weighted by Schmidt eigenvalues~\cite{ShiozakiHigher2025}:
\begin{align}
{\cal A}^{(0,2)}(\Delta^2)
:= 
\arg \Big( \mathrm{Tr}[\Lambda(\tau_0)^{2/3}\, X(\tau_0,\tau_1)\, \Lambda(\tau_1)^{2/3}\, X(\tau_1,\tau_2)\, \Lambda(\tau_2)^{2/3}\, X(\tau_2,\tau_0)] \Big)\,. 
\label{eq:def_HBConn}
\end{align}
Note that since the rectangular matrix $X(\Delta^1)$ is nearly an isometry~\footnote{Precisely, if $D(\tau_0)\ge D(\tau_1)$, then $X(\Delta^1)^\dagger X(\Delta^1)$ is nearly the identity matrix ${\bf 1}_{D(\tau_1)}$, and $X(\Delta^1) X(\Delta^1)^\dagger$ is nearly a rank-$D(\tau_0)$ projection matrix. The same applies to the case $D(\tau_0)\le D(\tau_1)$.}, it is sensitive to changes in the bond dimensions $D(\tau_0),D(\tau_1)$ (which are unphysical parameters). Therefore, it is necessary to weight appropriately by the Schmidt eigenvalues in the bond Hilbert space. The exponent $2/3$ on $\Lambda$ is chosen so as not to conflict with the normalization condition $\mathrm{Tr}[\Lambda^2]=1$ in the limit where $\tau_0,\tau_1,\tau_2$ coincide~\cite{OhyamaHigher2024}. From (\ref{eq:X01_dag}), under reversal of the orientation of a 2-simplex $\Delta^2$, ${\cal A}^{(0,2)}(\Delta^2)$ flips sign:
\begin{align}
{\cal A}^{(0,2)}(-\Delta^2) = -\, {\cal A}^{(0,2)}(\Delta^2)\,. 
\end{align}

Now we summarize the gauge ambiguities.  The vertex gauge transformation (\ref{eq:v_gauge}) induces the gauge transformation of the mixed transfer matrix:
\begin{align}
T(\Delta^1) \mapsto e^{-i\, d\chi^{(0,0)}(\Delta^1)}\, (W(\tau_0)\otimes W(\tau_1)^*)^\dagger\, T(\Delta^1)\, (W(\tau_0)\otimes W(\tau_1)^*)\,,
\label{eq:1d_gauge_tr_transfer_mat}
\end{align}
where $d\chi^{(0,0)}(\Delta^1) = \chi^{(0,0)}(\tau_1) - \chi^{(0,0)}(\tau_0)$. Since the overlap matrix $X$ is a solution to the linear eigenvalue equation (\ref{eq:X_def}), there is an \emph{edge gauge transformation}:
\begin{align}
X(\Delta^1) \mapsto e^{i\chi^{(0,1)}(\Delta^1)}\, X(\Delta^1)\,,
\label{eq:1d_edge_gauge}
\end{align}
with $\chi^{(0,1)}(-\Delta^1) = -\, \chi^{(0,1)}(\Delta^1)$. Under these gauge transformations, the overlap matrix $X$, the 1-cochain ${\cal A}^{(0,1)}$, and the 2-cochain ${\cal A}^{(0,2)}$ transform as follows:
\begin{align}
\begin{cases}
\pet{X(\Delta^1)} \mapsto e^{i\chi^{(0,1)}(\Delta^1)}\, (W(\tau_0)\otimes W(\tau_1)^*)^\dagger\, \pet{X(\Delta^1)}\,, \\
{\cal A}^{(0,1)}(\Delta^1) \mapsto {\cal A}^{(0,1)}(\Delta^1) - d\, \chi^{(0,0)}(\Delta^1)\,, \\
{\cal A}^{(0,2)}(\Delta^2) \mapsto {\cal A}^{(0,2)}(\Delta^2) + d\, \chi^{(0,1)}(\Delta^2)\,. 
\end{cases}
\label{eq:1d_gauge_tr_summary}
\end{align}
Using this structure, one can define the following two $\R/2\pi\Z$-valued gauge-invariant quantities:
\begin{itemize}
  \item The \emph{ordinary Berry phase} along a closed loop $\ell$:
  \begin{align}
  \gamma^{(1)}(\ell) := \sum_{\Delta^1 \in \ell} {\cal A}^{(0,1)}(\Delta^1)\,,
  \end{align}
  \item The \emph{higher Berry phase} on an oriented closed 2D surface $\Sigma^{(2)}$:
  \begin{align}
  \gamma^{(2)}(\Sigma^{(2)}) := \sum_{\Delta^2 \in \Sigma^{(2)}} {\cal A}^{(0,2)}(\Delta^2)\,.
  \end{align}
\end{itemize}
Furthermore, we define the corresponding Berry curvatures (fluxes) as follows:
\begin{align}
{\cal F}^{(2)}(\Delta^2) := \widetilde{d {\cal A}^{(0,1)}(\Delta^2)} \in \R\,, \\
{\cal F}^{(3)}(\Delta^3) := \widetilde{d {\cal A}^{(0,2)}(\Delta^3)} \in \R\,,
\end{align}
and define the corresponding topological numbers:
\begin{itemize}
  \item The Chern number for an oriented closed 2D surface $\Sigma^{(2)}$:
  \begin{align}
  \nu^{(2)}(\Sigma^{(2)}) := \frac{1}{2\pi} \sum_{\Delta^2 \in \Sigma^{(2)}} {\cal F}^{(2)}(\Delta^2) \in \Z\,,
  \end{align}  
  \item The DDKS number for an oriented closed 3D space $\Sigma^{(3)}$:
  \begin{align}
  \nu^{(3)}(\Sigma^{(3)}) := \frac{1}{2\pi} \sum_{\Delta^3 \in \Sigma^{(3)}} {\cal F}^{(3)}(\Delta^3) \in \Z\,.\label{eq:DDKS}
  \end{align}
\end{itemize}
For a numerical example of the DDKS number in this framework, see Ref.~\cite{ShiozakiHigher2025}.

Note: the fact that an ordinary 1-cochain Berry connection ${\cal A}^{(0,1)}$ can be defined from a family of MPS is a special feature resulting from the assumption of translational invariance. For a general family of short-range entangled states that is not necessarily translationally invariant, one expects that only the higher Berry connection ${\cal A}^{(0,2)}$ can be well-defined.

\subsection{$G$-Equivariance}

Now consider the case where a symmetry group $G$ (take $G$ discrete, though in Sec.~\ref{sec:thouless} we will consider some continuous subgroups) acts on the parameter space $\mathcal{M}$:
\begin{align}
\tau \mapsto g\, \tau, \quad g(h \tau ) = (gh)\tau,\quad 
\tau \in \mathcal{M}, \quad 
g,h \in G\,.
\end{align}
We require the triangulation to be compatible with the group action as in Sec.~\ref{sec:0d_group_action}. 

Below, we define the $G$-action on an injective MPS. As in Sec.~\ref{sec:0d}, we specify the unitary/antiunitary nature of $g \in G$ by a homomorphism $\phi: G \to \{\pm 1\}$ (we set $\phi_g = -1$ for antiunitary $g$). Furthermore, depending on $\phi_g$, we introduce the notation for matrix or complex conjugation as follows:
\begin{align}
M^{\phi_g} := \begin{cases}
M & (\phi_g = 1), \\
M^* & (\phi_g = -1)\,.
\end{cases}
\end{align}
Let $u_g \in U(n)$ be a linear representation of $G$:
\begin{align}
u_g\, u_h^{\phi_g} = u_{gh}, \quad g,h \in G\,.
\label{eq:u_gu_h}
\end{align}
The action of $G$ on an MPS is given by 
\begin{align}
\hat g\, \ket{\{A^i\}_i}
= \Big|\Big\{\sum_{j=1}^n [u_g]_{ij}\, (A^j)^{\phi_g}\Big\}_i\Big\rangle.
\label{eq:MPS_group_action_A}
\end{align}

Now, impose the following $G$-equivariance on the family of Hamiltonians $\hat H(\tau \in \mathcal{M})$:
\begin{align}
\hat g\, \hat H(\tau)\, \hat g^{-1} = \hat H(g \tau),\quad g \in G\,.
\end{align}
By equivariance, it is natural to assume that the bond dimensions satisfy:
\begin{align}
D(g\tau) = D(\tau),\quad g \in G\,.
\end{align}
Then, from Eqs.~(\ref{eq:MPS_group_action_A}) and the gauge transformation formulas (\ref{eq:gauge_tr}), (\ref{eq:La_gauge}), the MPSs obey the following $G$-equivariance:
\begin{align}
\sum_{j=1}^n [u_g]_{ij}\, A^j(\tau)^{\phi_g}
\sim A^i(g\tau)\,.
\end{align}
Here “$\sim$” means gauge equivalent. Therefore, there exists a unique $U(1)$ phase $e^{i{\cal A}^{(1,0)}_g(\tau)}$ and a unitary matrix $V_g(\tau) \in U(D(\tau))$ (unique up to a phase) such that:
\begin{align}
\Lambda(g\tau) = \Lambda(\tau), \quad 
V_g(\tau)\, \Lambda(\tau) = \Lambda(\tau)\, V_g(\tau), 
\end{align}
and
\begin{align}
\sum_{j=1}^n [u_g]_{ij}\, A^j(\tau)^{\phi_g}
=
e^{-i{\cal A}^{(1,0)}_g(\tau)}\, V_g(\tau)^\dagger\, A^i(g\tau)\, V_g(\tau),\quad g \in G,
\label{eq:1d_A_G_action}
\end{align}
or equivalently,
\begin{align}
A^i(g\tau)
=
e^{i{\cal A}^{(1,0)}_g(\tau)}\, V_g(\tau)\, \Big( \sum_{j=1}^n [u_g]_{ij}\, A^j(\tau)^{\phi_g} \Big)\, V_g(\tau)^\dagger,\quad g \in G\,.
\label{eq:g_str_A}
\end{align}
Here, for the identity element $e \in G$, we can set 
\begin{align}
V_e(\tau) = {\bf 1}_{D(\tau)}\,,
\label{eq:1d_V_e}
\end{align}
Eq.~(\ref{eq:1d_V_e}) will be assumed hereafter.

Under the vertex gauge transformation (\ref{eq:v_gauge}), ${\cal A}^{(1,0)}_g(\tau_0)$ and $V_g(\tau_0)$ transform as 
\begin{align}
\begin{cases}
{\cal A}^{(1,0)}_g(\tau_0) 
\mapsto {\cal A}^{(1,0)}_g(\tau_0) - (\delta \chi^{(0,0)})_g(\tau_0) = {\cal A}^{(1,0)}_g(\tau_0) - \phi_g\, \chi^{(0,0)}(\tau_0) + \chi^{(0,0)}(g\tau_0)\,,\\
V_g(\tau_0) \mapsto W(g\tau_0)^\dagger\, V_g(\tau_0)\, W(\tau_0)^{\phi_g}\,.
\end{cases}
\end{align}
Using the linearity of $u_g$ (\ref{eq:u_gu_h}), we derive from Eq.~(\ref{eq:g_str_A}) that
\begin{align}
\sum_{j=1}^n [u_{gh}]_{ij}\, A^j(\tau)^{\phi_{gh}}
&=\sum_{j=1}^n [u_g]_{ij}\, 
\Big(\sum_{k=1}^n [u_h]_{jk}\, A^k(\tau)^{\phi_h}\Big)^{\phi_g} \nonumber \\
&=\sum_{j=1}^n [u_g]_{ij} \Big( e^{-i{\cal A}^{(1,0)}_h(\tau)}\, V_h(\tau)^\dagger\, A^j(h\tau)\, V_h(\tau) \Big)^{\phi_g} \nonumber \\
&= e^{-i(\phi_g {\cal A}^{(1,0)}_h(\tau)+{\cal A}^{(1,0)}_g(h\tau))}\,
(V_h(\tau)^{\phi_g})^\dagger\,
V_g(h\tau)^\dagger\, 
A^i(gh\tau)\, 
V_g(h\tau)\, V_h(\tau)^{\phi_g}\,.
\end{align}
On the other hand, this must equal
\begin{align}
e^{-i{\cal A}^{(1,0)}_{gh}(\tau)}\, V_{gh}^\dag(\tau)\, A^i(gh\tau)\, V_{gh}(\tau)\,,
\end{align}
so by the uniqueness of $e^{-i{\cal A}^{(1,0)}_g(\tau)}$ and $V_g(\tau)$ up to phase, we obtain the cocycle condition
\begin{align}
(\delta {\cal A}^{(1,0)})_{g,h}(\tau)
= \phi_g\, {\cal A}^{(1,0)}_h(\tau) - {\cal A}^{(1,0)}_{gh}(\tau) + {\cal A}^{(1,0)}_g(h\tau) = 0, \quad 
g,h \in G, 
\end{align}
and the existence of a $(2,0)$-cochain $e^{i{\cal A}^{(2,0)}_{g,h}(\tau)} \in U(1)$ such that
\begin{align}
V_g(h\tau)\, V_h(\tau)^{\phi_g}
= e^{i{\cal A}^{(2,0)}_{g,h}(\tau)}\, V_{gh}(\tau), \quad g,h \in G\,,
\label{eq:1d_V_cocycle}
\end{align}
with ${\cal A}^{(2,0)}_{e,g}(\tau) = {\cal A}^{(2,0)}_{g,e}(\tau) = 0$.
From the two decompositions of $V_{ghk}(\tau)$,
\begin{align}
V_{ghk}(\tau) 
&= e^{-i{\cal A}^{(2,0)}_{g,hk}(\tau)}\, V_g(hk\tau)\, V_{hk}(\tau)^{\phi_g}\nonumber
  \\
&= e^{-i{\cal A}^{(2,0)}_{g,hk}(\tau)}\, V_g(hk\tau)\,\Big(
e^{-i{\cal A}^{(2,0)}_{h,k}(\tau)}\, V_h(k\tau)\, V_k(\tau)^{\phi_h}\Big)^{\phi_g}\nonumber \\
&= e^{-i{\cal A}^{(2,0)}_{g,hk}(\tau)}\, e^{-i\phi_g{\cal A}^{(2,0)}_{h,k}(\tau)}\, 
(V_h(\tau)^{\phi_g})^\dagger\,
V_g(hk\tau)^\dagger\, 
A^i(gh\tau)\, 
V_g(hk\tau)\, V_h(\tau)^{\phi_g}\,.
\end{align}
and
\begin{align}
V_{ghk}(\tau) 
&= e^{-i{\cal A}^{(2,0)}_{gh,k}(\tau)}\, 
V_{gh}(k\tau)\, V_k(\tau)^{\phi_{gh}}\nonumber \\
&= e^{-i{\cal A}^{(2,0)}_{gh,k}(\tau)}\, e^{-{\cal A}^{(2,0)}_{g,h}(k\tau)}\, 
V_g(hk\tau)\, V_h(k\tau)^{\phi_g}\, V_k(\tau)^{\phi_{gh}}\,,
\end{align}
we obtain the cocycle condition
\begin{align}
(\delta {\cal A}^{(2,0)})_{g,h,k}(\tau)
:= \phi_g\, {\cal A}^{(2,0)}_{h,k}(\tau) - {\cal A}^{(2,0)}_{gh,k}(\tau) + {\cal A}^{(2,0)}_{g,hk}(\tau) - {\cal A}^{(2,0)}_{g,h}(k\tau) = 0, \quad  g,h,k \in G.
\label{eq:cocycle_cond_omega}
\end{align}

The $G$-equivariance of the transfer matrix is obtained by direct computation:
\begin{align}
T(g\Delta^1)
&= \sum_{i=1}^n A^i(g\tau_0) \otimes A^i(g\tau_1)^* \nonumber \\
&=
e^{-i d{\cal A}^{(1,0)}_g(\Delta^1)}\, v_g(\Delta^1)\, T(\Delta^1)^{\phi_g}\, 
v_g(\Delta^1)^\dag 
\end{align}
with the unitary matrix 
\begin{align}
v_g(\Delta^1) = V_g(\tau_0) \otimes V_g(\tau_1)^*\,.
\end{align}
Note from Eq.~(\ref{eq:1d_V_cocycle}) the relation 
\begin{align}
v_g(h\Delta^1)\, v_h(\Delta^1)^{\phi_g} = e^{-i\, d{\cal A}^{(2,0)}_{g,h}(\Delta^1)}\, 
v_{gh}(\Delta^1)\,.
\end{align}
Substituting the equivariance of the transfer matrix into the definition equation of $X$ (\ref{eq:X_def}), we obtain
\begin{align}
T(g\Delta^1)\, v_g(\Delta^1)\, \pet{X(\Delta^1)}^{\phi_g}
= e^{-i\, d{\cal A}^{(1,0)}_g(\Delta^1)}\, \mu_{\rm max}(\Delta^1)^{\phi_g}\, 
v_g(\Delta^1)\, \pet{X(\Delta^1)}^{\phi_g}\,.
\end{align}
From this, for the $U(1)$ phase of the eigenvalue, we find a relation between the (1,0)- and (0,1)-cochains
\begin{align}
d{\cal A}^{(1,0)}_g(\Delta^1)
= \phi_g\, {\cal A}^{(0,1)}(\Delta^1) - {\cal A}^{(0,1)}(g\Delta^1)\,,
\end{align}
and for the eigenvector, we introduce a (1,1)-cochain via
\begin{align}
\pet{X(g\Delta^1)}\, e^{i{\cal A}^{(1,1)}_g(\Delta^1)}
= v_g(\Delta^1)\, \pet{X(\Delta^1)}^{\phi_g}\,,
\end{align}
i.e.,
\begin{align}
V_g(\tau_0)\, X(\Delta^1)^{\phi_g}\, V_g(\tau_1)^\dagger
= e^{i{\cal A}^{(1,1)}_g(\Delta^1)}\, X(g\Delta^1)\,.\label{eq:1d_X_G_action}
\end{align}
For the identity element $e \in G$, we set ${\cal A}^{(1,1)}_e(\Delta^1) = 0$.
From Eq.~(\ref{eq:1d_X_G_action}), we derive the $G$-equivariance of the higher Berry connection~(\ref{eq:def_HBConn}):
\begin{align}
{\cal A}^{(0,2)}(g\Delta^2)
&= \arg \Big( \mathrm{Tr}[\,\Lambda(g\tau_0)^{2/3}\, X(g\tau_0,g\tau_1)\, \Lambda(g\tau_1)^{2/3}\, X(g\tau_1,g\tau_2)\, \Lambda(g\tau_2)^{2/3}\, X(g\tau_2,g\tau_0)\,] \Big)\nonumber \\
&= -\, d{\cal A}^{(1,1)}_g(\Delta^2) + \phi_g\, {\cal A}^{(0,2)}(\Delta^2)\,.
\end{align}
Comparing two expressions 
\begin{align}
\pet{X(gh\Delta^1)} 
= e^{-i{\cal A}^{(1,1)}_{gh}(\Delta^1)}\, v_{gh}(\Delta^1)\, \pet{X(\Delta^1)}^{\phi_{gh}} 
\end{align}
and 
\begin{align}
\pet{X(gh\Delta^1)} 
&= e^{-i{\cal A}^{(1,1)}_{g}(h\Delta^1)}\, v_{g}(h\Delta^1)\, \pet{X(h\Delta^1)}^{\phi_g} \nonumber \\
&= e^{-i{\cal A}^{(1,1)}_{g}(h\Delta^1)}\, v_{g}(h\Delta^1)\, \Big(e^{-i{\cal A}^{(1,1)}_{h}(\Delta^1)}\, v_h(\Delta^1)\, \pet{X(\Delta^1)}^{\phi_h}\Big)^{\phi_g} \nonumber \\
&= e^{-i{\cal A}^{(1,1)}_{g}(h\Delta^1)}\, e^{-i\phi_g{\cal A}^{(1,1)}_{h}(\Delta^1)}\, v_{g}(h\Delta^1)\, v_h(\Delta^1)^{\phi_g}\, \pet{X(\Delta^1)}^{\phi_{gh}}\,,
\end{align}
we get the cocycle condition
\begin{align}
(\delta {\cal A}^{(1,1)})_{g,h}(\Delta^1)
= \phi_g\, {\cal A}^{(1,1)}_h(\Delta^1) - {\cal A}^{(1,1)}_{gh}(\Delta^1) + {\cal A}^{(1,1)}_g(h\Delta^1)
= -\, d{\cal A}^{(2,0)}_{g,h}(\Delta^1)\,.
\end{align}

Finally, summarize the gauge transformations of the newly introduced cochains ${\cal A}^{(1,0)}$, ${\cal A}^{(1,1)}$, and ${\cal A}^{(2,0)}$. The $U(1)$ coefficient of the matrix $V_g(\tau)$ is ambiguous. Under a vertex gauge transformation depending on $G$,
\begin{align}
V_g(\tau) \mapsto e^{i\chi^{(1,0)}_g(\tau)} V_g(\tau),\quad \chi^{(1,0)}_e(\tau)=0,
\end{align}
the $(2,0)$ and $(1,1)$ cochains transform as follows:
\begin{align}
&{\cal A}^{(2,0)}_{g,h}(\tau)
\mapsto
{\cal A}^{(2,0)}_{g,h}(\tau)
+ (\delta \chi^{(1,0)})_{g,h}(\tau)
= 
{\cal A}^{(2,0)}_{g,h}(\tau)
+ \phi_g \chi^{(1,0)}_h(\tau)- \chi^{(1,0)}_{gh}(\tau) + \chi^{(1,0)}_g(h\tau), \\
&{\cal A}^{(1,1)}_g(\Delta^1)
\mapsto 
{\cal A}^{(1,1)}_g(\Delta^1) - d\chi^{(1,0)}_g(\Delta^1)
= {\cal A}^{(1,1)}_g(\Delta^1) - \chi^{(1,0)}_g(\tau_1)+\chi^{(1,0)}_g(\tau_0),
\end{align}
respectively. In addition, the edge gauge transformation~(\ref{eq:1d_edge_gauge}) induces the transformation
\begin{align}
{\cal A}^{(1,1)}_g(\Delta^1) 
\mapsto {\cal A}^{(1,1)}_g(\Delta^1) + (\delta \chi^{(0,1)})_{g}(\Delta^1)
= {\cal A}^{(1,1)}_g(\Delta^1) + \phi_g\chi^{(0,1)}(\Delta^1)- \chi^{(0,1)}(g\Delta^1).
\end{align}

Here is the summary of the gauge transformations and cocycle conditions.

For the $(0,1)$ and $(1,0)$ cochains originating from translational invariance:
\begin{align}
\begin{cases}
{\cal A}^{(0,1)} \mapsto {\cal A}^{(0,1)} - d\, \chi^{(0,0)}, \\
{\cal A}^{(1,0)} \mapsto {\cal A}^{(1,0)} - \delta \chi^{(0,0)}, 
\end{cases}\label{eq:1d_0d_gauge}
\end{align}
and the cocycle conditions:
\begin{align}
\begin{cases}
d{\cal A}^{(0,1)} \equiv {\cal F}^{(2)}\quad (\bmod\, 2\pi), \\
\delta {\cal A}^{(0,1)} + d\, {\cal A}^{(1,0)} = 0,\\
\delta {\cal A}^{(1,0)} = 0\,.
\end{cases}\label{eq:1d_0d_cocycle}
\end{align}

For the $(0,2),(1,1),(2,0)$ cochains inherent to the 1D system:
\begin{align}
\begin{cases}
{\cal A}^{(0,2)} \mapsto {\cal A}^{(0,2)} + d\, \chi^{(0,1)}, \\
{\cal A}^{(1,1)} \mapsto {\cal A}^{(1,1)} + \delta \chi^{(0,1)} - d\, \chi^{(1,0)}, \\
{\cal A}^{(2,0)} \mapsto {\cal A}^{(2,0)} + \delta \chi^{(1,0)}\,,
\end{cases}\label{eq:1d_A1_gauge_tr}
\end{align}
and the cocycle conditions:
\begin{align}
\begin{cases}
d\, {\cal A}^{(0,2)} \equiv {\cal F}^{(3)}\quad (\bmod\, 2\pi), \\
\delta {\cal A}^{(0,2)} - d\, {\cal A}^{(1,1)} = 0, \label{eq:1d_cocycle_cond_0211}\\
\delta {\cal A}^{(1,1)} + d\, {\cal A}^{(2,0)} = 0, \\
\delta {\cal A}^{(2,0)} = 0\,.
\end{cases}
\end{align}
(Here, the sign and assignment of degrees follow the $G$-simplicial complex; see Appendix~\ref{app:double_complex} for details.)

\subsubsection{Comment on inversion-type symmetry action}
An inversion-type group action that flips the position $x \mapsto -x$ acts on an MPS via transpose:
\begin{align}
    \sum_j [u_g]_{ij} (A^j)^\top.
\end{align}
This is in the left canonical form, and the transfer matrix $T$ is mapped to the transpose $T^\top$ (with a similarity transformation). As a result, the higher Berry connection defined using the right eigenvector $X$ of the transfer matrix (see Eq.~(\ref{eq:def_HBConn})) is mapped, under spatial inversion, to the higher Berry connection constructed using the left eigenvector. In general, no clear relation can be found between the higher Berry connections defined using the right and left eigenvectors, and no nice structure was found. For this reason, inversion-type group actions are not considered in this paper.

\subsection{Constraint on DDKS number and higher Berry phase under group action}

The $G$-action on the higher Berry flux is, from $\delta d\, {\cal A}^{(0,2)}=0$,
\begin{align}
{\cal F}^{(3)}(g\Delta^3) = \phi_g\, {\cal F}^{(3)}(\Delta^3)\,.
\end{align}
Therefore, if there exists an antiunitary $g \in G$ such that a closed 3D surface $\Sigma^{(3)}$ is invariant (as a set, including orientation) under the $g$-action, or if there exists a unitary $g \in G$ that reverses the orientation of a closed surface $\Sigma^{(3)}$, then the DDKS number is zero:
\begin{align}
\exists\, g \in G \ \text{s.t.}\quad 
g\, \Sigma^{(3)} = -\,\phi_g\, \Sigma^{(3)} \quad \text{then} \quad 
\nu^{(3)}(\Sigma^{(3)}) = 0\,.\label{eq:1d_DDnumber_constraint}
\end{align}
For example, under time-reversal symmetry ($\phi_g=-1$) that leaves the parameter point invariant, the DDKS number is always zero.

Similarly, the $G$-action on the higher Berry phase is
\begin{align}
\gamma^{(2)}(g \Sigma^{(2)}) = \phi_g\, \gamma^{(2)}(\Sigma^{(2)}),\quad g \in G,
\end{align}
so,
\begin{align}
\exists\, g \in G \ \text{s.t.}\quad 
g\, \Sigma^{(2)} = -\,\phi_g\, \Sigma^{(2)} \quad \text{then} \quad 
\gamma^{(2)}(\Sigma^{(2)}) \in \{0,\pi\}\,. 
\label{eq:1d_BP_quantized_OP}
\end{align}
For example, under time-reversal symmetry ($\phi_g=-1$) that leaves the parameter point invariant, the higher Berry phase is quantized~\footnote{
The existence of this $\Z_2$ topological number is consistent with the $\Omega$-spectrum structure of invertible states. In a quantum mechanical system with time-reversal symmetry, a family of time-reversal symmetric pure states over $S^1$ can be $\Z_2$-nontrivial, distinguished by the first Stiefel-Whitney class $w_1$. This is nothing but the $\pi$-Berry phase known in band theory. The $\Omega$-spectrum implies that in a 1D spin system, a $\Z_2$ number should be definable for a time-reversal symmetric family over $S^2$.}.

\section{Family topological invariants and group actions}
\label{sec:MPS_inv_with_G}

In this section, given a family on the parameter space ${\cal M}$ of MPS and a group action, we discuss what topological numbers can be defined and under what settings a fixed point formula can be obtained. Since the translational symmetry induced 1-cochain $({\cal A}^{(0,1)}, {\cal A}^{(1,0)})$ and the Berry flux ${\cal F}^{(2)}$ have the same gauge transformation property (\ref{eq:1d_0d_gauge}) and cocycle condition (\ref{eq:1d_0d_cocycle}) as those in the quantum mechanical system in Sec.~\ref{sec:0d}, the construction of topological invariants and the fixed point formulas can be obtained in exactly the same way. Therefore, in this section, we will discuss only the topological invariants derived from the 2-cochain $({\cal A}^{(0,2)}, {\cal A}^{(1,1)}, {\cal A}^{(2,0)})$ and higher Berry flux ${\cal F}^{(3)}$ that are unique to the one-dimensional quantum spin system. In the following, we denote the pairing between a $(p,q)$-cochain and a $q$-chain $C^q$ as
\begin{align}
\left({\cal A}^{(p,q)},C^q\right) = {\cal A}^{(p,q)}(C^q) := \sum_{\Delta^q \in C^q} {\cal A}^{(p,q)}(\Delta^q).
\end{align} 

\subsection{SPT invariants}
First, consider the case where the parameter space is a single point ${\tau_*}$ and the group $G$ does not move ${\tau_*}$, i.e., it acts as symmetry. In this case, only ${\cal A}^{(2,0)}(\tau_*)$ is nonzero, and the equivalence class under the gauge transformation
\begin{align}
{\cal A}^{(2,0)}_{g,h}(\tau_*) \mapsto {\cal A}^{(2,0)}_{g,h}(\tau_*) + \phi_g \chi^{(1,0)}_h(\tau_*) - \chi^{(1,0)}_{gh}(\tau_*)+\chi^{(1,0)}_g(\tau_*)
\end{align}
takes values in the second group cohomology $H_{\rm group}^2(G,\R/2\pi \Z)$. The SPT invariants are of the following three types~\cite{ShiozakiMatrix2017}:
\begin{align}
&\mu^{\rm T}_{g,h}(\tau_*) := {\cal A}^{(2,0)}_{g,h}(\tau_*) - {\cal A}^{(2,0)}_{h,g}(\tau_*)\quad \mbox{for } \phi_g=\phi_h=1, gh=hg, \label{eq:Z2inv_RP2}\\
&\mu^{\rm RP}_a(\tau_*):= {\cal A}^{(2,0)}_{a,a}(\tau_*)\quad \mbox{for }\phi_a=-1, a^2=e, \\
&\mu^{\rm KB}_{a,g}(\tau_*):= {\cal A}^{(2,0)}_{a,g^{-1}}(\tau_*)+{\cal A}^{(2,0)}_{g,g^{-1}}(\tau_*)-{\cal A}^{(2,0)}_{g,a}(\tau*)\quad \mbox{for }\phi_a=-1, \phi_g=1, ag^{-1}=ga.
\end{align}
They correspond to the partition function on a torus, real projective plane, and Klein bottle, respectively. These are familiar invariants for detecting the bosonic SPT phase in one dimension. In the following, they will play a role as invariants at the fixed points. 

\subsection{A microscopic model with nontrivial DDKS invariant}
A parameter family with an odd DDKS number $\nu^{(3)}(S^3)$ on the three-dimensional sphere is found to be nontrivial over its lower-dimensional sub-spheres $S^2$, $S^1$, and $S^0$ (i.e., two points). Before entering the main discussion, we introduce an explicit microscopic model and examine its possible $G$-equivariance. 

We consider a spin chain model where two spin-$S$ degrees of freedom reside on each site $j \in \mathbb{Z}$, with corresponding spin operators denoted by $\hat{\bm{s}}^L_j$ and $\hat{\bm{s}}^R_j$. A point in the parameter space $S^3$ is expressed in Cartesian coordinates as $\bm{n} = (n_0, n_1, n_2, n_3) \in \mathbb{R}^4$, satisfying $\bm{n}^2 = 1$. We consider the following Hamiltonian~\cite{WenFlow2023}:
\begin{align}
    \hat H(\bm{n})
    =\bm{n} \cdot \sum_{j \in \mathbb{Z}} (-\hat{\bm{s}}^L_j + \hat{\bm{s}}^R_j)
    +|n_0| \sum_{j \in \mathbb{Z}} 
    \begin{cases}
        \hat{\bm{s}}^L_j\cdot \hat{\bm{s}}^R_j & (n_0\leq 0) \\
        \hat{\bm{s}}^R_j\cdot \hat{\bm{s}}^L_{j+1} & (n_0\geq 0) \\
    \end{cases}.
    \label{eq:1d_model}
\end{align}
This model reduces to a two-spin problem at every point $\bm{n} \in S^3$, and it is easy to verify that the ground state is non-degenerate and the excited states are gapped~\footnote{
The model in Eq.~(\ref{eq:1d_model}) corresponds to a special case $\eta = \pm 1$ of the following more general model:
\begin{align}
    \hat H = J \sum_{j\in \mathbb{Z}} \left(1+\eta (-1)^j\right) \hat{\bm{s}}_j \cdot \hat{\bm{s}}_{j+1} + \bm{h} \cdot \sum_{j \in \mathbb{Z}} (-1)^j \hat{\bm{s}}_j.
    \label{eq:1d_generic_model}
\end{align}
While the model~(\ref{eq:1d_generic_model}) breaks the translation symmetry under $j\mapsto j+1$, the model in Eq.~(\ref{eq:1d_model}) is obtained by grouping sites $(2j-1, 2j)$ into a single site.
}. The DDKS invariant takes the quantized value $\nu^{(3)}(S^3) = 2S$~\footnote{With the same discussion as \cite{WenFlow2023}, this model pumps Chern number $2S$ at the boundary.}. 

Let us introduce the time-reversal operator 
\begin{align}
    \hat T = \prod_j\left(e^{i \pi (\hat s^{L,y}_j+\hat s^{R,y}_j)}\right){\cal K},   
\end{align}
and the spin rotation operator around the axis $\bm{m}$ by angle $\theta$ 
\begin{align}
    \hat q_{\bm{m},\theta} = e^{i \theta \bm{m} \cdot (\hat{\bm{s}}^L_j+\hat{\bm{s}}^R_j)}.
\end{align}
Then, the model~(\ref{eq:1d_model}) has the following $SO(3) \times \mathbb{Z}_2^T$ equivariance:
\begin{align}
    &\hat T \hat H(\bm{n}) \hat T^{-1} = \hat H(n_0,-n_1,-n_2,-n_3),\\
    &\hat q_{\bm{m},\theta} \hat H(\bm{n}) \hat q_{\bm{m},\theta}^{-1} = \hat H(n_0,(n_1,n_2,n_3)R_{\bm{m},\theta}),
\end{align}
where $R_{\bm{m},\theta}$ is the SO(3) rotation matrix acting from the right on the row vector $(n_1,n_2,n_3)$. Therefore, the model~(\ref{eq:1d_model}) is equivariant under any subgroup $G$ of $SO(3) \times \mathbb{Z}_2^T$. Below, we list several examples that will be addressed later. We denote the rotation axes in spin space by $\hat x = (1,0,0), \hat y = (0,1,0), \hat z = (0,0,1)$, and introduce the notation $C_{2\mu} = q_{\hat \mu,\pi}$ for $\pi$-rotations.

\begin{itemize}
    \item The composition $C_{2z}T$ leads to a $\mathbb{Z}_2$ equivariance:
    \begin{align}
        \wh{C_{2z}T} \hat H(\bm{n}) \wh{C_{2z}T}^{-1} = \hat H(n_0,n_1,n_2,-n_3).
        \label{eq:sym_c2zT}
    \end{align}
    The $C_{2z}T$-invariant subspace forms a 2-sphere.

    \item $U(1)_z$ rotations about the $z$-axis, or their subgroups:
    \begin{align}
        \hat q_{\hat z,\theta} \hat H(\bm{n}) \hat q_{\hat z,\theta}^{-1}
        = \hat H(n_0,\cos \theta n_1 + \sin \theta n_2,-\sin \theta n_1+\cos \theta n_2,n_3).
    \end{align}
    The $U(1)_z$-invariant subspace forms a circle $S^1$.

    \item $\mathbb{Z}_2 \times \mathbb{Z}_2$ spin rotations:
    \begin{align}
        &\hat C_{2x} \hat H(\bm{n}) \hat C_{2x}^{-1}
        = \hat H(n_0,n_1 ,-n_2,-n_3), \\
        &\hat C_{2y} \hat H(\bm{n}) \hat C_{2y}^{-1}
        = \hat H(n_0,-n_1 ,n_2,-n_3).
    \end{align}
    The $\mathbb{Z}_2 \times \mathbb{Z}_2$-invariant subspace consists of two points: $(\pm 1,0,0,0)$.
\end{itemize}

A crucial fact is that, at the two points $(1,0,0,0)$ and $(-1,0,0,0)$, the SPT invariants differ relatively when the spin is a half-integer. In the model~(\ref{eq:1d_model}), the point $\bm{n}=(-1,0,0,0)$ corresponds to a trivial model containing only on-site terms, while $\bm{n}=(1,0,0,0)$ corresponds to a nontrivial model. That is,
\begin{align}
    &\mu^{\rm RP}_{T}(-1,0,0,0) = 0, \quad \mu^{\rm RP}_{T}(1,0,0,0) = 2 \pi S, \\
    &\mu^{\rm T}_{C_{2x},C_{2y}}(-1,0,0,0)=0, \quad  \mu^{\rm T}_{C_{2x},C_{2y}}(1,0,0,0)=2 \pi S.
\end{align}

\subsection{One-parameter family: Thouless pump}
\label{sec:thouless}

We consider the subgroup that is unitary and preserves the parameter space pointwise:
\begin{align}
    G_0 = \{g \in G \mid \phi_g = 1,\; g \tau = \tau \text{ for } \tau \in {\cal M} \}.
\end{align}
We allow $G_0$ to include continuous groups in this section. For such $G_0$, the cocycle condition~(\ref{eq:1d_cocycle_cond_0211}) and gauge transformation~(\ref{eq:1d_A1_gauge_tr}) of the $(1,1)$-cochain reduce to
\begin{align}
&d {\cal A}^{(1,1)}_g = 0, \\
&{\cal A}^{(1,1)}_g \mapsto {\cal A}^{(1,1)}_g - d\chi^{(1,1)}_g,
\quad g \in G_0.
\end{align}
Thus, for each $g \in G_0$, one obtains a gauge-invariant and quantized quantity defined for a loop $\ell$ in the parameter space:
\begin{align}
    \eta_g(\ell) := \left( {\cal A}^{(1,1)}_g, \ell \right) \in \mathbb{R}/2\pi \mathbb{Z}, \quad g \in G_0.
\end{align}
Due to the flatness condition $d{\cal A}^{(1,1)}_g = 0$, $\eta_g$ is invariant under continuous deformations of the loop $\ell$, i.e., $\eta_{g \in G_0}(\ell)$ depends only on the homotopy class $[\ell] \in [S^1,{\cal M}]$ and is a topological invariant of the family. The allowed values of $\eta_g$ satisfy
\begin{align}
    \eta_{h}(\ell) - \eta_{gh}(\ell) + \eta_g(\ell)
    = \left( (\delta {\cal A}^{(1,1)})_{g,h}, \ell \right)
    = \left( -d{\cal A}^{(1,1)}_{g,h}, \ell \right)
    = 0,
\end{align}
and hence $e^{i\eta}$ defines a one-dimensional representation of $G_0$ for each homotopy class of loops:
\begin{align}
    e^{i\eta}: [S^1, {\cal M}] \to \mathrm{Hom}(G_0, U(1)).
\end{align}
Thus, the “$G_0$ charge” $e^{i\eta(\ell)}$ is assigned to each loop.

In particular, when $G_0 = U(1)$, this setting reduces to the original Thouless pump. Parameterizing the group elements of $U(1)$ by $\theta \in [0, 2\pi]$, the integer-valued $U(1)$ charge is given by
\begin{align}
    N(\ell) = \frac{1}{2\pi} \frac{d \eta_\theta(\ell)}{d\theta} \in \mathbb{Z},
\end{align}
which means that the adiabatic time evolution along the loop $\ell$ pumps $U(1)$ charge by an amount $N(\ell)$.

\subsubsection{Realization as an MPS soliton}

The topological number of a family over $S^1$ can also be formulated as a $G_0$ charge localized on a “soliton,” where the local Hamiltonian in real space depends on the adiabatic parameter. In the context of MPS, this corresponds concretely to giving an $x$-dependent MPS as follows. Let the loop $\ell$ be discretized as $\ell = (\tau_1, \tau_2) + \cdots + (\tau_N, \tau_1)$ and introduce 
\begin{align}
    \ket{\ell} = \sum_{i_1,\dots,i_N=1}^n \mathrm{tr}\left[A^{i_1}(\tau_1)X(\tau_1,\tau_2) \cdots A^{i_N}(\tau_N)X(\tau_N,\tau_1)\right]\ket{i_1,\dots,i_N}.
\end{align}
The state $\ket{\ell}$ is invariant under the gauge transformation (\ref{eq:v_gauge}) up to a $U(1)$ phase, and hence is well-defined as a physical state. The $G_0$ charge of the state $\ket{\ell}$, using (\ref{eq:1d_A_G_action}) and (\ref{eq:1d_X_G_action}), is given by
\begin{align}
    \Braket{\ell|\hat g|\ell}
    = e^{-i\sum_{x=1}^N {\cal A}^{(1,0)}_g(\tau_x)} \times e^{i\eta_g(\ell)}, \quad g \in G_0.
\end{align}
The first factor corresponds to the $G_0$ charge per site and is unrelated to the topological number of the family over the loop $\ell$. To eliminate this contribution, one may, for instance, consider the MPS at a single point in the parameter space, i.e., a trivial loop such as $\tau_1 \in \ell$:
\begin{align}
    \ket{\tau_1} = \sum_{i_1,\dots,i_N=1}^n \mathrm{tr}\left[A^{i_1}(\tau_1) \cdots A^{i_N}(\tau_1)\right]\ket{i_1,\dots,i_N},
\end{align}
whose $G_0$ charge is
\begin{align}
    \Braket{\tau_1|\hat g|\tau_1} = e^{-i\sum_{x=1}^N {\cal A}^{(1,0)}_g(\tau_x)}, \quad g \in G_0.
\end{align}
The ratio of the two quantities gives the desired result.

Let us comment on the relation to the ground state of a Hamiltonian with spatial modulation. Denote the Hamiltonian at parameter point $\tau_x$ as
\begin{align}
    \hat H(\tau_x) = \sum_{j=1}^N \hat h_j(\tau_x),
\end{align}
where $\hat h_j(\tau_x)$ is an interaction term depending on $\tau_x$ and supported near site $j$. Note that $\tau_x$ is a parameter and $\hat H(\tau_x)$ is translationally invariant. Correspondingly, one can consider a non-uniform Hamiltonian that varies in space along the loop $\ell$:
\begin{align}
    \hat H(\ell) = \sum_{j=1}^N \hat h_j(\tau_j),
\end{align}
where the local term $\hat h_j(\tau_j)$ depends on the parameter $\tau_j$ assigned to site $j$. The ground state of this non-uniform Hamiltonian $\hat H(\ell)$ is not exactly $\ket{\ell}$, but if the spatial variation is sufficiently slow compared to the energy gap (i.e., for sufficiently large $N$), the state $\ket{\ell}$ may approximate the ground state of $\hat H(\ell)$ well. In this sense, $\ket{\ell}$ can be viewed as the ground state in the “semiclassical approximation” of $\hat H(\ell)$.

\subsubsection{Fixed-point formula for Thouless pump}

In addition to the symmetry of the group $G_0$, consider the situation where there exists a group element $h \in G$ such that the loop $\ell$ is given by the difference between an arc $C$ and its image under $h$, that is, $\ell = C - hC$ (see Fig.~\ref{fig:0d_fixed_pt}[a]). Note the following identities for $g \in G_0$:
\begin{align}
    &(\delta {\cal A}^{(1,1)})_{g,h}(\Delta^1) = {\cal A}^{(1,1)}_h(\Delta^1) - {\cal A}^{(1,1)}_{gh}(\Delta^1) + {\cal A}^{(1,1)}_g(h\Delta^1), \\
    &(\delta {\cal A}^{(1,1)})_{h,g}(\Delta^1) = \phi_h {\cal A}^{(1,1)}_g(\Delta^1) - {\cal A}^{(1,1)}_{hg}(\Delta^1) + {\cal A}^{(1,1)}_h(\Delta^1).
\end{align}
When $h$ is unitary ($\phi_h = 1$) and $gh = hg$, we obtain
\begin{align}
    (\delta {\cal A}^{(1,1)})_{h,g}(\Delta^1) - (\delta {\cal A}^{(1,1)})_{g,h}(\Delta^1)
    = {\cal A}^{(1,1)}_g(\Delta^1) - {\cal A}^{(1,1)}_g(h\Delta^1).
\end{align}
In this case, the pump invariant $\eta_g(\ell)$ for elements $g \in G_0$ that commute with $h$, i.e., $gh = hg$, is completely determined by the SPT invariants at the endpoints $\partial C = Q - P$ of the arc $C$:
\begin{align}
    \eta_g(\ell) 
    &= \left( {\cal A}^{(1,1)}_{g,h}, C - hC \right)
    = \left( (\delta {\cal A}^{(1,1)})_{h,g} - (\delta {\cal A}^{(1,1)})_{g,h}, C \right)
    = \left( -d {\cal A}^{(2,0)}_{h,g} + d {\cal A}^{(2,0)}_{g,h}, C \right) \nonumber\\
    &= \mu^{\rm T}_{g,h}(Q) - \mu^{\rm T}_{g,h}(P).
    \label{eq:thouless_fixed_pt_formula}
\end{align}

For example, consider the case of the Haldane phase protected by the unitary $\Z_2[C_{2x}] \times \Z_2[C_{2y}]$ symmetry, which is characterized by the nontrivial invariant $\mu^{\rm T}_{C_{2x},C_{2y}} = \pi$. There must exist a phase transition point between nontrivial and trivial phases. If one breaks one of the symmetries, say $\Z_2[C_{2y}]$, then an adiabatic path $C$ can exist. The combination of this adiabatic path $C$ and its image under $\sigma_2$ forms the loop $\ell = C - \sigma_2 C$. According to the fixed-point formula~(\ref{eq:thouless_fixed_pt_formula}), the pump invariant $\eta_{\sigma_1}(\ell)$ associated with the unbroken symmetry $\Z_2[\sigma_1]$ along the loop $\ell$ must take the nontrivial value $\eta_{\sigma_1}(\ell) = \pi$. 
See also prior works~\cite{BergQuantized2011,RossiniTopological2013,KunoInteractioninduced2020}.

\subsection{Two-parameter family: $\pi$-higher Berry phase}
According to the constraint~(\ref{eq:1d_BP_quantized_OP}), when there exists an antiunitary symmetry $a \in G$ that leaves each point in parameter space invariant, i.e., $a \tau = \tau$ for all $\tau \in \mathcal{M}$, with $\phi_a = -1$, the higher Berry phase $\gamma^{(2)}(\Sigma^{(2)})$ is quantized to $\Z_2$ values $\{0, \pi\}$. Hence, in the presence of such a symmetry $a$, there exists a codimension-3 defect in parameter space characterized by a $\pi$-quantized higher Berry phase.

\subsubsection{Model example}

As a model exhibiting a nontrivial $\pi$-quantized higher Berry phase, consider the half-integer spin model~(\ref{eq:1d_model}) restricted to the $n_3 = 0$ subspace, where the $\Z_2$ symmetry $C_{2z}T$ (see Eq.~\ref{eq:sym_c2zT}) is preserved. Since this is a codimension-3 defect, it is stable against any $C_{2z}T$-invariant perturbation, such as an Ising interaction $J_z \sum_{j} (\hat{s}^{L,z}_j \hat{s}^{R,z}_j + \hat{s}^{R,z}_j \hat{s}^{L,z}_{j+1})$.

\subsubsection{Fixed-point formula for $\pi$-higher Berry phase}

More generally, when there exists an antiunitary group action $b \in G$ with $\phi_b = -1$ that leaves the closed two-dimensional surface $\Sigma^{(2)}$ (on which the higher Berry phase $\gamma^{(2)}(\Sigma^{(2)})$ is defined) invariant as a set, and preserves its orientation, then by Eq.~(\ref{eq:1d_BP_quantized_OP}), the higher Berry phase is quantized to values in $\{0,\pi\}$. When $b$ acts nontrivially on $\Sigma^{(2)}$, the $\pi$-quantized higher Berry phase is expected to be determined solely by the fixed points of the action $b$.

For simplicity, consider the two-sphere $S^2 \subset \mathcal{M}$, with coordinates $\bm{n} = (n_0, n_1, n_2)$ satisfying $\bm{n}^2 = 1$. Let $b$ act as a “$180^\circ$ rotation about the $n_0$ axis”:
\begin{align}
    b: (n_0, n_1, n_2) \mapsto (n_0, -n_1, -n_2)
    \label{eq:1d_b_action}
\end{align}
and assume $b^2 = e$ for the reason that will be explained shortly. Take the upper hemisphere $D = \{\bm{n} \in S^2 \mid n_3 \geq 0\}$ as an independent integration region for the higher Berry phase, and the semicircular arc $C = \{\bm{n} \in S^2 \mid n_2 \geq 0, n_3 = 0\}$ as an independent region on the boundary of $D$. Noting that $(\delta {\cal A}^{(0,2)})_{b}(\Delta^2) = - {\cal A}^{(0,2)}(\Delta^2) - {\cal A}^{(0,2)}(b\Delta^2)$, we have
\begin{align}
    \gamma^{(2)}(S^2) 
    &= ({\cal A}^{(0,2)}, D + bD)
    = (-(\delta {\cal A}^{(0,2)})_{b}, D)
    = (-d {\cal A}^{(1,1)}_{b}, D)
    = (- {\cal A}^{(1,1)}_{b}, C - bC).
\end{align}
Furthermore, using 
$(\delta {\cal A}^{(1,1)})_{b,b}(\Delta^1) = - {\cal A}^{(1,1)}_{b}(\Delta^1) - {\cal A}^{(1,1)}_{b^2}(\Delta^1) + {\cal A}^{(1,1)}_{b}(b\Delta^1)$, and noting that ${\cal A}^{(1,1)}_{e} = 0$ when $b^2 = e$, we obtain
\begin{align}
    \gamma^{(2)}(S^2)
    = ((\delta{\cal A}^{(1,1)})_{b,b}, C)
    = (-d{\cal A}^{(2,0)}_{b,b}, C)
    = \mu^{\mathrm{RP}}_{b}(P) - \mu^{\mathrm{RP}}_{b}(Q),
    \label{eq:1d_Z2BP_fixed_pt}
\end{align}
meaning that the higher Berry phase is determined by the difference of the $\Z_2$ invariants~(\ref{eq:Z2inv_RP2}) at the fixed points $P$ and $Q$.

If $b$ is regarded as a time-reversal operation, then the meaning of the fixed-point formula~(\ref{eq:1d_Z2BP_fixed_pt}) is as follows: along the fixed-point line of the $b$-action in parameter space, given by $(x,0,0) \in \R^3$ with $x \in \R$, there must exist a phase transition point between the Haldane phase ($\mu^{\mathrm{RP}}_b = \pi$) and the trivial phase ($\mu^{\mathrm{RP}}_b = 0$). However, once the $b$-symmetry is broken, one can find an adiabatic path connecting the Haldane and trivial phases. Since this path can be taken in two ``independent" directions, it forms a closed sphere $S^2$ in parameter space surrounding the transition point. In such a case, the higher Berry phase $\gamma^{(2)}(S^2)$ on this $S^2$ must be equal to $\pi$.

In the model~(\ref{eq:1d_model}), the elements $a$ and $b$ correspond to $C_{2z}T$ and $T$, respectively. When $C_{2z}T$-equivariance is present, the higher Berry phase $\gamma^{(2)}(S^2)$ is quantized in the subspace $n_3 = 0$. If the additional $T$-equivariance is also imposed, then by the fixed-point formula~(\ref{eq:1d_Z2BP_fixed_pt}), one obtains $\gamma^{(2)}(S^2) = S \times 2\pi$.

\subsection{Three-parameter family: DDKS number}
The DDKS number~(\ref{eq:DDKS}) is an integer-valued topological invariant defined for three-dimensional closed manifolds, independent of any symmetry or group action. In this section, we consider the case where the three-dimensional closed manifold is \( S^3 \), and examine the relationship between the DDKS number and the topological invariants on subspaces invariant under group actions. We introduce the Cartesian coordinates of \( S^3 \) as \( \bm{n} = (n_0, n_1, n_2, n_3) \) with \( \bm{n}^2 = 1 \).

\subsubsection{$C_{2z}T$-equivariance: $\pi$-higher Berry phase}
Consider the case where the following antiunitary \( \mathbb{Z}_2 \) action exists on \( S^3 \):
\begin{align}
    C_{2z}T: \bm{n} \mapsto (n_0,n_1,n_2,-n_3),\quad \phi_{C_{2z}T}=-1.
\end{align}
Since \( C_{2z}T \) is antiunitary and reverses the orientation of \( S^3 \), the higher Berry curvature is invariant under the \( C_{2z}T \) action. Therefore, the integration of the DDKS number can be restricted to the independent region \( D^3 = \{\bm{n} \in S^3 \mid n_3\geq 0\} \), yielding
\begin{align}
    2\pi \nu^{(3)}(S^3) 
    = \left({\cal F}^{(3)}, D^3-C_{2z}T D^3\right) 
    = 2  \left({\cal F}^{(3)}, D^3\right).
\end{align}
Thus,
\begin{align}
    \pi \nu^{(3)}(S^3) 
    \underset{\bmod 2\pi}{\equiv} \left(d {\cal A}^{(0,2)}, D^3\right)
    = \left({\cal A}^{(0,2)}, \p D^3\right).\label{eq:DDKS_Berry_rel}
\end{align}
Therefore, the $\pi$-higher Berry phase on the \( C_{2z}T \)-invariant two-sphere \( \p D^3 = \{\bm{n} \in S^3 \mid n_3=0\} \) gives the parity of the DDKS number.

\subsubsection{$C_{nz}$-equivariance: Pump invariant}
Consider the case where the following unitary \( \mathbb{Z}_n \) action exists on \( S^3 \):
\begin{align}
    C_{nz}:\bm{n}\mapsto  (n_0,\cos \frac{2\pi}{n} n_1 + \sin \frac{2\pi}{n} n_2,-\sin \frac{2\pi}{n} n_1+\cos \frac{2\pi}{n} n_2,n_3), \quad \phi_{C_{nz}}=1.
\end{align}
Since \( C_{nz} \) preserves the orientation of \( S^3 \) and is unitary, the integration of the DDKS number can be restricted to the independent region \( D^3 = \{\bm{n} \in S^3 \mid  n_1\geq \cos \frac{2\pi}{n},n_2\geq 0\} \). The boundary of \( D^3 \) consists of a contractible 2D region \( D^2 = \{\bm{n}\in S^3 \mid n_1\geq 0,n_2=0\} \) and its image under \( C_{nz} \), i.e., \( \p D^3 = D^2 - C_{nz}D^2 \). The boundary of \( D^2 \) is a loop invariant under \( C_{nz} \), \( \p D^2 = \{\bm{n} \in S^3 \mid n_1=n_2=0\} \). Then,
\begin{align}
    \frac{2\pi}{n} \nu^{(3)}(S^3) 
    &\underset{\bmod 2\pi}{\equiv} \left(d{\cal A}^{(0,2)}, D^3\right) 
    = \left({\cal A}^{(0,2)}, D^2-C_{nz}D^2\right) 
    = \left((\delta {\cal A}^{(0,2)})_{C_{nz}}, D^2\right) \nonumber\\
    &= \left(d {\cal A}^{(1,1)}_{C_{nz}}, D^2\right) 
    = \left( {\cal A}^{(1,1)}_{C_{nz}},\p D^2\right) 
    = \eta_{C_{nz}}(\p D^2),\label{eq:DDKS_pump_rel}
\end{align}
establishing the mod-\(n\) relation between the DDKS number and the \( \mathbb{Z}_n \) Thouless pump invariant.

\subsubsection{$T$-equivariance: Fixed point formula (1)}
\label{sec:DD_fixed_pt_formula_Z2T}
As an example of the fixed point formula for the DDKS invariant, we consider the case where a $\Z_2$ group acts on $S^3$ as
\begin{align}
    T: \bm{n} \mapsto (n_0,-n_1,-n_2,-n_3),
\end{align}
and $T$ is an antiunitary operation with $\phi_T = -1$. The only fixed points are $(\pm 1,0,0,0)$. Since $T$ reverses the orientation of $S^3$, no restriction is imposed on the DDKS invariant. Let us define the $q$-dimensional disks $D^q$ and fixed points as follows:
\begin{align}
    &D^3 = \{\bm{n} \in S^3 \mid n_3 \geq 0\}, \label{eq:1d_D3}\\
    &D^2 = \{\bm{n} \in S^3 \mid n_2 \geq 0, n_3 = 0\}, \\
    &D^1 = \{\bm{n} \in S^3 \mid n_1 \geq 0, n_2 = n_3 = 0\}, \\
    &P_\pm = (\pm 1,0,0,0). \label{eq:1d_D0}
\end{align}
Note the chain relations:
\begin{align}
    &\p D^3 = D^2 - T D^2, \\
    &\p D^2 = D^1 + T D^1, \\
    &\p D^1 = P_- - P_+.
\end{align}
Since the independent integration domain of the DDKS invariant $\nu^{(3)}(S^3)$ is $D^3$, we obtain by straightforward calculation:
\begin{align}
    \pi \nu^{(3)}(S^3)
    &= \left({\cal F}^{(0,3)}, D^3\right)
    \underset{\bmod 2\pi}{=} \left(d {\cal A}^{(0,2)}, D^3\right)
    = \left({\cal A}^{(0,2)}, \p D^3\right)
    = \left({\cal A}^{(0,2)}, D^2 + T D^2\right) \nonumber \\
    &= \left(-(\delta {\cal A}^{(0,2)})_T, D^2\right)
    = \left(-d {\cal A}^{(1,1)}_T, D^2\right)
    = \left(-{\cal A}^{(1,1)}_T, D^1 - T D^1\right)
    = \left((\delta {\cal A}^{(1,1)})_{T,T}, D^1\right) \nonumber \\
    &= \left(-d{\cal A}^{(2,0)}_{T,T}, D^1\right)
    = \left(-{\cal A}^{(2,0)}_{T,T}, P_- - P_+\right)
    = \mu^{\rm RP}_T(P_+) - \mu^{\rm RP}_T(P_-). \label{eq:1d_T_fixed_pt_formula}
\end{align}
That is, the parity of the DDKS invariant coincides with the difference of the $\Z_2$ SPT invariants $\mu^{\rm RP}_T$ evaluated at the fixed points $(\pm 1,0,0,0)$.

\subsubsection{$C_{2x} \times C_{2y}$-equivariance: Fixed point formula (2)}

It is known that the Haldane phase is also protected by a unitary $\Z_2 \times \Z_2$ symmetry.
As another example of the fixed point formula for the DDKS invariant, we consider the action of the $\Z_2 \times \Z_2$ group on $S^3$ given by
\begin{align}
    C_{2x}: \bm{n} \mapsto (n_0,n_1,-n_2,-n_3), \\
    C_{2y}: \bm{n} \mapsto (n_0,-n_1,n_2,-n_3),
\end{align}
where both transformations are unitary: $\phi_{C_{2x}}=\phi_{C_{2y}}=1$. As in the previous section, the only fixed points under the $\Z_2 \times \Z_2$ action are $P_\pm = (\pm 1,0,0,0)$. Since the independent integration domain for the DDKS invariant $\nu^{(3)}(S^3)$ is one fourth of $S^3$, a formula modulo 4 is anticipated.

Let us take the independent domain of $S^3$ to be the 3-disk
\begin{align}
    D^3_{*++*} = \{\bm{n} \in S^3 \mid n_1,n_2 \geq 0\}.
\end{align}
Its boundary is separated into two parts:
\begin{align}
    &\partial D^3_{*++*} = -D^2_{*+0*} + D^2_{*0+*}, \\
    &D^2_{*+0*} = \{\bm{n} \in S^3 \mid n_1 \geq 0, n_2=0\}, \\
    &D^2_{*0+*} = \{\bm{n} \in S^3 \mid n_1=0, n_2\geq 0\}.
\end{align}
Each of these can be expressed as a sum of an independent region and its group image:
\begin{align}
    &D^2_{*+0*} = D^2_{*+0+} - C_{2x} D^2_{*+0+}, \quad D^2_{*+0+} = \{\bm{n} \in S^3 \mid n_1 \geq 0, n_2=0, n_3 \geq 0\}, \\
    &D^2_{*0+*} = D^2_{*0+-} - C_{2y} D^2_{*0+-}, \quad D^2_{*0+-} = \{\bm{n} \in S^3 \mid n_1=0, n_2 \geq 0, n_3\leq 0\}.
\end{align}
The boundaries of $D^2_{*+0+}$ and $D^2_{*0+-}$ are
\begin{align}
    &\partial D^2_{*+0+} = D^1_{*+00}-D^1_{*00+},\\
    &\partial D^2_{*0+-} = D^1_{*00-}-D^1_{*0+0} = C_{2y} D^1_{*00+}-D^1_{*0+0},\\
    &D^1_{*00+} = \{\bm{n} \in S^3 \mid n_1,n_2=0, n_3\geq 0\}, \\
    &D^1_{*+00} = \{\bm{n} \in S^3 \mid n_2,n_3=0, n_1\geq 0\}, \\
    &D^1_{*0+0} = \{\bm{n} \in S^3 \mid n_1,n_3=0, n_2\geq 0\}.
\end{align}
A straightforward computation yields the mod 4 formula
\begin{align}
    \frac{\pi}{2} \nu^{(3)}(S^3)
    &= \left(d {\cal A}^{(0,2)},D^3_{*++*}\right)\nonumber\\
    &= \left({\cal A}^{(0,2)},-D^2_{*+0+}+C_{2x} D^2_{*+0+} + D^2_{*0+-}-C_{2y} D^2_{*0+-}\right)\nonumber \\
    &= \left((\delta {\cal A}^{(0,2)})_{C_{2x}},-D^2_{*+0+}\right)+\left((\delta {\cal A}^{(0,2)})_{C_{2y}},D^2_{*0+-}\right)\nonumber \\
    &= \left(d {\cal A}^{(1,1)}_{C_{2x}},-D^2_{*+0+}\right)+\left(d {\cal A}^{(1,1)}_{C_{2y}},D^2_{*0+-}\right)\nonumber \\
    &= \left({\cal A}^{(1,1)}_{C_{2x}},-D^1_{*+00}+D^1_{*00+}\right)+\left({\cal A}^{(1,1)}_{C_{2y}},C_{2y} D^1_{*00+}-D^1_{*0+0}\right) \nonumber\\
    &= \left({\cal A}^{(1,1)}_{C_{2x}},-D^1_{*+00}\right)+\left({\cal A}^{(1,1)}_{C_{2y}},-D^1_{*0+0}\right)
    +\left({\cal A}^{(1,1)}_{C_{2x}C_{2y}}+(\delta {\cal A}^{(1,1)})_{C_{2x},C_{2y}},D^1_{*00+}\right)\nonumber\\
    &= \left({\cal A}^{(1,1)}_{C_{2x}},-D^1_{*+00}\right)+\left({\cal A}^{(1,1)}_{C_{2y}},-D^1_{*0+0}\right)
    +\left({\cal A}^{(1,1)}_{C_{2x}C_{2y}},D^1_{*00+}\right)
    +{\cal A}^{(2,0)}_{C_{2x},C_{2y}}(P_-)-{\cal A}^{(2,0)}_{C_{2x},C_{2y}}(P_+).
    \label{1d:Z2Z2_mod_4}
\end{align}
It is easy to verify that this expression is gauge invariant under the gauge transformation~(\ref{eq:1d_A1_gauge_tr}). The integration path $C=-D^1_{*+00}-D^1_{*0+0}+D^1_{*00+}$ consists of three arcs intersecting at the fixed points $P_\pm$ and thus $C$ is not a cycle and has boundary $\partial C = -P_+ + P_-$. This boundary gauge redundancy is corrected by the boundary term ${\cal A}^{(2,0)}_{C_{2x},C_{2y}}(P_-)-{\cal A}^{(2,0)}_{C_{2x},C_{2y}}(P_+)$, making the whole expression gauge invariant.

Now, although the mod 4 formula~(\ref{1d:Z2Z2_mod_4}) involves integrals over one-dimensional paths, the following fixed point formula modulo 2 is obtained: Using 
\begin{align}
    &(\delta {\cal A}^{(1,1)})_{C_{2x},C_{2x}}(D^1_{*+00})
    = 2 {\cal A}^{(1,1)}_{C_{2x}}(D^1_{*+00}), \\
    &(\delta {\cal A}^{(1,1)})_{C_{2y},C_{2y}}(D^1_{*0+0})
    = 2 {\cal A}^{(1,1)}_{C_{2y}}(D^1_{*0+0}), \\
    &(\delta {\cal A}^{(1,1)})_{C_{2x}C_{2y},C_{2x}C_{2y}}(D^1_{*00+})
    = 2 {\cal A}^{(1,1)}_{C_{2x}C_{2y}}(D^1_{*00+}), 
\end{align}
we obtain
\begin{align}
    \pi \nu^{(3)}(S^3)
    &= \left(2{\cal A}^{(1,1)}_{C_{2x}},-D^1_{*+00}\right)+\left(2{\cal A}^{(1,1)}_{C_{2y}},-D^1_{*0+0}\right)
    +\left(2{\cal A}^{(1,1)}_{C_{2x}C_{2y}},D^1_{*00+}\right)
    +2{\cal A}^{(2,0)}_{C_{2x},C_{2y}}(P_-)-2{\cal A}^{(2,0)}_{C_{2x},C_{2y}}(P_+)\nonumber \\
    &= \left((\delta {\cal A}^{(1,1)})_{C_{2x},C_{2x}},-D^1_{*+00}\right)
    +\left((\delta {\cal A}^{(1,1)})_{C_{2y},C_{2y}},-D^1_{*0+0}\right)
    +\left((\delta {\cal A}^{(1,1)})_{C_{2x}C_{2y},C_{2x}C_{2y}},D^1_{*00+}\right)\nonumber \\
    &\quad +2{\cal A}^{(2,0)}_{C_{2x},C_{2y}}(P_-)-2{\cal A}^{(2,0)}_{C_{2x},C_{2y}}(P_+)\nonumber \\
    &= \left(-d{\cal A}^{(1,1)}_{C_{2x},C_{2x}},-D^1_{*+00}\right)
    +\left(-d {\cal A}^{(1,1)}_{C_{2y},C_{2y}},-D^1_{*0+0}\right)
    +\left(-d{\cal A}^{(1,1)}_{C_{2x}C_{2y},C_{2x}C_{2y}},D^1_{*00+}\right)\nonumber \\
    &\quad +2{\cal A}^{(2,0)}_{C_{2x},C_{2y}}(P_-)-2{\cal A}^{(2,0)}_{C_{2x},C_{2y}}(P_+)\nonumber \\
    &= \left({\cal A}^{(2,0)}_{C_{2x},C_{2x}}(P)+{\cal A}^{(2,0)}_{C_{2y},C_{2y}}(P)
    -{\cal A}^{(2,0)}_{C_{2x}C_{2y},C_{2x}C_{2y}}(P)
    -2{\cal A}^{(2,0)}_{C_{2x},C_{2y}}(P)\right)\Big|_{P=P_-}^{P=P_+} \nonumber \\
    &= \left({\cal A}^{(2,0)}_{C_{2x},C_{2y}}(P)-{\cal A}^{(2,0)}_{C_{2y},C_{2x}}(P)\right)\Big|_{P=P_-}^{P=P_+} \nonumber \\
    &= \mu^{\rm T}_{C_{2x},C_{2y}}(P_+) - \mu^{\rm T}_{C_{2x},C_{2y}}(P_-).
    \label{eq:1d_z2z2_fixed_pt_formula}
\end{align}
Thus, as in the previous section, the parity of the DDKS invariant coincides with the difference of $\Z_2$ topological invariants evaluated at the fixed points $(\pm 1,0,0,0)$.

Together with Sec.~\ref{sec:DD_fixed_pt_formula_Z2T}, this proved that the phase transition between the Haldane phase and the trivial phase, protected by time-reversal symmetry or unitary $\Z_2 \times \Z_2$ symmetry, behaves as a source of higher Berry curvature.

\subsection{Topological invariants and geometric phases defined by group actions}

Neither the higher Berry phase $\gamma^{(2)}(\Sigma^{(2)})$ nor the DDKS number $\nu^{(3)}(\Sigma^{(3)})$ requires $G$-equivariance for their definitions.
In this section, we introduce certain types of higher Berry phases and topological invariants that arise only in the presence of $G$-equivariance.

\subsubsection{Free $G$-action and Higher Berry Phase}

Let us consider the case where a unitary group element $\sigma_1 \in G$ with $\phi_{\sigma_1} = 1$ acts freely on the parameter space $S^1 \times \mathbb{R}$ of a two-dimensional cylinder as
\begin{align}
    \sigma_1: (\theta, t) \mapsto (\theta, t+1).
\end{align}
Then, a gauge-invariant higher Berry phase on a finite cylinder of unit length in the $\mathbb{R}$-direction can be defined using the action of $\sigma_1$ as
\begin{align}
    \gamma^{(2)}\left(S^1 \times [0,1]; \sigma_1\right)
    :=
    \left({\cal A}^{(0,2)}, S^1 \times [0,1]\right) + \left({\cal A}^{(1,1)}_{\sigma_1}, S^1 \times \{0\}\right).
\end{align}
Indeed, under a gauge transformation~\eqref{eq:1d_A1_gauge_tr}, this quantity transforms as
\begin{align}
    \gamma^{(2)}\left(S^1 \times [0,1]\right)
    \mapsto \gamma^{(2)}\left(S^1 \times [0,1]\right)
    + \left(d \chi^{(0,1)}, S^1 \times [0,1]\right)
    + \left((\delta \chi^{(0,1)})_{\sigma_1} - d \chi^{(1,0)}_{\sigma_1}, S^1 \times \{0\}\right),
\end{align}
but the gauge-dependent terms cancel out.

Similarly, consider the planar parameter space $\mathbb{R}^2$ on which two commuting unitary group elements $\sigma_1, \sigma_2 \in G$, with $\sigma_1 \sigma_2 = \sigma_2 \sigma_1$ and $\phi_{\sigma_1} = \phi_{\sigma_2} = 1$, act freely as
\begin{align}
    &\sigma_1: (s, t) \mapsto (s+1, t), \\
    &\sigma_2: (s, t) \mapsto (s, t+1).
\end{align}
In this case, a gauge-invariant higher Berry phase on the unit plaquette $[0,1]^2$ is defined as
\begin{align}
    \gamma^{(2)}\left([0,1]^2; \sigma_1, \sigma_2\right)
    &:=
    \left({\cal A}^{(0,2)}, [0,1]^2\right)
    + \left({\cal A}^{(1,1)}_{\sigma_1}, [0,1] \times \{0\}\right)
    - \left({\cal A}^{(1,1)}_{\sigma_2}, \{0\} \times [0,1]\right) \nonumber\\
    &\quad + {\cal A}^{(2,0)}_{\sigma_1, \sigma_2}(0,0) - {\cal A}^{(2,0)}_{\sigma_2, \sigma_1}(0,0).
\end{align}
It is easy to verify that this expression is invariant under the gauge transformation~\eqref{eq:1d_A1_gauge_tr}.

\subsubsection{Free $\mathbb{Z}_2$ Action and a $\mathbb{Z}_2$ Topological Invariant}
\label{sec:1d_new_z2}

Consider the case where the $\mathbb{Z}_2$ group acts freely on the parameter space $S^3$ (a three-dimensional sphere), that is, the $\mathbb{Z}_2$ element $\sigma$ acts on the Cartesian coordinates $\bm{n} = (n_0, n_1, n_2, n_3)$ of $S^3$ (with $\bm{n}^2 = 1$) as
\begin{align}
    \sigma: \bm{n} \mapsto -\bm{n},
\end{align}
and the action is antiunitary, i.e., $\phi_{\sigma} = -1$. Since the action $\sigma$ preserves the orientation of $S^3$, the DDKS number on $S^3$ vanishes due to~(\ref{eq:1d_DDnumber_constraint}). On the other hand, the following $\mathbb{Z}_2$-valued invariant can be defined:
\begin{align}
    \xi(S^3;\sigma)
    = \frac{1}{2}\left( {\cal F}^{(0,3)}, D^3 \right)
    - \left( {\cal A}^{(0,2)}, D^2 \right)
    - \left( {\cal A}^{(1,1)}_\sigma, D^1 \right)
    - {\cal A}^{(2,0)}_{\sigma,\sigma}(P_+).
\end{align}
Here, $D^3, D^2, D^1, P_+$ are defined in the same way as in~\eqref{eq:1d_D3}–\eqref{eq:1d_D0}. That $\xi(S^3; \sigma)$ takes values in $\mathbb{Z}_2 = \{0, \pi\}$ is directly verified as follows. Noting the boundary relations of the chains:
\begin{align}
    \partial D^3 = D^2 - \sigma D^2, \quad 
    \partial D^2 = D^1 + \sigma D^1, \quad 
    \partial D^1 = \sigma P_+ - P_+,
\end{align}
we compute:
\begin{align}
    2\xi(S^3;\sigma)
    &= \left(d {\cal A}^{(0,2)}, D^3\right)
    - 2\left({\cal A}^{(0,2)}, D^2\right)
    - 2\left({\cal A}^{(1,1)}_\sigma, D^1\right)
    - 2{\cal A}^{(2,0)}_{\sigma,\sigma}(P_+) \nonumber \\
    &= \left({\cal A}^{(0,2)}, D^2 - \sigma D^2\right)
    - 2\left({\cal A}^{(0,2)}, D^2\right)
    - 2\left({\cal A}^{(1,1)}_\sigma, D^1\right)
    - 2{\cal A}^{(2,0)}_{\sigma,\sigma}(P_+) \nonumber \\
    &= \left((\delta {\cal A}^{(0,2)})_\sigma, D^2\right)
    - 2\left({\cal A}^{(1,1)}_\sigma, D^1\right)
    - 2{\cal A}^{(2,0)}_{\sigma,\sigma}(P_+) \nonumber \\
    &= \left(d {\cal A}^{(1,1)}_\sigma, D^2\right)
    - 2\left({\cal A}^{(1,1)}_\sigma, D^1\right)
    - 2{\cal A}^{(2,0)}_{\sigma,\sigma}(P_+) \nonumber \\
    &= \left({\cal A}^{(1,1)}_\sigma, D^1 + \sigma D^1\right)
    - 2\left({\cal A}^{(1,1)}_\sigma, D^1\right)
    - 2{\cal A}^{(2,0)}_{\sigma,\sigma}(P_+) \nonumber \\
    &= \left((\delta {\cal A}^{(1,1)})_{\sigma,\sigma}, D^1 \right)
    - 2{\cal A}^{(2,0)}_{\sigma,\sigma}(P_+) \nonumber \\
    &= \left(-d {\cal A}^{(2,0)}_{\sigma,\sigma}, D^1 \right)
    - 2{\cal A}^{(2,0)}_{\sigma,\sigma}(P_+) \nonumber \\
    &= \left(-{\cal A}^{(2,0)}_{\sigma,\sigma}, \sigma P_+ - P_+ \right)
    - 2{\cal A}^{(2,0)}_{\sigma,\sigma}(P_+) \nonumber \\
    &= (\delta {\cal A}^{(2,0)})_{\sigma,\sigma,\sigma}(P_+) = 0.
\end{align}
Here we used ${\cal A}^{(1,1)}_e(\Delta^1)=0$ and ${\cal A}^{(2,0)}_{e,\sigma}(\Delta^2) = {\cal A}^{(2,0)}_{\sigma,e}(\Delta^2) = 0$. Gauge invariance is similarly established.

If there exists no symmetry that leaves any point inside $S^3$ invariant, then the only allowed type of defect structure is a codimension-4 defect carrying a DDKS number. Since the DDKS number flips sign under the $\sigma$ action, it follows that when the $\mathbb{Z}_2$ invariant $\xi(S^3; \sigma)$ takes the nontrivial value $\pi$, the interior of $S^3$ generically hosts a pair of codimension-4 defects with DDKS numbers $+1$ and $-1$ related under the $\sigma$ action. At the fixed point of the $\mathbb{Z}_2$ action, namely the center of the sphere $S^3$, $\tau = (0,0,0,0) \in \mathbb{R}^4$, the pair annihilation of DDKS defects with values $+1$ and $-1$ is prohibited by the $\mathbb{Z}_2$ equivariance. Thus, a pair of such defects is topologically stable.

\section{Discussion: defect structures in the parameter space}
\label{sec:defect_str}

Finally, we summarize general properties of defect structures that may arise in the parameter space of one-dimensional quantum systems.

In this paper, a “non-defective” parameter point refers to a point $\tau$ at which the Hamiltonian $\hat{H}(\tau)$ is gapped and does not exhibit SSB. Hence, a “defect” corresponds to a point where the system is either gapless or SSB.

\subsection{Symmetry}

Firstly, let us focus on the subgroup $G_0 = \{ g \in G \mid g\tau = \tau \}$ of the symmetry group $G$ that preserves a given parameter point $\tau$. The following types of topological defects can stably exist under the symmetry $G_0$:

\begin{itemize}
    \item If $H^2_{\mathrm{group}}(G_0, \mathbb{R}/2\pi\mathbb{Z})$ is nontrivial: codimension-1 SPT defects can exist.
    \item If $H^1_{\mathrm{group}}(G_0, \mathbb{R}/2\pi\mathbb{Z})$ is nontrivial: codimension-2 pump defects can exist.
    \item If antiunitary symmetries are present: the higher Berry phase $\gamma^{(2)}(S^2) \in \{0,\pi\}$ is quantized, and codimension-3 defects can exist.
    \item If no antiunitary symmetries are present: the DDKS number $\nu^{(3)}(S^3) \in \mathbb{Z}$ leads to codimension-4 defects.
\end{itemize}
Based solely on logical possibilities (ignoring physical constraints that may further restrict realizability), the following general rules hold:

\begin{itemize}
  \item[(i)] If a topological charge $\nu$ characterizing a defect can be decomposed into small charges as $\nu = \nu_1 + \nu_2$ with $|\nu_1|, |\nu_2| < |\nu|$, then the corresponding defect can be split into two or more defects.
  \item[(ii)] When both codimension-$p$ and codimension-$q$ defects ($p < q$) are allowed under a given symmetry, the lower-codimension defect (codimension $p$) is generally realized. Furthermore, since a codimension-$q$ defect is defined on $S^{q-1}$, it may enclose a closed submanifold composed of codimension-$p$ defects (e.g., a conformal manifold) that carry the topological charge of the codimension-$q$ defect.
\end{itemize}

As an example of rule (i), a codimension-4 defect with DDKS number $\nu^{(3)}(S^3) = 2$ can be decomposed into two separated defects in the parameter space, each carrying $\nu = 1$.

As for rule (ii), in the presence of time-reversal symmetry, codimension-3 defects characterized by $\gamma^{(2)}(S^2) = \pi$ cannot generally be realized without fine-tuning; instead, they appear as closed surfaces formed by codimension-1 defects inside $S^2$. Similarly, under a unitary $\mathbb{Z}_2$ symmetry, the closed surface of codimension-2 defects corresponding to a Thouless pump may exist inside $S^3$, collectively carrying a nontrivial DDKS number.

Such general rules for defect structures are well-known in the context of Fermi surface topology in band theory, where lower-codimension defects emerge as Fermi surfaces under generic Bloch Hamiltonians respecting symmetry. On the other hand, in one-dimensional quantum many-body systems, the extent to which such rules apply remains an open question, as it may involve SSB defects and renormalization group flows.

\subsection{$G$-Equivariance}

Let us consider the case where the symmetry group $G_0$ that leaves a parameter point invariant is a subgroup of a larger group $G$ with an equivariant structure. As discussed in Sec.~\ref{sec:MPS_inv_with_G}, topological invariants defined under the lower symmetry $G_0$ may be related to those defined in the $G$-invariant subspace under the full group action. On the other hand, as shown in Sec.~\ref{sec:1d_new_z2}, there also exist topological invariants that are defined only in the presence of $G$-equivariance.

We do not repeat the discussion of the latter type of defect structures here, since they were commented on at the end of Sec.~\ref{sec:1d_new_z2}. Instead, we elaborate on the former case.

In general, suppose we have a subgroup inclusion $H \subset G$. Whether a codimension-$q$ defect defined in the $G$-invariant subspace ${\cal M}^G = \{ \tau \in {\cal M} \mid g\tau = \tau,\ g \in G \}$ can be stably extended into the $H$-invariant subspace ${\cal M}^H = \{ \tau \in {\cal M} \mid g\tau = \tau,\ g \in H \}$ is determined by the compatibility between the sets of topological invariants $\{ \nu_{G,i}(S^{q-1}) \}_i$ defined under $G$ and $\{ \nu_{H,i}(S^{q-1}) \}_i$ defined under $H$. Since the $G$-symmetry implies the $H$-symmetry, there exists a homomorphism between topological invariants 
\begin{align}
f: \{ \nu_{G,i}(S^{q-1}) \}_i \to \{ \nu_{H,i}(S^{q-1}) \}_i,    
\end{align}
and the kernel of this map, i.e., those defect charges $x$ such that $f(x) = 0$, corresponds to defects that are unstable under the $H$-symmetry and thus disappear in ${\cal M}^H$. In contrast, if $f(x) \neq 0$, then the same codimension-$q$ defect structure remains stable.

\subsubsection{Example}

Let us consider the chain of subgroups of the equivariant structure of the model in Eq.~\eqref{eq:1d_model}:
\begin{align}
    \{e\} \subset G'' = \Z_2[C_{2z}T] \subset G' = \Z_2[C_{2z}T] \times \Z_2[C_{2y}T] \subset G = \Z_2[C_{2z}T] \times \Z_2[C_{2y}T] \times \Z_2[T].
\end{align}
We take the parameter space to be a four-dimensional vector $\bm{n} \in \mathbb{R}^4$. The $G$-invariant subspaces are given by:
\begin{align}
    &(\mathbb{R}^4)^G = \{(x_0, 0, 0, 0) \in \mathbb{R}^4 \mid x_0 \in \mathbb{R}\}, \\
    &(\mathbb{R}^4)^{G'} = \{(x_0, x_1, 0, 0) \in \mathbb{R}^4 \mid x_0, x_1 \in \mathbb{R}\}, \\
    &(\mathbb{R}^4)^{G''} = \{(x_0, x_1, x_2, 0) \in \mathbb{R}^4 \mid x_0, x_1, x_2 \in \mathbb{R}\}.
\end{align}
On the line $(\mathbb{R}^4)^G$, four $\mathbb{Z}_2$ SPT invariants are defined: $(\mu^{\rm RP}_{C_{2z}T}, \mu^{\rm RP}_{C_{2y}T}, \mu^{\rm RP}_{T}, \mu^{\rm T}_{C_{2z},C_{2x}})$. On the plane $(\mathbb{R}^4)^{G'}$, two $\mathbb{Z}_2$ SPT invariants $(\mu^{\rm RP}_{C_{2z}T}, \mu^{\rm RP}_{C_{2y}T})$ and one $\mathbb{Z}_2$ pump invariant $\eta_{C_{2x}}$ are defined. On the 3D space $(\mathbb{R}^4)^{G''}$, one $\mathbb{Z}_2$ SPT invariant $\mu^{\rm RP}_{C_{2z}T}$ and a quantized higher Berry phase $\gamma^{(2)}$ are defined. Accordingly, the defect structure in the full parameter space $\mathbb{R}^4$ depends on the topological charges on the line $(\mathbb{R}^4)^G$ as follows:
\begin{itemize}
    \item For $(\mu^{\rm RP}_{C_{2z}T}, \mu^{\rm RP}_{C_{2y}T}, \mu^{\rm RP}_{T}, \mu^{\rm T}_{C_{2z},C_{2x}}) = (0, 0, \pi, 0)$ or $(0, 0, 0, \pi)$, the defect does not extend along the $\mathbb{R}^4$ directions and becomes a codimension-4 defect, where the DDKS number $\nu^{(3)}(S^3)$ is defined. This is consistent with the fixed-point formulas \eqref{eq:1d_T_fixed_pt_formula} and \eqref{eq:1d_z2z2_fixed_pt_formula}.
    
    \item For $(\mu^{\rm RP}_{C_{2z}T}, \mu^{\rm RP}_{C_{2y}T}, \mu^{\rm RP}_{T}, \mu^{\rm T}_{C_{2z},C_{2x}}) = (0, \pi, 0, 0)$, the defect extends into $(\mathbb{R}^4)^{G'}$ and becomes a codimension-2 defect.
    
    \item For $(\mu^{\rm RP}_{C_{2z}T}, \mu^{\rm RP}_{C_{2y}T}, \mu^{\rm RP}_{T}, \mu^{\rm T}_{C_{2z},C_{2x}}) = (\pi, 0, 0, 0)$, the defect extends into $(\mathbb{R}^4)^{G''}$ and becomes a codimension-3 defect.
\end{itemize}
Moreover, since the pump invariant $\eta_{C_{2x}}$ under symmetry $G'$ is not defined under $G''$, we obtain:
\begin{itemize}
    \item A point defect in the plane $(\mathbb{R}^4)^{G'}$ with $\eta_{C_{2x}} = \pi$ cannot extend into $\mathbb{R}^4$ and thus becomes a codimension-4 defect, where the DDKS number is defined. This is consistent with the relation \eqref{eq:DDKS_pump_rel} between the DDKS number and the pump invariant.
\end{itemize}
Finally, the quantized higher Berry phase $\gamma^{(2)}$ under symmetry $G''$ is no longer quantized once the symmetry is broken. Thus:
\begin{itemize}
    \item A point defect in the three-dimensional space $(\mathbb{R}^4)^{G''}$ with $\gamma^{(2)} = \pi$ cannot extend into $\mathbb{R}^4$ and hence becomes a codimension-4 defect, where the DDKS number is defined. This is consistent with the relation \eqref{eq:DDKS_Berry_rel} between the DDKS number and the quantized higher Berry phase.
\end{itemize}
We remark that the above hierarchical structure is consistent with the nonlinear $\sigma$-model description of SPT phases protected by symmetries defined on the sphere target space~\cite{BiClassification2015}.

\section{Conclusion}

In this work, we have developed a discrete formulation of higher Berry connection and curvature that incorporates the $G$-equivariant structure of the symmetry group $G$, and systematically analyzed the structure of topological defects in the parameter space of one-dimensional quantum spin systems.

In Sec.~\ref{sec:1d}, we constructed a framework to incorporate $G$-equivariance into the discrete formulation of the higher Berry connection. We explicitly formulated the gauge transformation rules under group actions for the definitions of higher Berry connections and curvatures, and confirmed that the resulting constructions on the discretized parameter space are naturally described in terms of $G$-simplicial complexes. This framework enables a systematic discussion of various topological invariants, including their behavior under equivariant symmetry and the construction of topological invariants that require equivariance for their definition.

In Sec.~\ref{sec:MPS_inv_with_G}, we derived the relation between the DDKS number and the topological invariants defined on the invariant subspaces of the group action. In particular, we derived a fixed point formula in the case where the group action has isolated fixed points. Using this result, we explicitly demonstrated that the quantum critical point between the Haldane phase and the trivial phase manifests as an emergence of higher Berry curvature.

In Sec.~\ref{sec:defect_str}, we systematically discussed the structure of topological defects in the theory space (i.e., parameter space) of quantum spin systems. In particular, we illustrated with examples how defect structures are constrained and organized according to the subgroup relations among symmetries.

\section*{Acknowledgements}
We thank Keisuke Totsuka for helpful discussions. 
We were supported by JST CREST Grant No. JPMJCR19T2, and JSPS KAKENHI Grant No. JP22H05118 and JP23H01097.

\appendix
\section{Simplicial $G$-complex}
\label{app:double_complex}

Let ${\cal M}$ be a simplicial complex and $A$ an abelian group. A $q$-simplex is denoted in shorthand as $\Delta^q = (\tau_0,\dots,\tau_q)$. The $j$-th face of a $q$-simplex is written as
\begin{align}
\partial_j \Delta^{q} := (\tau_0,\dots,\tau_{j-1},\tau_{j+1},\dots,\tau_{q}).
\end{align}
The coboundary operator $d: C^q({\cal M}, A) \to C^{q+1}({\cal M}, A)$ acting on the cochain complex $C^q({\cal M}, A)$ is defined by
\begin{align}
(df)(\Delta^{q+1}) = \sum_{j=0}^{q+1} (-1)^j f(\partial_j \Delta^{q+1}).
\end{align}
Let $G$ be a discrete group with a left action on ${\cal M}$ written as $(g, \tau) \mapsto g \tau$. Let $G$ also act on $A$ from the left, denoted by $(g, a) \mapsto g.a$. Then, left and right actions of $G$ on the cochain complex $C^q({\cal M}, A)$ are defined by
\begin{align}
(g \cdot f)(\Delta^q) := g. \big(f(\Delta^q)\big), \quad 
(f \cdot g)(\Delta^q) := f(g\Delta^q), \quad f \in C^q({\cal M}, A),
\end{align}
where we write $g \Delta^q := (g \tau_0, \dots, g \tau_q)$ as shorthand.
Based on this structure, we define the double complex combining group cochains and spatial cochains as
\begin{align}
C^{p,q}_G({\cal M},A) := C^p(G, C^q({\cal M}, A)).
\end{align}
For $f \in C^{p,q}$, the differential in the spatial direction $d: C^{p,q}_G({\cal M},A) \to C^{p,q+1}_G({\cal M},A)$ is given by the standard cochain differential:
\begin{align}
(df)_{g_1,\dots,g_p}(\Delta^{q+1}) = \sum_{j=0}^{q+1} (-1)^j f_{g_1,\dots,g_p}(\partial_j \Delta^{q+1}).
\end{align}
On the other hand, the differential in the group direction $\delta: C^{p,q}_G({\cal M},A) \to C^{p+1,q}_G({\cal M},A)$ is defined by the group cochain differential as
\begin{align}
(\delta f)_{g_1,\dots,g_{p+1}} = 
&\ g_1 \cdot f_{g_2,\dots,g_{p+1}} 
- f_{g_1 g_2, g_3,\dots,g_{p+1}} 
+ \cdots \nonumber\\
&+ (-1)^{p-1} f_{g_1,\dots,g_{p-1}, g_p g_{p+1}} 
+ (-1)^p f_{g_1,\dots,g_p}\cdot g_{p+1}.
\end{align}
In low degrees, this gives:
\begin{align}
&(\delta f)_g(\Delta^q) = g. \big(f(\Delta^q)\big) - f(g\Delta^q), \quad f \in C^{0,q}_G({\cal M},A), \\
&(\delta f)_{g,h}(\Delta^q) = g. \big(f_h(\Delta^q)\big) - f_{gh}(\Delta^q) + f_g(h\Delta^q), \quad f \in C^{1,q}_G({\cal M},A), \\
&(\delta f)_{g,h,k}(\Delta^q) = g. \big(f_{h,k}(\Delta^q)\big) - f_{gh,k}(\Delta^q) + f_{g,hk}(\Delta^q) - f_{g,h}(k\Delta^q), \quad f \in C^{2,q}_G({\cal M},A).
\end{align}
It is important to note that the differentials commute:
\begin{align}
d \delta = \delta d.
\end{align}

The total differential on the entire double complex
\begin{align}
D: \bigoplus_{p+q=n} C^{p,q}_G({\cal M},A) \to \bigoplus_{p+q=n+1} C^{p,q}_G({\cal M},A)
\end{align}
is defined as
\begin{align}
D := \delta + (-1)^p d.
\end{align}
That is, for an $n$-cochain
\begin{align}
f^n = (f^{0,n}, f^{1,n-1}, \dots, f^{n,0}), \quad f^{p,q} \in C^{p,q}_G({\cal M},A),
\end{align}
the total differential is given by
\begin{align}
Df^n = (d f^{0,n},\ \delta f^{0,n} - d f^{1,n-1},\ \dots,\ \delta f^{n-1,1} + (-1)^n d f^{n,0},\ \delta f^{n,0}).
\end{align}
This total differential satisfies $D^2 = 0$. Thus, we define the $D$-coboundaries and $D$-cocycles by
\begin{align}
&B^n_G({\cal M},A) := \{ D f^{n-1} \mid f^{n-1} \in \bigoplus_{p+q=n-1} C^{p,q}_G({\cal M},A) \},\\
&Z^n_G({\cal M},A) := \{ f^n \in \bigoplus_{p+q=n} C^{p,q}_G({\cal M},A) \mid D f^n = 0 \},
\end{align}
and the equivariant cohomology group is defined by
\begin{align}
H^n_G({\cal M},A) := Z^n_G({\cal M},A) / B^n_G({\cal M},A).
\end{align}

\subsection{Families of pure states}

The gauge transformation and cocycle condition for $G$-equivariant families of pure states, as discussed in Section~\ref{sec:0d}, can be summarized from the perspective of the double complex as follows.

Let the coefficient abelian group be $A = \R/2\pi\Z$, and define the left action of the group $G$ on $\R/2\pi\Z$ via the homomorphism $\phi: G \to \Z_2$ (cf.~Eq.~\eqref{eq:0d_phi_g}) that specifies whether each $g \in G$ is unitary or anti-unitary:
\begin{align}
g.\theta = \phi_g \theta,\quad \theta \in \R/2\pi \Z, \quad g \in G. \label{eq:app_left_action_co}
\end{align}
The discrete Berry connection~\eqref{eq:0d_DBC} and the $U(1)$ phase~\eqref{eq:0d_G_phase} define a pair of cochains of bidegree $(0,1)$ and $(1,0)$:
\begin{align}
{\sf A} = (A,\alpha) \in C^{0,1}_G({\cal M},\R/2\pi \Z) \oplus C^{1,0}_G({\cal M},\R/2\pi \Z).
\end{align}
The gauge transformation~\eqref{eq:0d_gauge_tr} is generated by a $(0,0)$-cochain
\begin{align}
\chi \in C^{0,0}_G({\cal M},\R/2\pi \Z),
\end{align}
and is compactly written as
\begin{align}
{\sf A} \mapsto {\sf A} + D \chi.
\end{align}
The Berry curvature is given by a $(0,2)$-cocycle with real coefficients:
\begin{align}
F \in C^{0,2}_G({\cal M},\R),\quad dF=0.
\end{align}
The cocycle condition for the cochain ${\sf A}$ is written as
\begin{align}
D {\sf A} \equiv F \quad \bmod 2\pi.
\end{align}

\subsection{Families of MPS}

The gauge transformation and cocycle condition for families of injective MPS with $G$-equivariance, as discussed in Section~\ref{sec:1d}, can similarly be formulated using the double complex.

As in the previous subsection, let the coefficient abelian group be $A = \R/2\pi \Z$ and define the left $G$-action on $\R/2\pi \Z$ via Eq.~\eqref{eq:app_left_action_co}.
The $(0,1)$ and $(1,0)$ cochains are defined as:
\begin{align}
{\sf A}^{(1)} = ({\cal A}^{(0,1)},{\cal A}^{(1,0)}) \in C^{0,1}_G({\cal M},\R/2\pi \Z) \oplus C^{1,0}_G({\cal M},\R/2\pi \Z),
\end{align}
and transform under gauge transformation as
\begin{align}
{\sf A}^{(1)} \mapsto {\sf A}^{(1)} - D \chi^{(0,0)},\quad \chi^{(0,0)} \in C^{0,0}_G({\cal M},\R/2\pi \Z).
\end{align}
The Berry curvature is a $(0,2)$-cocycle with real coefficients:
\begin{align}
{\cal F}^{(2)} \in C^{0,2}_G({\cal M},\R),\quad d{\cal F}^{(2)}=0.
\end{align}
The cocycle condition is given by
\begin{align}
D {\sf A}^{(1)} \equiv {\cal F}^{(2)} \quad \bmod 2\pi.
\end{align}
The cochains of bidegree $(0,2)$, $(1,1)$, and $(2,0)$ are collectively organized as a degree-2 $D$-cochain:
\begin{align}
{\sf A}^{(2)} = ({\cal A}^{(0,2)},{\cal A}^{(1,1)},{\cal A}^{(2,0)}) \in C^{0,2}_G({\cal M},\R/2\pi \Z) \oplus C^{1,1}_G({\cal M},\R/2\pi \Z) \oplus C^{2,0}_G({\cal M},\R/2\pi \Z).
\end{align}
The corresponding gauge transformation is
\begin{align}
{\sf A}^{(2)} \mapsto {\sf A}^{(2)} + D {\sf X}^{(1)},\quad 
{\sf X}^{(1)} = (\chi^{(0,1)},\chi^{(1,0)}) \in C^{0,1}_G({\cal M},\R/2\pi \Z)\oplus C^{1,0}_G({\cal M},\R/2\pi \Z).
\end{align}
The higher Berry curvature is a $(0,3)$-cocycle with real coefficients:
\begin{align}
{\cal F}^{(3)} \in C^{0,3}_G({\cal M},\R),\quad d{\cal F}^{(3)}=0,
\end{align}
and the cocycle condition is
\begin{align}
D {\sf A}^{(2)} \equiv {\cal F}^{(3)} \quad \bmod 2\pi.
\end{align}

\bibliography{refs}

\begin{thebibliography}{56}%
\makeatletter
\providecommand \@ifxundefined [1]{%
 \@ifx{#1\undefined}
}%
\providecommand \@ifnum [1]{%
 \ifnum #1\expandafter \@firstoftwo
 \else \expandafter \@secondoftwo
 \fi
}%
\providecommand \@ifx [1]{%
 \ifx #1\expandafter \@firstoftwo
 \else \expandafter \@secondoftwo
 \fi
}%
\providecommand \natexlab [1]{#1}%
\providecommand \enquote  [1]{``#1''}%
\providecommand \bibnamefont  [1]{#1}%
\providecommand \bibfnamefont [1]{#1}%
\providecommand \citenamefont [1]{#1}%
\providecommand \href@noop [0]{\@secondoftwo}%
\providecommand \href [0]{\begingroup \@sanitize@url \@href}%
\providecommand \@href[1]{\@@startlink{#1}\@@href}%
\providecommand \@@href[1]{\endgroup#1\@@endlink}%
\providecommand \@sanitize@url [0]{\catcode `\\12\catcode `\$12\catcode
  `\&12\catcode `\#12\catcode `\^12\catcode `\_12\catcode `\%12\relax}%
\providecommand \@@startlink[1]{}%
\providecommand \@@endlink[0]{}%
\providecommand \url  [0]{\begingroup\@sanitize@url \@url }%
\providecommand \@url [1]{\endgroup\@href {#1}{\urlprefix }}%
\providecommand \urlprefix  [0]{URL }%
\providecommand \Eprint [0]{\href }%
\providecommand \doibase [0]{http://dx.doi.org/}%
\providecommand \selectlanguage [0]{\@gobble}%
\providecommand \bibinfo  [0]{\@secondoftwo}%
\providecommand \bibfield  [0]{\@secondoftwo}%
\providecommand \translation [1]{[#1]}%
\providecommand \BibitemOpen [0]{}%
\providecommand \bibitemStop [0]{}%
\providecommand \bibitemNoStop [0]{.\EOS\space}%
\providecommand \EOS [0]{\spacefactor3000\relax}%
\providecommand \BibitemShut  [1]{\csname bibitem#1\endcsname}%
\let\auto@bib@innerbib\@empty
\bibitem [{\citenamefont {Teo}\ and\ \citenamefont
  {Kane}(2010)}]{TeoTopological2010}%
  \BibitemOpen
  \bibfield  {author} {\bibinfo {author} {\bibfnamefont {J.~C.~Y.}\
  \bibnamefont {Teo}}\ and\ \bibinfo {author} {\bibfnamefont {C.~L.}\
  \bibnamefont {Kane}},\ }\href {\doibase 10.1103/PhysRevB.82.115120}
  {\bibfield  {journal} {\bibinfo  {journal} {Physical Review B}\ }\textbf
  {\bibinfo {volume} {82}},\ \bibinfo {pages} {115120} (\bibinfo {year}
  {2010})}\BibitemShut {NoStop}%
\bibitem [{\citenamefont {Kitaev}(2011)}]{KitaevTopological2011}%
  \BibitemOpen
  \bibfield  {author} {\bibinfo {author} {\bibfnamefont {A.}~\bibnamefont
  {Kitaev}},\ }\href {http://scgp.stonybrook.edu/archives/1087} {\enquote
  {\bibinfo {title} {Toward a topological classification of manybody quantum
  states with short-range entanglement},}\ } (\bibinfo {year}
  {2011})\BibitemShut {NoStop}%
\bibitem [{\citenamefont {Kitaev}(2013)}]{KitaevClassification2013}%
  \BibitemOpen
  \bibfield  {author} {\bibinfo {author} {\bibfnamefont {A.}~\bibnamefont
  {Kitaev}},\ }\href {http://scgp.stonybrook.edu/archives/16180} {\enquote
  {\bibinfo {title} {On the classification of short-range entangled states},}\
  } (\bibinfo {year} {2013})\BibitemShut {NoStop}%
\bibitem [{\citenamefont {Kitaev}(2015)}]{KitaevHomotopytheoretic2015}%
  \BibitemOpen
  \bibfield  {author} {\bibinfo {author} {\bibfnamefont {A.}~\bibnamefont
  {Kitaev}},\ }\href
  {https://www.ipam.ucla.edu/programs/workshops/symmetry-and-topology-in-quantum-matter/}
  {\enquote {\bibinfo {title} {Homotopy-theoretic approach to spt phases in
  action: {{Z16}} classification of threedimensional superconductors},}\ }
  (\bibinfo {year} {2015})\BibitemShut {NoStop}%
\bibitem [{\citenamefont {Kubota}(2025)}]{KubotaStable2025}%
  \BibitemOpen
  \bibfield  {author} {\bibinfo {author} {\bibfnamefont {Y.}~\bibnamefont
  {Kubota}},\ }\href {\doibase 10.48550/arXiv.2503.12618} {\enquote {\bibinfo
  {title} {Stable homotopy theory of invertible gapped quantum spin systems
  {{I}}: {{Kitaev}}'s \${{{\textohm}}}\$-spectrum},}\ } (\bibinfo {year}
  {2025}),\ \Eprint {http://arxiv.org/abs/2503.12618} {arXiv:2503.12618
  [math-ph]} \BibitemShut {NoStop}%
\bibitem [{\citenamefont {Gaiotto}\ and\ \citenamefont
  {{Johnson-Freyd}}(2019)}]{GaiottoSymmetry2019}%
  \BibitemOpen
  \bibfield  {author} {\bibinfo {author} {\bibfnamefont {D.}~\bibnamefont
  {Gaiotto}}\ and\ \bibinfo {author} {\bibfnamefont {T.}~\bibnamefont
  {{Johnson-Freyd}}},\ }\href {\doibase 10.1007/JHEP05(2019)007} {\bibfield
  {journal} {\bibinfo  {journal} {Journal of High Energy Physics}\ }\textbf
  {\bibinfo {volume} {2019}},\ \bibinfo {pages} {7} (\bibinfo {year}
  {2019})}\BibitemShut {NoStop}%
\bibitem [{\citenamefont {Xiong}(2018)}]{XiongMinimalist2018}%
  \BibitemOpen
  \bibfield  {author} {\bibinfo {author} {\bibfnamefont {C.~Z.}\ \bibnamefont
  {Xiong}},\ }\href {\doibase 10.1088/1751-8121/aae0b1} {\bibfield  {journal}
  {\bibinfo  {journal} {Journal of Physics A: Mathematical and Theoretical}\
  }\textbf {\bibinfo {volume} {51}},\ \bibinfo {pages} {445001} (\bibinfo
  {year} {2018})}\BibitemShut {NoStop}%
\bibitem [{\citenamefont {Roy}\ and\ \citenamefont
  {Harper}(2017)}]{RoyFloquet2017}%
  \BibitemOpen
  \bibfield  {author} {\bibinfo {author} {\bibfnamefont {R.}~\bibnamefont
  {Roy}}\ and\ \bibinfo {author} {\bibfnamefont {F.}~\bibnamefont {Harper}},\
  }\href {\doibase 10.1103/PhysRevB.95.195128} {\bibfield  {journal} {\bibinfo
  {journal} {Physical Review B}\ }\textbf {\bibinfo {volume} {95}},\ \bibinfo
  {pages} {195128} (\bibinfo {year} {2017})}\BibitemShut {NoStop}%
\bibitem [{\citenamefont {Kikuchi}\ and\ \citenamefont
  {Tanizaki}(2017)}]{KikuchiGlobal2017}%
  \BibitemOpen
  \bibfield  {author} {\bibinfo {author} {\bibfnamefont {Y.}~\bibnamefont
  {Kikuchi}}\ and\ \bibinfo {author} {\bibfnamefont {Y.}~\bibnamefont
  {Tanizaki}},\ }\href {\doibase 10.1093/ptep/ptx148} {\bibfield  {journal}
  {\bibinfo  {journal} {Progress of Theoretical and Experimental Physics}\
  }\textbf {\bibinfo {volume} {2017}},\ \bibinfo {pages} {113B05} (\bibinfo
  {year} {2017})}\BibitemShut {NoStop}%
\bibitem [{\citenamefont {C{\'o}rdova}\ \emph
  {et~al.}(2020{\natexlab{a}})\citenamefont {C{\'o}rdova}, \citenamefont
  {Freed}, \citenamefont {Lam},\ and\ \citenamefont
  {Seiberg}}]{CordovaAnomalies2020}%
  \BibitemOpen
  \bibfield  {author} {\bibinfo {author} {\bibfnamefont {C.}~\bibnamefont
  {C{\'o}rdova}}, \bibinfo {author} {\bibfnamefont {D.}~\bibnamefont {Freed}},
  \bibinfo {author} {\bibfnamefont {H.~T.}\ \bibnamefont {Lam}}, \ and\
  \bibinfo {author} {\bibfnamefont {N.}~\bibnamefont {Seiberg}},\ }\href
  {\doibase 10.21468/SciPostPhys.8.1.001} {\bibfield  {journal} {\bibinfo
  {journal} {SciPost Physics}\ }\textbf {\bibinfo {volume} {8}},\ \bibinfo
  {pages} {001} (\bibinfo {year} {2020}{\natexlab{a}})}\BibitemShut {NoStop}%
\bibitem [{\citenamefont {C{\'o}rdova}\ \emph
  {et~al.}(2020{\natexlab{b}})\citenamefont {C{\'o}rdova}, \citenamefont
  {Freed}, \citenamefont {Lam},\ and\ \citenamefont
  {Seiberg}}]{CordovaAnomalies2020a}%
  \BibitemOpen
  \bibfield  {author} {\bibinfo {author} {\bibfnamefont {C.}~\bibnamefont
  {C{\'o}rdova}}, \bibinfo {author} {\bibfnamefont {D.}~\bibnamefont {Freed}},
  \bibinfo {author} {\bibfnamefont {H.~T.}\ \bibnamefont {Lam}}, \ and\
  \bibinfo {author} {\bibfnamefont {N.}~\bibnamefont {Seiberg}},\ }\href
  {\doibase 10.21468/SciPostPhys.8.1.002} {\bibfield  {journal} {\bibinfo
  {journal} {SciPost Physics}\ }\textbf {\bibinfo {volume} {8}},\ \bibinfo
  {pages} {002} (\bibinfo {year} {2020}{\natexlab{b}})}\BibitemShut {NoStop}%
\bibitem [{\citenamefont {Hsin}\ \emph {et~al.}(2020)\citenamefont {Hsin},
  \citenamefont {Kapustin},\ and\ \citenamefont {Thorngren}}]{HsinBerry2020}%
  \BibitemOpen
  \bibfield  {author} {\bibinfo {author} {\bibfnamefont {P.-S.}\ \bibnamefont
  {Hsin}}, \bibinfo {author} {\bibfnamefont {A.}~\bibnamefont {Kapustin}}, \
  and\ \bibinfo {author} {\bibfnamefont {R.}~\bibnamefont {Thorngren}},\ }\href
  {\doibase 10.1103/PhysRevB.102.245113} {\bibfield  {journal} {\bibinfo
  {journal} {Physical Review B}\ }\textbf {\bibinfo {volume} {102}},\ \bibinfo
  {pages} {245113} (\bibinfo {year} {2020})}\BibitemShut {NoStop}%
\bibitem [{\citenamefont {Kapustin}\ and\ \citenamefont
  {Spodyneiko}(2020{\natexlab{a}})}]{KapustinHigherdimensional2020a}%
  \BibitemOpen
  \bibfield  {author} {\bibinfo {author} {\bibfnamefont {A.}~\bibnamefont
  {Kapustin}}\ and\ \bibinfo {author} {\bibfnamefont {L.}~\bibnamefont
  {Spodyneiko}},\ }\href {\doibase 10.48550/arXiv.2003.09519} {\enquote
  {\bibinfo {title} {Higher-dimensional generalizations of the {{Thouless}}
  charge pump},}\ } (\bibinfo {year} {2020}{\natexlab{a}}),\ \Eprint
  {http://arxiv.org/abs/2003.09519} {arXiv:2003.09519 [cond-mat]} \BibitemShut
  {NoStop}%
\bibitem [{\citenamefont {Shiozaki}(2022)}]{ShiozakiAdiabatic2022}%
  \BibitemOpen
  \bibfield  {author} {\bibinfo {author} {\bibfnamefont {K.}~\bibnamefont
  {Shiozaki}},\ }\href {\doibase 10.1103/PhysRevB.106.125108} {\bibfield
  {journal} {\bibinfo  {journal} {Physical Review B}\ }\textbf {\bibinfo
  {volume} {106}},\ \bibinfo {pages} {125108} (\bibinfo {year}
  {2022})}\BibitemShut {NoStop}%
\bibitem [{\citenamefont {Yao}\ \emph {et~al.}(2022)\citenamefont {Yao},
  \citenamefont {Oshikawa},\ and\ \citenamefont
  {Furusaki}}]{YaoGappability2022}%
  \BibitemOpen
  \bibfield  {author} {\bibinfo {author} {\bibfnamefont {Y.}~\bibnamefont
  {Yao}}, \bibinfo {author} {\bibfnamefont {M.}~\bibnamefont {Oshikawa}}, \
  and\ \bibinfo {author} {\bibfnamefont {A.}~\bibnamefont {Furusaki}},\ }\href
  {\doibase 10.1103/PhysRevLett.129.017204} {\bibfield  {journal} {\bibinfo
  {journal} {Physical Review Letters}\ }\textbf {\bibinfo {volume} {129}},\
  \bibinfo {pages} {017204} (\bibinfo {year} {2022})}\BibitemShut {NoStop}%
\bibitem [{\citenamefont {Choi}\ and\ \citenamefont
  {Ohmori}(2022)}]{ChoiHigher2022}%
  \BibitemOpen
  \bibfield  {author} {\bibinfo {author} {\bibfnamefont {Y.}~\bibnamefont
  {Choi}}\ and\ \bibinfo {author} {\bibfnamefont {K.}~\bibnamefont {Ohmori}},\
  }\href {\doibase 10.1007/JHEP09(2022)022} {\bibfield  {journal} {\bibinfo
  {journal} {Journal of High Energy Physics}\ }\textbf {\bibinfo {volume}
  {2022}},\ \bibinfo {pages} {22} (\bibinfo {year} {2022})}\BibitemShut
  {NoStop}%
\bibitem [{\citenamefont {Bachmann}\ \emph {et~al.}(2024)\citenamefont
  {Bachmann}, \citenamefont {De~Roeck}, \citenamefont {Fraas},\ and\
  \citenamefont {Jappens}}]{BachmannClassification2024}%
  \BibitemOpen
  \bibfield  {author} {\bibinfo {author} {\bibfnamefont {S.}~\bibnamefont
  {Bachmann}}, \bibinfo {author} {\bibfnamefont {W.}~\bibnamefont {De~Roeck}},
  \bibinfo {author} {\bibfnamefont {M.}~\bibnamefont {Fraas}}, \ and\ \bibinfo
  {author} {\bibfnamefont {T.}~\bibnamefont {Jappens}},\ }\href {\doibase
  10.1007/s00220-024-05010-w} {\bibfield  {journal} {\bibinfo  {journal}
  {Communications in Mathematical Physics}\ }\textbf {\bibinfo {volume}
  {405}},\ \bibinfo {pages} {157} (\bibinfo {year} {2024})}\BibitemShut
  {NoStop}%
\bibitem [{\citenamefont {Ohyama}\ \emph {et~al.}(2024)\citenamefont {Ohyama},
  \citenamefont {Terashima},\ and\ \citenamefont
  {Shiozaki}}]{OhyamaDiscrete2024}%
  \BibitemOpen
  \bibfield  {author} {\bibinfo {author} {\bibfnamefont {S.}~\bibnamefont
  {Ohyama}}, \bibinfo {author} {\bibfnamefont {Y.}~\bibnamefont {Terashima}}, \
  and\ \bibinfo {author} {\bibfnamefont {K.}~\bibnamefont {Shiozaki}},\ }\href
  {\doibase 10.1103/PhysRevB.110.035114} {\bibfield  {journal} {\bibinfo
  {journal} {Physical Review B}\ }\textbf {\bibinfo {volume} {110}},\ \bibinfo
  {pages} {035114} (\bibinfo {year} {2024})}\BibitemShut {NoStop}%
\bibitem [{\citenamefont {Debray}\ \emph {et~al.}(2024)\citenamefont {Debray},
  \citenamefont {Devalapurkar}, \citenamefont {Krulewski}, \citenamefont {Liu},
  \citenamefont {{Pacheco-Tallaj}},\ and\ \citenamefont
  {Thorngren}}]{DebrayLong2024}%
  \BibitemOpen
  \bibfield  {author} {\bibinfo {author} {\bibfnamefont {A.}~\bibnamefont
  {Debray}}, \bibinfo {author} {\bibfnamefont {S.~K.}\ \bibnamefont
  {Devalapurkar}}, \bibinfo {author} {\bibfnamefont {C.}~\bibnamefont
  {Krulewski}}, \bibinfo {author} {\bibfnamefont {Y.~L.}\ \bibnamefont {Liu}},
  \bibinfo {author} {\bibfnamefont {N.}~\bibnamefont {{Pacheco-Tallaj}}}, \
  and\ \bibinfo {author} {\bibfnamefont {R.}~\bibnamefont {Thorngren}},\ }\href
  {\doibase 10.48550/arXiv.2309.16749} {\enquote {\bibinfo {title} {A {{Long
  Exact Sequence}} in {{Symmetry Breaking}}: Order parameter constraints,
  defect anomaly-matching, and higher {{Berry}} phases},}\ } (\bibinfo {year}
  {2024}),\ \Eprint {http://arxiv.org/abs/2309.16749} {arXiv:2309.16749
  [hep-th]} \BibitemShut {NoStop}%
\bibitem [{\citenamefont {Beaudry}\ \emph {et~al.}(2024)\citenamefont
  {Beaudry}, \citenamefont {Hermele}, \citenamefont {Moreno}, \citenamefont
  {Pflaum}, \citenamefont {Qi},\ and\ \citenamefont
  {Spiegel}}]{BeaudryHomotopical2024}%
  \BibitemOpen
  \bibfield  {author} {\bibinfo {author} {\bibfnamefont {A.}~\bibnamefont
  {Beaudry}}, \bibinfo {author} {\bibfnamefont {M.}~\bibnamefont {Hermele}},
  \bibinfo {author} {\bibfnamefont {J.}~\bibnamefont {Moreno}}, \bibinfo
  {author} {\bibfnamefont {M.~J.}\ \bibnamefont {Pflaum}}, \bibinfo {author}
  {\bibfnamefont {M.}~\bibnamefont {Qi}}, \ and\ \bibinfo {author}
  {\bibfnamefont {D.~D.}\ \bibnamefont {Spiegel}},\ }\href {\doibase
  10.1142/S0129055X24600031} {\bibfield  {journal} {\bibinfo  {journal}
  {Reviews in Mathematical Physics}\ }\textbf {\bibinfo {volume} {36}},\
  \bibinfo {pages} {2460003} (\bibinfo {year} {2024})}\BibitemShut {NoStop}%
\bibitem [{\citenamefont {Beaudry}\ \emph {et~al.}(2025)\citenamefont
  {Beaudry}, \citenamefont {Hermele}, \citenamefont {Pflaum}, \citenamefont
  {Qi}, \citenamefont {Spiegel},\ and\ \citenamefont
  {Stephen}}]{BeaudryClassifying2025}%
  \BibitemOpen
  \bibfield  {author} {\bibinfo {author} {\bibfnamefont {A.}~\bibnamefont
  {Beaudry}}, \bibinfo {author} {\bibfnamefont {M.}~\bibnamefont {Hermele}},
  \bibinfo {author} {\bibfnamefont {M.~J.}\ \bibnamefont {Pflaum}}, \bibinfo
  {author} {\bibfnamefont {M.}~\bibnamefont {Qi}}, \bibinfo {author}
  {\bibfnamefont {D.~D.}\ \bibnamefont {Spiegel}}, \ and\ \bibinfo {author}
  {\bibfnamefont {D.~T.}\ \bibnamefont {Stephen}},\ }\href {\doibase
  10.48550/arXiv.2501.14241} {\enquote {\bibinfo {title} {A {{Classifying
  Space}} for {{Phases}} of {{Matrix Product States}}},}\ } (\bibinfo {year}
  {2025}),\ \Eprint {http://arxiv.org/abs/2501.14241} {arXiv:2501.14241
  [math-ph]} \BibitemShut {NoStop}%
\bibitem [{\citenamefont {Manjunath}\ and\ \citenamefont
  {Else}(2025)}]{ManjunathAnomalous2025}%
  \BibitemOpen
  \bibfield  {author} {\bibinfo {author} {\bibfnamefont {N.}~\bibnamefont
  {Manjunath}}\ and\ \bibinfo {author} {\bibfnamefont {D.~V.}\ \bibnamefont
  {Else}},\ }\href {\doibase 10.1103/PhysRevB.111.125151} {\bibfield  {journal}
  {\bibinfo  {journal} {Physical Review B}\ }\textbf {\bibinfo {volume}
  {111}},\ \bibinfo {pages} {125151} (\bibinfo {year} {2025})}\BibitemShut
  {NoStop}%
\bibitem [{\citenamefont {Jones}\ \emph {et~al.}(2025)\citenamefont {Jones},
  \citenamefont {Thorngren}, \citenamefont {Verresen},\ and\ \citenamefont
  {Prakash}}]{JonesCharge2025}%
  \BibitemOpen
  \bibfield  {author} {\bibinfo {author} {\bibfnamefont {N.~G.}\ \bibnamefont
  {Jones}}, \bibinfo {author} {\bibfnamefont {R.}~\bibnamefont {Thorngren}},
  \bibinfo {author} {\bibfnamefont {R.}~\bibnamefont {Verresen}}, \ and\
  \bibinfo {author} {\bibfnamefont {A.}~\bibnamefont {Prakash}},\ }\href
  {\doibase 10.48550/arXiv.2507.00995} {\enquote {\bibinfo {title} {Charge
  pumps, pivot {{Hamiltonians}} and symmetry-protected topological phases},}\ }
  (\bibinfo {year} {2025}),\ \Eprint {http://arxiv.org/abs/2507.00995}
  {arXiv:2507.00995 [cond-mat]} \BibitemShut {NoStop}%
\bibitem [{\citenamefont {Thouless}(1983)}]{ThoulessQuantization1983}%
  \BibitemOpen
  \bibfield  {author} {\bibinfo {author} {\bibfnamefont {D.~J.}\ \bibnamefont
  {Thouless}},\ }\href {\doibase 10.1103/PhysRevB.27.6083} {\bibfield
  {journal} {\bibinfo  {journal} {Physical Review B}\ }\textbf {\bibinfo
  {volume} {27}},\ \bibinfo {pages} {6083} (\bibinfo {year}
  {1983})}\BibitemShut {NoStop}%
\bibitem [{\citenamefont {Kapustin}\ and\ \citenamefont
  {Spodyneiko}(2020{\natexlab{b}})}]{KapustinHigherdimensional2020}%
  \BibitemOpen
  \bibfield  {author} {\bibinfo {author} {\bibfnamefont {A.}~\bibnamefont
  {Kapustin}}\ and\ \bibinfo {author} {\bibfnamefont {L.}~\bibnamefont
  {Spodyneiko}},\ }\href {\doibase 10.1103/PhysRevB.101.235130} {\bibfield
  {journal} {\bibinfo  {journal} {Physical Review B}\ }\textbf {\bibinfo
  {volume} {101}},\ \bibinfo {pages} {235130} (\bibinfo {year}
  {2020}{\natexlab{b}})}\BibitemShut {NoStop}%
\bibitem [{\citenamefont {Kapustin}\ and\ \citenamefont
  {Sopenko}(2022)}]{KapustinLocal2022}%
  \BibitemOpen
  \bibfield  {author} {\bibinfo {author} {\bibfnamefont {A.}~\bibnamefont
  {Kapustin}}\ and\ \bibinfo {author} {\bibfnamefont {N.}~\bibnamefont
  {Sopenko}},\ }\href {\doibase 10.1063/5.0085964} {\bibfield  {journal}
  {\bibinfo  {journal} {Journal of Mathematical Physics}\ }\textbf {\bibinfo
  {volume} {63}},\ \bibinfo {pages} {091903} (\bibinfo {year}
  {2022})}\BibitemShut {NoStop}%
\bibitem [{\citenamefont {Wen}\ \emph {et~al.}(2023)\citenamefont {Wen},
  \citenamefont {Qi}, \citenamefont {Beaudry}, \citenamefont {Moreno},
  \citenamefont {Pflaum}, \citenamefont {Spiegel}, \citenamefont {Vishwanath},\
  and\ \citenamefont {Hermele}}]{WenFlow2023}%
  \BibitemOpen
  \bibfield  {author} {\bibinfo {author} {\bibfnamefont {X.}~\bibnamefont
  {Wen}}, \bibinfo {author} {\bibfnamefont {M.}~\bibnamefont {Qi}}, \bibinfo
  {author} {\bibfnamefont {A.}~\bibnamefont {Beaudry}}, \bibinfo {author}
  {\bibfnamefont {J.}~\bibnamefont {Moreno}}, \bibinfo {author} {\bibfnamefont
  {M.~J.}\ \bibnamefont {Pflaum}}, \bibinfo {author} {\bibfnamefont
  {D.}~\bibnamefont {Spiegel}}, \bibinfo {author} {\bibfnamefont
  {A.}~\bibnamefont {Vishwanath}}, \ and\ \bibinfo {author} {\bibfnamefont
  {M.}~\bibnamefont {Hermele}},\ }\href {\doibase 10.1103/PhysRevB.108.125147}
  {\bibfield  {journal} {\bibinfo  {journal} {Physical Review B}\ }\textbf
  {\bibinfo {volume} {108}},\ \bibinfo {pages} {125147} (\bibinfo {year}
  {2023})}\BibitemShut {NoStop}%
\bibitem [{\citenamefont {Qi}\ \emph {et~al.}(2025)\citenamefont {Qi},
  \citenamefont {Stephen}, \citenamefont {Wen}, \citenamefont {Spiegel},
  \citenamefont {Pflaum}, \citenamefont {Beaudry},\ and\ \citenamefont
  {Hermele}}]{QiCharting2025}%
  \BibitemOpen
  \bibfield  {author} {\bibinfo {author} {\bibfnamefont {M.}~\bibnamefont
  {Qi}}, \bibinfo {author} {\bibfnamefont {D.}~\bibnamefont {Stephen}},
  \bibinfo {author} {\bibfnamefont {X.}~\bibnamefont {Wen}}, \bibinfo {author}
  {\bibfnamefont {D.}~\bibnamefont {Spiegel}}, \bibinfo {author} {\bibfnamefont
  {M.~J.}\ \bibnamefont {Pflaum}}, \bibinfo {author} {\bibfnamefont
  {A.}~\bibnamefont {Beaudry}}, \ and\ \bibinfo {author} {\bibfnamefont
  {M.}~\bibnamefont {Hermele}},\ }\href {\doibase
  10.21468/SciPostPhys.18.5.168} {\bibfield  {journal} {\bibinfo  {journal}
  {SciPost Physics}\ }\textbf {\bibinfo {volume} {18}},\ \bibinfo {pages} {168}
  (\bibinfo {year} {2025})}\BibitemShut {NoStop}%
\bibitem [{\citenamefont {Ohyama}\ and\ \citenamefont
  {Ryu}(2024)}]{OhyamaHigher2024}%
  \BibitemOpen
  \bibfield  {author} {\bibinfo {author} {\bibfnamefont {S.}~\bibnamefont
  {Ohyama}}\ and\ \bibinfo {author} {\bibfnamefont {S.}~\bibnamefont {Ryu}},\
  }\href {\doibase 10.1103/PhysRevB.109.115152} {\bibfield  {journal} {\bibinfo
   {journal} {Physical Review B}\ }\textbf {\bibinfo {volume} {109}},\ \bibinfo
  {pages} {115152} (\bibinfo {year} {2024})}\BibitemShut {NoStop}%
\bibitem [{\citenamefont {Shiozaki}\ \emph {et~al.}(2025)\citenamefont
  {Shiozaki}, \citenamefont {Heinsdorf},\ and\ \citenamefont
  {Ohyama}}]{ShiozakiHigher2025}%
  \BibitemOpen
  \bibfield  {author} {\bibinfo {author} {\bibfnamefont {K.}~\bibnamefont
  {Shiozaki}}, \bibinfo {author} {\bibfnamefont {N.}~\bibnamefont {Heinsdorf}},
  \ and\ \bibinfo {author} {\bibfnamefont {S.}~\bibnamefont {Ohyama}},\ }\href
  {\doibase 10.1103/1cbf-kxny} {\bibfield  {journal} {\bibinfo  {journal}
  {Physical Review B}\ }\textbf {\bibinfo {volume} {112}},\ \bibinfo {pages}
  {035154} (\bibinfo {year} {2025})}\BibitemShut {NoStop}%
\bibitem [{\citenamefont {Ohyama}\ and\ \citenamefont
  {Ryu}(2025{\natexlab{a}})}]{OhyamaHigher2025}%
  \BibitemOpen
  \bibfield  {author} {\bibinfo {author} {\bibfnamefont {S.}~\bibnamefont
  {Ohyama}}\ and\ \bibinfo {author} {\bibfnamefont {S.}~\bibnamefont {Ryu}},\
  }\href {\doibase 10.1103/PhysRevB.111.035121} {\bibfield  {journal} {\bibinfo
   {journal} {Physical Review B}\ }\textbf {\bibinfo {volume} {111}},\ \bibinfo
  {pages} {035121} (\bibinfo {year} {2025}{\natexlab{a}})}\BibitemShut
  {NoStop}%
\bibitem [{\citenamefont {Ohyama}\ and\ \citenamefont
  {Ryu}(2025{\natexlab{b}})}]{OhyamaHigher2025a}%
  \BibitemOpen
  \bibfield  {author} {\bibinfo {author} {\bibfnamefont {S.}~\bibnamefont
  {Ohyama}}\ and\ \bibinfo {author} {\bibfnamefont {S.}~\bibnamefont {Ryu}},\
  }\href {\doibase 10.1103/PhysRevB.111.045112} {\bibfield  {journal} {\bibinfo
   {journal} {Physical Review B}\ }\textbf {\bibinfo {volume} {111}},\ \bibinfo
  {pages} {045112} (\bibinfo {year} {2025}{\natexlab{b}})}\BibitemShut
  {NoStop}%
\bibitem [{\citenamefont {Sommer}\ \emph
  {et~al.}(2025{\natexlab{a}})\citenamefont {Sommer}, \citenamefont
  {Vishwanath},\ and\ \citenamefont {Wen}}]{SommerHigher2025}%
  \BibitemOpen
  \bibfield  {author} {\bibinfo {author} {\bibfnamefont {O.~E.}\ \bibnamefont
  {Sommer}}, \bibinfo {author} {\bibfnamefont {A.}~\bibnamefont {Vishwanath}},
  \ and\ \bibinfo {author} {\bibfnamefont {X.}~\bibnamefont {Wen}},\ }\href
  {\doibase 10.1103/PhysRevB.111.155110} {\bibfield  {journal} {\bibinfo
  {journal} {Physical Review B}\ }\textbf {\bibinfo {volume} {111}},\ \bibinfo
  {pages} {155110} (\bibinfo {year} {2025}{\natexlab{a}})}\BibitemShut
  {NoStop}%
\bibitem [{\citenamefont {Sommer}\ \emph
  {et~al.}(2025{\natexlab{b}})\citenamefont {Sommer}, \citenamefont {Wen},\
  and\ \citenamefont {Vishwanath}}]{SommerHigher2025a}%
  \BibitemOpen
  \bibfield  {author} {\bibinfo {author} {\bibfnamefont {O.~E.}\ \bibnamefont
  {Sommer}}, \bibinfo {author} {\bibfnamefont {X.}~\bibnamefont {Wen}}, \ and\
  \bibinfo {author} {\bibfnamefont {A.}~\bibnamefont {Vishwanath}},\ }\href
  {\doibase 10.1103/PhysRevLett.134.146601} {\bibfield  {journal} {\bibinfo
  {journal} {Physical Review Letters}\ }\textbf {\bibinfo {volume} {134}},\
  \bibinfo {pages} {146601} (\bibinfo {year} {2025}{\natexlab{b}})}\BibitemShut
  {NoStop}%
\bibitem [{\citenamefont {Wen}(2025)}]{WenSpace2025}%
  \BibitemOpen
  \bibfield  {author} {\bibinfo {author} {\bibfnamefont {X.}~\bibnamefont
  {Wen}},\ }\href {\doibase 10.48550/arXiv.2507.12546} {\enquote {\bibinfo
  {title} {Space of conformal boundary conditions from the view of higher
  {{Berry}} phase: {{Flow}} of {{Berry}} curvature in parametrized
  {{BCFTs}}},}\ } (\bibinfo {year} {2025}),\ \Eprint
  {http://arxiv.org/abs/2507.12546} {arXiv:2507.12546 [hep-th]} \BibitemShut
  {NoStop}%
\bibitem [{\citenamefont {Choi}\ \emph {et~al.}(2025)\citenamefont {Choi},
  \citenamefont {Ha}, \citenamefont {Kim}, \citenamefont {Kusuki},
  \citenamefont {Ohyama},\ and\ \citenamefont {Ryu}}]{ChoiHigher2025}%
  \BibitemOpen
  \bibfield  {author} {\bibinfo {author} {\bibfnamefont {Y.}~\bibnamefont
  {Choi}}, \bibinfo {author} {\bibfnamefont {H.}~\bibnamefont {Ha}}, \bibinfo
  {author} {\bibfnamefont {D.}~\bibnamefont {Kim}}, \bibinfo {author}
  {\bibfnamefont {Y.}~\bibnamefont {Kusuki}}, \bibinfo {author} {\bibfnamefont
  {S.}~\bibnamefont {Ohyama}}, \ and\ \bibinfo {author} {\bibfnamefont
  {S.}~\bibnamefont {Ryu}},\ }\href {\doibase 10.48550/arXiv.2507.12525}
  {\enquote {\bibinfo {title} {Higher {{Structures}} on {{Boundary Conformal
  Manifolds}}: {{Higher Berry Phase}} and {{Boundary Conformal Field
  Theory}}},}\ } (\bibinfo {year} {2025}),\ \Eprint
  {http://arxiv.org/abs/2507.12525} {arXiv:2507.12525 [hep-th]} \BibitemShut
  {NoStop}%
\bibitem [{\citenamefont {Kitaev}(2019)}]{KitaevDifferential2019}%
  \BibitemOpen
  \bibfield  {author} {\bibinfo {author} {\bibfnamefont {A.}~\bibnamefont
  {Kitaev}},\ }\href {https://web.ma.utexas. edu/topqft/talkslides/kitaev.pdf}
  {\enquote {\bibinfo {title} {Differential forms on the space of statistical
  mechanics models},}\ } (\bibinfo {year} {(2019)})\BibitemShut {NoStop}%
\bibitem [{\citenamefont {Viennot}\ and\ \citenamefont
  {Lages}(2011)}]{ViennotNew2011}%
  \BibitemOpen
  \bibfield  {author} {\bibinfo {author} {\bibfnamefont {D.}~\bibnamefont
  {Viennot}}\ and\ \bibinfo {author} {\bibfnamefont {J.}~\bibnamefont
  {Lages}},\ }\href {\doibase 10.1088/1751-8113/44/36/365301} {\bibfield
  {journal} {\bibinfo  {journal} {Journal of Physics A: Mathematical and
  Theoretical}\ }\textbf {\bibinfo {volume} {44}},\ \bibinfo {pages} {365301}
  (\bibinfo {year} {2011})}\BibitemShut {NoStop}%
\bibitem [{\citenamefont {Pollmann}\ \emph {et~al.}(2012)\citenamefont
  {Pollmann}, \citenamefont {Berg}, \citenamefont {Turner},\ and\ \citenamefont
  {Oshikawa}}]{PollmannSymmetry2012}%
  \BibitemOpen
  \bibfield  {author} {\bibinfo {author} {\bibfnamefont {F.}~\bibnamefont
  {Pollmann}}, \bibinfo {author} {\bibfnamefont {E.}~\bibnamefont {Berg}},
  \bibinfo {author} {\bibfnamefont {A.~M.}\ \bibnamefont {Turner}}, \ and\
  \bibinfo {author} {\bibfnamefont {M.}~\bibnamefont {Oshikawa}},\ }\href
  {\doibase 10.1103/PhysRevB.85.075125} {\bibfield  {journal} {\bibinfo
  {journal} {Physical Review B}\ }\textbf {\bibinfo {volume} {85}},\ \bibinfo
  {pages} {075125} (\bibinfo {year} {2012})}\BibitemShut {NoStop}%
\bibitem [{\citenamefont {Chen}\ \emph {et~al.}(2011)\citenamefont {Chen},
  \citenamefont {Gu},\ and\ \citenamefont {Wen}}]{ChenClassification2011}%
  \BibitemOpen
  \bibfield  {author} {\bibinfo {author} {\bibfnamefont {X.}~\bibnamefont
  {Chen}}, \bibinfo {author} {\bibfnamefont {Z.-C.}\ \bibnamefont {Gu}}, \ and\
  \bibinfo {author} {\bibfnamefont {X.-G.}\ \bibnamefont {Wen}},\ }\href
  {\doibase 10.1103/PhysRevB.83.035107} {\bibfield  {journal} {\bibinfo
  {journal} {Physical Review B}\ }\textbf {\bibinfo {volume} {83}},\ \bibinfo
  {pages} {035107} (\bibinfo {year} {2011})}\BibitemShut {NoStop}%
\bibitem [{\citenamefont {Schuch}\ \emph {et~al.}(2011)\citenamefont {Schuch},
  \citenamefont {{P{\'e}rez-Garc{\'i}a}},\ and\ \citenamefont
  {Cirac}}]{SchuchClassifying2011}%
  \BibitemOpen
  \bibfield  {author} {\bibinfo {author} {\bibfnamefont {N.}~\bibnamefont
  {Schuch}}, \bibinfo {author} {\bibfnamefont {D.}~\bibnamefont
  {{P{\'e}rez-Garc{\'i}a}}}, \ and\ \bibinfo {author} {\bibfnamefont
  {I.}~\bibnamefont {Cirac}},\ }\href {\doibase 10.1103/PhysRevB.84.165139}
  {\bibfield  {journal} {\bibinfo  {journal} {Physical Review B}\ }\textbf
  {\bibinfo {volume} {84}},\ \bibinfo {pages} {165139} (\bibinfo {year}
  {2011})}\BibitemShut {NoStop}%
\bibitem [{\citenamefont {Pollmann}\ and\ \citenamefont
  {Turner}(2012)}]{PollmannDetection2012}%
  \BibitemOpen
  \bibfield  {author} {\bibinfo {author} {\bibfnamefont {F.}~\bibnamefont
  {Pollmann}}\ and\ \bibinfo {author} {\bibfnamefont {A.~M.}\ \bibnamefont
  {Turner}},\ }\href {\doibase 10.1103/PhysRevB.86.125441} {\bibfield
  {journal} {\bibinfo  {journal} {Physical Review B}\ }\textbf {\bibinfo
  {volume} {86}},\ \bibinfo {pages} {125441} (\bibinfo {year}
  {2012})}\BibitemShut {NoStop}%
\bibitem [{\citenamefont {{Perez-Garcia}}\ \emph {et~al.}(2007)\citenamefont
  {{Perez-Garcia}}, \citenamefont {Verstraete}, \citenamefont {Wolf},\ and\
  \citenamefont {Cirac}}]{Perez-GarciaMatrix2007}%
  \BibitemOpen
  \bibfield  {author} {\bibinfo {author} {\bibfnamefont {D.}~\bibnamefont
  {{Perez-Garcia}}}, \bibinfo {author} {\bibfnamefont {F.}~\bibnamefont
  {Verstraete}}, \bibinfo {author} {\bibfnamefont {M.}~\bibnamefont {Wolf}}, \
  and\ \bibinfo {author} {\bibfnamefont {J.}~\bibnamefont {Cirac}},\ }\href
  {\doibase 10.26421/qic7.5-6-1} {\bibfield  {journal} {\bibinfo  {journal}
  {Quantum Information and Computation}\ }\textbf {\bibinfo {volume} {7}},\
  \bibinfo {pages} {401} (\bibinfo {year} {2007})}\BibitemShut {NoStop}%
\bibitem [{\citenamefont {Freed}\ and\ \citenamefont
  {Moore}(2013)}]{FreedTwisted2013}%
  \BibitemOpen
  \bibfield  {author} {\bibinfo {author} {\bibfnamefont {D.~S.}\ \bibnamefont
  {Freed}}\ and\ \bibinfo {author} {\bibfnamefont {G.~W.}\ \bibnamefont
  {Moore}},\ }\href {\doibase 10.1007/s00023-013-0236-x} {\bibfield  {journal}
  {\bibinfo  {journal} {Annales Henri Poincar{\'e}}\ }\textbf {\bibinfo
  {volume} {14}},\ \bibinfo {pages} {1927} (\bibinfo {year}
  {2013})}\BibitemShut {NoStop}%
\bibitem [{\citenamefont {Thiang}(2016)}]{ThiangKTheoretic2016}%
  \BibitemOpen
  \bibfield  {author} {\bibinfo {author} {\bibfnamefont {G.~C.}\ \bibnamefont
  {Thiang}},\ }\href {\doibase 10.1007/s00023-015-0418-9} {\bibfield  {journal}
  {\bibinfo  {journal} {Annales Henri Poincar{\'e}}\ }\textbf {\bibinfo
  {volume} {17}},\ \bibinfo {pages} {757} (\bibinfo {year} {2016})}\BibitemShut
  {NoStop}%
\bibitem [{\citenamefont {Shiozaki}\ \emph {et~al.}(2017)\citenamefont
  {Shiozaki}, \citenamefont {Sato},\ and\ \citenamefont
  {Gomi}}]{ShiozakiTopological2017}%
  \BibitemOpen
  \bibfield  {author} {\bibinfo {author} {\bibfnamefont {K.}~\bibnamefont
  {Shiozaki}}, \bibinfo {author} {\bibfnamefont {M.}~\bibnamefont {Sato}}, \
  and\ \bibinfo {author} {\bibfnamefont {K.}~\bibnamefont {Gomi}},\ }\href
  {\doibase 10.1103/PhysRevB.95.235425} {\bibfield  {journal} {\bibinfo
  {journal} {Physical Review B}\ }\textbf {\bibinfo {volume} {95}},\ \bibinfo
  {pages} {235425} (\bibinfo {year} {2017})}\BibitemShut {NoStop}%
\bibitem [{\citenamefont {Fukui}\ \emph {et~al.}(2005)\citenamefont {Fukui},
  \citenamefont {Hatsugai},\ and\ \citenamefont {Suzuki}}]{FukuiChern2005}%
  \BibitemOpen
  \bibfield  {author} {\bibinfo {author} {\bibfnamefont {T.}~\bibnamefont
  {Fukui}}, \bibinfo {author} {\bibfnamefont {Y.}~\bibnamefont {Hatsugai}}, \
  and\ \bibinfo {author} {\bibfnamefont {H.}~\bibnamefont {Suzuki}},\ }\href
  {\doibase 10.1143/JPSJ.74.1674} {\bibfield  {journal} {\bibinfo  {journal}
  {Journal of the Physical Society of Japan}\ }\textbf {\bibinfo {volume}
  {74}},\ \bibinfo {pages} {1674} (\bibinfo {year} {2005})}\BibitemShut
  {NoStop}%
\bibitem [{\citenamefont {Fang}\ \emph {et~al.}(2012)\citenamefont {Fang},
  \citenamefont {Gilbert},\ and\ \citenamefont {Bernevig}}]{FangBulk2012}%
  \BibitemOpen
  \bibfield  {author} {\bibinfo {author} {\bibfnamefont {C.}~\bibnamefont
  {Fang}}, \bibinfo {author} {\bibfnamefont {M.~J.}\ \bibnamefont {Gilbert}}, \
  and\ \bibinfo {author} {\bibfnamefont {B.~A.}\ \bibnamefont {Bernevig}},\
  }\href {\doibase 10.1103/PhysRevB.86.115112} {\bibfield  {journal} {\bibinfo
  {journal} {Physical Review B}\ }\textbf {\bibinfo {volume} {86}},\ \bibinfo
  {pages} {115112} (\bibinfo {year} {2012})}\BibitemShut {NoStop}%
\bibitem [{\citenamefont {Freed}(1986)}]{FreedDeterminants1986}%
  \BibitemOpen
  \bibfield  {author} {\bibinfo {author} {\bibfnamefont {D.~S.}\ \bibnamefont
  {Freed}},\ }\href {\doibase 10.1007/BF01221001} {\bibfield  {journal}
  {\bibinfo  {journal} {Communications in Mathematical Physics}\ }\textbf
  {\bibinfo {volume} {107}},\ \bibinfo {pages} {483} (\bibinfo {year}
  {1986})}\BibitemShut {NoStop}%
\bibitem [{\citenamefont {Cirac}\ \emph {et~al.}(2021)\citenamefont {Cirac},
  \citenamefont {{P{\'e}rez-Garc{\'i}a}}, \citenamefont {Schuch},\ and\
  \citenamefont {Verstraete}}]{CiracMatrix2021}%
  \BibitemOpen
  \bibfield  {author} {\bibinfo {author} {\bibfnamefont {J.~I.}\ \bibnamefont
  {Cirac}}, \bibinfo {author} {\bibfnamefont {D.}~\bibnamefont
  {{P{\'e}rez-Garc{\'i}a}}}, \bibinfo {author} {\bibfnamefont {N.}~\bibnamefont
  {Schuch}}, \ and\ \bibinfo {author} {\bibfnamefont {F.}~\bibnamefont
  {Verstraete}},\ }\href {\doibase 10.1103/RevModPhys.93.045003} {\bibfield
  {journal} {\bibinfo  {journal} {Reviews of Modern Physics}\ }\textbf
  {\bibinfo {volume} {93}},\ \bibinfo {pages} {045003} (\bibinfo {year}
  {2021})}\BibitemShut {NoStop}%
\bibitem [{\citenamefont {{P{\'e}rez-Garc{\'i}a}}\ \emph
  {et~al.}(2008)\citenamefont {{P{\'e}rez-Garc{\'i}a}}, \citenamefont {Wolf},
  \citenamefont {Sanz}, \citenamefont {Verstraete},\ and\ \citenamefont
  {Cirac}}]{Perez-GarciaString2008}%
  \BibitemOpen
  \bibfield  {author} {\bibinfo {author} {\bibfnamefont {D.}~\bibnamefont
  {{P{\'e}rez-Garc{\'i}a}}}, \bibinfo {author} {\bibfnamefont {M.~M.}\
  \bibnamefont {Wolf}}, \bibinfo {author} {\bibfnamefont {M.}~\bibnamefont
  {Sanz}}, \bibinfo {author} {\bibfnamefont {F.}~\bibnamefont {Verstraete}}, \
  and\ \bibinfo {author} {\bibfnamefont {J.~I.}\ \bibnamefont {Cirac}},\ }\href
  {\doibase 10.1103/PhysRevLett.100.167202} {\bibfield  {journal} {\bibinfo
  {journal} {Physical Review Letters}\ }\textbf {\bibinfo {volume} {100}},\
  \bibinfo {pages} {167202} (\bibinfo {year} {2008})}\BibitemShut {NoStop}%
\bibitem [{\citenamefont {Shiozaki}\ and\ \citenamefont
  {Ryu}(2017)}]{ShiozakiMatrix2017}%
  \BibitemOpen
  \bibfield  {author} {\bibinfo {author} {\bibfnamefont {K.}~\bibnamefont
  {Shiozaki}}\ and\ \bibinfo {author} {\bibfnamefont {S.}~\bibnamefont {Ryu}},\
  }\href {\doibase 10.1007/JHEP04(2017)100} {\bibfield  {journal} {\bibinfo
  {journal} {Journal of High Energy Physics}\ }\textbf {\bibinfo {volume}
  {2017}},\ \bibinfo {pages} {100} (\bibinfo {year} {2017})}\BibitemShut
  {NoStop}%
\bibitem [{\citenamefont {Berg}\ \emph {et~al.}(2011)\citenamefont {Berg},
  \citenamefont {Levin},\ and\ \citenamefont {Altman}}]{BergQuantized2011}%
  \BibitemOpen
  \bibfield  {author} {\bibinfo {author} {\bibfnamefont {E.}~\bibnamefont
  {Berg}}, \bibinfo {author} {\bibfnamefont {M.}~\bibnamefont {Levin}}, \ and\
  \bibinfo {author} {\bibfnamefont {E.}~\bibnamefont {Altman}},\ }\href
  {\doibase 10.1103/PhysRevLett.106.110405} {\bibfield  {journal} {\bibinfo
  {journal} {Physical Review Letters}\ }\textbf {\bibinfo {volume} {106}},\
  \bibinfo {pages} {110405} (\bibinfo {year} {2011})}\BibitemShut {NoStop}%
\bibitem [{\citenamefont {Rossini}\ \emph {et~al.}(2013)\citenamefont
  {Rossini}, \citenamefont {Gibertini}, \citenamefont {Giovannetti},\ and\
  \citenamefont {Fazio}}]{RossiniTopological2013}%
  \BibitemOpen
  \bibfield  {author} {\bibinfo {author} {\bibfnamefont {D.}~\bibnamefont
  {Rossini}}, \bibinfo {author} {\bibfnamefont {M.}~\bibnamefont {Gibertini}},
  \bibinfo {author} {\bibfnamefont {V.}~\bibnamefont {Giovannetti}}, \ and\
  \bibinfo {author} {\bibfnamefont {R.}~\bibnamefont {Fazio}},\ }\href
  {\doibase 10.1103/PhysRevB.87.085131} {\bibfield  {journal} {\bibinfo
  {journal} {Physical Review B}\ }\textbf {\bibinfo {volume} {87}},\ \bibinfo
  {pages} {085131} (\bibinfo {year} {2013})}\BibitemShut {NoStop}%
\bibitem [{\citenamefont {Kuno}\ and\ \citenamefont
  {Hatsugai}(2020)}]{KunoInteractioninduced2020}%
  \BibitemOpen
  \bibfield  {author} {\bibinfo {author} {\bibfnamefont {Y.}~\bibnamefont
  {Kuno}}\ and\ \bibinfo {author} {\bibfnamefont {Y.}~\bibnamefont
  {Hatsugai}},\ }\href {\doibase 10.1103/PhysRevResearch.2.042024} {\bibfield
  {journal} {\bibinfo  {journal} {Physical Review Research}\ }\textbf {\bibinfo
  {volume} {2}},\ \bibinfo {pages} {042024} (\bibinfo {year}
  {2020})}\BibitemShut {NoStop}%
\bibitem [{\citenamefont {Bi}\ \emph {et~al.}(2015)\citenamefont {Bi},
  \citenamefont {Rasmussen}, \citenamefont {Slagle},\ and\ \citenamefont
  {Xu}}]{BiClassification2015}%
  \BibitemOpen
  \bibfield  {author} {\bibinfo {author} {\bibfnamefont {Z.}~\bibnamefont
  {Bi}}, \bibinfo {author} {\bibfnamefont {A.}~\bibnamefont {Rasmussen}},
  \bibinfo {author} {\bibfnamefont {K.}~\bibnamefont {Slagle}}, \ and\ \bibinfo
  {author} {\bibfnamefont {C.}~\bibnamefont {Xu}},\ }\href {\doibase
  10.1103/PhysRevB.91.134404} {\bibfield  {journal} {\bibinfo  {journal}
  {Physical Review B}\ }\textbf {\bibinfo {volume} {91}},\ \bibinfo {pages}
  {134404} (\bibinfo {year} {2015})}\BibitemShut {NoStop}%
\end{thebibliography}%

\end{document}